\documentclass[12pt,a4paper]{article}
\pdfoutput=1
\usepackage{jheppub,psfrag,slashed,cancel,lscape,caption,array,graphicx,subcaption}
\usepackage[utf8]{inputenc}
\usepackage{amsmath}
\usepackage{mathtools}
\usepackage{mathrsfs}
\usepackage{multirow}
\usepackage{booktabs} 

\DeclareFontFamily{OT1}{pzc}{}
\DeclareFontShape{OT1}{pzc}{m}{it}{<-> s * [1.350] pzcmi7t}{}
\DeclareMathAlphabet{\mathpzc}{OT1}{pzc}{m}{it}

\definecolor{myblue}{rgb}{0,0,0.7}

\definecolor{mygreen}{rgb}{0,0.4,0}

\definecolor{myred}{rgb}{0.4,0,0}

\def\cN{\mathcal{N}}
\def\cO{\mathcal{O}}

\def\cV{\mathcal{V}}
\def\cH{\mathcal{H}}

\def\cT{\mathcal{T}}
\def\cM{\mathcal{M}}
\def\cA{\mathcal{A}}
\def\tf{\tilde{f}}

\def\d{\mathrm{d}}

\def\nn{\nonumber}
\def\etat{\tilde{\eta}}
\def\etab{\bar{\eta}}
\def\mt{\widetilde{m}}
\def\lambdat{\tilde{\lambda}}
\newcommand{\widebar}{\overline}

\DeclareMathOperator{\tr}{\rm tr}

\def\eps{\epsilon}

\def\eqn#1{eq.~(\ref{#1})}

\def\sec#1{section~{\ref{#1}}}

\def\braket#1{\langle #1 \rangle}

\def\usegraph#1#2{\includegraphics[scale=1.0,trim=0 #1 0 0]{graphs/#2.pdf}}

\preprint{UUITP-19/17 \\
\phantom{~} \hfill NORDITA 2017-062 \\
\phantom{~} \hfill Edinburgh 2017/12 }

\title{Two-loop supersymmetric QCD and half-maximal supergravity amplitudes}

\author[a,b]{Henrik Johansson,}
\author[a]{Gregor K\"{a}lin}
\author[a,c]{and Gustav Mogull}
\affiliation[a]{Department of Physics and Astronomy, Uppsala University, 75108 Uppsala, Sweden}
\affiliation[b]{Nordita, Stockholm University and KTH Royal Institute of Technology,\\ Roslagstullsbacken 23, 10691 Stockholm, Sweden}
\affiliation[c]{Higgs Centre for Theoretical Physics, School of Physics and Astronomy,\\ The University of Edinburgh, Edinburgh EH9 3FD, Scotland, UK}
\emailAdd{henrik.johansson@physics.uu.se}
\emailAdd{gregor.kaelin@physics.uu.se}
\emailAdd{g.mogull@ed.ac.uk}

\abstract{Using the duality between color and kinematics, we construct two-loop four-point scattering amplitudes in $\cN=2$ super-Yang-Mills (SYM) theory coupled to $N_f$ fundamental hypermultiplets.  Our results are valid in $D\le 6$ dimensions, where the upper bound corresponds to six-dimensional chiral $\cN=(1,0)$ SYM theory.  By exploiting a close connection with $\cN=4$ SYM theory --- and, equivalently, six-dimensional $\cN=(1,1)$ SYM theory --- we find compact integrands with four-dimensional external vectors in both the maximally-helicity-violating (MHV) and all-chiral-vector sectors.  Via the double-copy construction corresponding $D$-dimensional half-maximal supergravity amplitudes with external graviton multiplets are obtained in the MHV and all-chiral sectors.  Appropriately tuning $N_f$ enables us to consider both pure and matter-coupled supergravity, with arbitrary numbers of vector multiplets in $D=4$.  As a bonus, we obtain the integrands of the genuinely six-dimensional supergravities with $\cN=(1,1)$ and $\cN=(2,0)$ supersymmetry. Finally, we extract the potential ultraviolet divergence of half-maximal supergravity in $D=5-2\eps$ and show that it non-trivially cancels out as expected.}


\begin{document}
\maketitle
\flushbottom

\section{Introduction}

It is by now well established that a variety of gauge and gravity theories are perturbatively related through the so-called double copy. In terms of the asymptotic states a squaring relation between gauge theory and gravity follows readily from representation theory;  that such a structure is preserved by the interactions is a remarkable fact first brought to light by the Kawai-Lewellen-Tye (KLT) relations~\cite{Kawai:1985xq} between open and closed string amplitudes.  More recently, the double copy has been understood to arise due to the Bern-Carrasco-Johansson (BCJ) duality between color and kinematics~\cite{Bern:2008qj, Bern:2010ue} that is present in many familiar gauge theories. This realization has opened the path towards constructions of scattering amplitudes in a multitude of different gravity theories~\cite{Bern:2010ue, Carrasco:2012ca, Chiodaroli:2013upa, Johansson:2014zca, Chiodaroli:2014xia, Chiodaroli:2015rdg, Chiodaroli:2015wal, Anastasiou:2016csv} starting from the much simpler gauge-theory amplitudes. 

The duality between color and kinematics refers to the observation that many familiar gauge theories have a hidden kinematic structure that mirrors that of the gauge-group color structure~\cite{Bern:2008qj,Bern:2010ue}. Amplitudes in, for example, pure Yang-Mills or super-Yang-Mills (SYM) theories in generic dimensions can be brought to forms where the kinematic numerators of individual diagrams obey Jacobi relations in complete analogue with the color factors of the same diagrams. A natural expectation is that one (or several) unknown kinematic Lie algebras underlie the duality~\cite{Monteiro:2011pc, Cheung:2016prv}. At tree level the duality is known to be equivalent to the existence of BCJ relations between partial amplitudes~\cite{Bern:2008qj}, which in turn have been proven for pure SYM theories through a variety of different techniques~\cite{BjerrumBohr:2009rd,Stieberger:2009hq,Feng:2010my, Chen:2011jxa, Barreiro:2013dpa}. As of yet there is no proof of the duality at loop level, although a number of calculations have established its presence up to four loops in the maximally-supersymmetric $\cN=4$ SYM theory~\cite{Bern:2010ue,Carrasco:2011mn,Bern:2011rj,Bern:2012uf,Bjerrum-Bohr:2013iza, He:2015wgf, Mafra:2015mja}, up to two loops for pure YM~\cite{Bern:2013yya}, and at one loop in matter-coupled YM theories with reduced supersymmetry~\cite{Carrasco:2012ca, Nohle:2013bfa, Chiodaroli:2013upa, Johansson:2014zca, Chiodaroli:2014xia, Berg:2016wux}.

Color-kinematics duality has been shown to be present in weakly-coupled quantum chromodynamics (QCD) and generalizations thereof~\cite{Johansson:2014zca,Johansson:2015oia}; this includes Yang-Mills theory coupled to massive quarks in any dimension, and corresponding supersymmetric extensions. The BCJ amplitude relations for QCD were worked out in ref.~\cite{Johansson:2015oia} and proven in ref.~\cite{delaCruz:2015dpa}. Using the duality for practical calculations in QCD phenomenology is a promising avenue as it poses strong constraints on the diagrammatic form of an amplitude, interweaving planar and non-planar contributions to the point of trivializing certain steps of the calculation~\cite{Badger:2015lda, Mogull:2015adi, Chester:2016ojq}. In the limit of massless quarks the (super-)QCD amplitudes are a crucial ingredient in the double-copy construction of pure $\cN=0,1,2,3$ supergravities in four dimensions \cite{Johansson:2014zca}, as well as for the symmetric, magical and homogeneous $\cN=2$ supergravities~\cite{Chiodaroli:2015wal}, and twin supergravities \cite{Anastasiou:2016csv}. 

The double copy is most straightforwardly understood as replacing the color factors in a gauge-theory amplitude by corresponding kinematic numerators~\cite{Bern:2008qj,Bern:2010ue}. When color-kinematics duality is present this replacement respects the Lie-algebraic structure of the amplitude, and furthermore enhances the gauge symmetry to diffeomorphism symmetry~\cite{Chiodaroli:2016jqw, Chiodaroli:2017ngp, Arkani-Hamed:2016rak}. Additional enhanced global symmetries typically also arise~\cite{Borsten:2013bp,Anastasiou:2013cya,Anastasiou:2013hba,Anastasiou:2014qba,Anastasiou:2015vba}. The resulting amplitudes are expressed in terms of Feynman-like diagrams with two copies of kinematical numerators, and describe scattering of spin-2 states in some gravitational theory. The double copy gives valid gravity amplitudes even if the kinematic numerators are drawn from two different gauge theories~\cite{Bern:2008qj}, as long as the two theories satisfy the same type of kinematic Lie algebra~\cite{Chiodaroli:2014xia} (with at least one copy manifestly so~\cite{Bern:2010ue,Bern:2010yg}). This flexibility of combining pairs of different gauge theories has given rise to a cornucopia of double-copy constructions~\cite{Bern:2010ue, Bern:2011rj,BoucherVeronneau:2011qv,Broedel:2012rc,Carrasco:2012ca, Bern:2013uka, Chiodaroli:2013upa, Bern:2014sna,Bern:2014lha, Johansson:2014zca, Chiodaroli:2014xia, Chiodaroli:2015rdg, Chiodaroli:2015wal, Anastasiou:2016csv, Bern:2017yxu}. 

More recently the double copy and color-kinematics duality has been observed to extend to effective theories such as the non-linear-sigma model (chiral Lagrangian), (Dirac-)Born-Infeld, Volkov-Akulov and special galileon theory~\cite{Chen:2013fya,Cheung:2014dqa,Cachazo:2014xea,Du:2016tbc,Cheung:2016prv,Carrasco:2016ldy}. Likewise, the double copy and duality show up in novel relations involving and interconnecting string- and field-theory amplitudes and effective theories~\cite{Mafra:2011nw,Mafra:2012kh,Broedel:2013tta,Huang:2016tag, Carrasco:2016ldy,Carrasco:2016ygv,Tourkine:2016bak,Hohenegger:2017kqy,He:2016mzd, He:2017spx, Cheung:2017ems}. The double copy has been extended beyond perturbation theory to classical solutions involving black holes~\cite{Monteiro:2014cda, Luna:2015paa, Ridgway:2015fdl, Cardoso:2016amd} and gravitational-wave radiation~\cite{Luna:2016due,Luna:2016hge, Goldberger:2016iau, Goldberger:2017frp}. 

In the work by Ochirov and one of the current authors~\cite{Johansson:2014zca} a detailed prescription was given for removing unwanted axion-dilaton-like states that appear in a naive double-copy construction of loop-level amplitudes of pure $\cN=0,1,2,3$ supergravities in $D=4$ dimensions. The prescription calls for the introduction of compensating ghost states that are obtained by tensoring fundamental matter multiplets of the two gauge theories entering the double copy. For consistency of the double copy, the kinematic numerators of the gauge theories need to obey Jacobi identities and commutation relations mirroring the adjoint- and fundamental-representation color algebra. The gauge theories are thus different varieties of massless (super-)QCD theories with a tunable parameter $N_f$ corresponding the number of quark flavors. Specifically, the gravitational matter is turned into ghosts by choosing $N_f=-1$ and $N_f=1$ for the two gauge theories, respectively. The prescription was shown to correctly reproduce the one-loop amplitudes in pure $\cN=0,1,2,3,4$ supergravity. 

In this paper, we obtain the two-loop amplitudes in $\cN=2$ super-QCD (SQCD) which are needed for the computation of pure supergravity amplitudes at this loop order. Using color-kinematics duality we compute the four-vector amplitude at two loops with contributions from $N_f$ internal hypermultiplets transforming in the fundamental representation. The amplitudes are valid for any gauge group $G$ and for any dimension $D\le6$. Dimensional regularization, in $D=4-2\epsilon$ dimensions, requires that the integrands are correct even for $D>4$ dimensions, and $D=6$ is the maximal uplift where the theory exists. To define the six-dimensional chiral theory, $\cN=(1,0)$ SYM coupled to hypers, we make use of its close relationship to the maximally supersymmetric $\cN=(1,1)$ SYM theory, whose tree amplitudes are conveniently written down using the six-dimensional version of spinor-helicity notation~\cite{Cheung:2009dc}.  

Using the two-loop amplitudes in $\cN=2$ SQCD, which obey color-kinematics duality, we compute the $(\cN=2, N_f) \otimes (\cN=2, N_f=1)$ double copy and obtain two-loop amplitudes in $\cN=4$ supergravity coupled to $2(N_f+1)$ vector multiplets. For the choice $N_f=-1$ we obtain pure $\cN=4$ supergravity amplitudes. It should be noted that pure $\cN=4$ supergravity amplitudes can alternatively be obtained from a $(\cN=4) \otimes (\cN=0)$ double copy~\cite{Bern:2011rj,BoucherVeronneau:2011qv,Bern:2012cd,Bern:2012gh,Bern:2013qca,Bern:2013uka,Bern:2014lha,Bern:2017lpv}, without the need of removing extraneous states. Indeed, the two-loop four-point $\cN=4$ supergravity amplitude was first computed this way in ref.~\cite{BoucherVeronneau:2011qv}, and with additional vector multiplets in ref.~\cite{Bern:2013qca}.

Our construction of the half-maximal supergravity amplitudes has several crucial virtues compared to previous work. Obtaining the $\cN=2$ SQCD amplitude in the process is an obvious bonus; this amplitude is a stepping stone for computing the pure $\cN=2,3$ supergravity two-loop amplitudes, as explained in ref.~\cite{Johansson:2014zca}. Furthermore, in $D=6$ the chiral nature of the $\cN=(1,0)$ SYM theory allows for the double-copy construction of two inequivalent half-maximal supergravity amplitudes: $\cN=(1,1)$ and $\cN=(2,0)$ supergravity. Only the former is equivalent to the $(\cN=4) \otimes (\cN=0)$ construction. The latter theory, pure $\cN=(2,0)$ supergravity, has a chiral gravitational anomaly that shows up at one loop, thus rendering the double-copy construction of the two-loop amplitude potentially inconsistent in $D=6$. However, it is known that the anomaly can be canceled by adding 21 self-dual $\cN=(2,0)$ tensor multiplets to the theory~\cite{Seiberg:1988pf}. This corresponds to $N_f=10$ in the current context. 

We extract the ultraviolet (UV) divergences of the half-maximal supergravity two-loop integrals in $D=4$ and $D=5$. In $D=4$ the integrals are manifestly free of divergences by inspection of the power counting, whereas in $D=5$ the individual integrals do diverge but after summing over all contributions to the amplitude divergences cancel out; thus exhibiting so-called enhanced UV cancellations~\cite{Bern:2014sna, Bern:2017lpv}. This non-trivially agrees with the calculations of the UV divergence of the same amplitudes in ref.~\cite{Bern:2013qca}, and thus provides an independent cross-check of our construction. While we do not compute the $D=6-2\epsilon$ divergences, we note that the pure $\cN=(1,1)$ theory should have a divergence starting already at one loop, thus the two-loop amplitude has both $1/\epsilon^2$ and $1/\epsilon$ poles~\cite{Bern:2013qca}. 

This paper is organized as follows: in \sec{sec:review}, we review color-kinematics duality and the double copy with special attention on the fundamental matter case, and pure supergravities.  In \sec{sec:treesAndCuts}, we introduce the tree amplitudes and needed two-loop cuts for $\cN=2$ SQCD both in four and six dimensions.  In \sec{sec:SQCDcalc}, the details of the calculation of the two-loop numerators of SQCD are presented.  In \sec{sec:HMSG}, the corresponding supergravity amplitudes are assembled and the UV divergence in five dimensions is computed. Conclusions are presented in \sec{sec:conclusions} and the explicit $\cN=2$ SQCD numerators and color factors are given in appendix \ref{sec:MHV_sol}.  

\section{Review}
\label{sec:review}

Here we review color-kinematics duality in supersymmetric Yang-Mills (SYM) theory and its extension to fundamental matter, following closely the discussion in ref.~\cite{Johansson:2014zca}.  We also show how the duality may be used to compute amplitudes in a wide variety of (super-)gravity theories using the double copy prescription.

\subsection{Color-kinematics duality with fundamental matter}

We begin by writing $L$-loop (supersymmetric) Yang-Mills amplitudes with fundamental matter as sums of trivalent graphs:
\begin{equation}
  \label{eq:colorKinDualAmp}
  \cA_m^{(L)}=i^{L-1}g_{\text{YM}}^{m+2L-2}\sum_{\text{cubic graphs }\Gamma_i}\int\!\frac{\d^{LD}\ell}{(2\pi)^{LD}}\frac{1}{S_i} \frac{n_i c_i}{D_i},
\end{equation}
where $g_\text{YM}$ is the coupling.  The structure of the gauge group $G$ allows two types of trivalent vertices: pure-adjoint, and those with a particle in each of the adjoint, fundamental and anti-fundamental representations.  The color factors $c_i$ may thus be expressed as products of structure constants~$\tf^{abc}=\tr([T^a,T^b]T^c)$ and generators~$T^a_{i\bar\jmath}$, normalized such that $\tr({T^aT^b}) = \delta^{ab}$.  Diagrammatically, these can be represented as
\begin{align}
  \tf^{abc}=c\bigg(\!\usegraph{14}{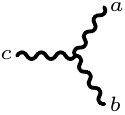}\bigg), &&
  T^a_{i\bar\jmath}=c\bigg(\!\usegraph{14}{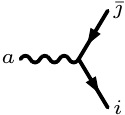}\bigg).
\end{align}
$S_i$ and $D_i$ are the usual symmetry factors and products of propagators respectively.  Finally, kinematic numerators $n_i$ collect kinematic information about the graphs.

The algebraic structure of the gauge group gives rise to linear relationships between the color factors.  In the adjoint representation these are Jacobi identities between the structure constants,\footnote{This may also be viewed as a commutation relation in the adjoint representation $(T^a_\text{adj})^{bc}=\tf^{bac}$.}
\begin{align}\label{eq:jacobi1}
  \begin{aligned}
    \tf^{a_1a_2b}\tf^{ba_3a_4}&=\tf^{a_4a_1b}\tf^{ba_2a_3}-\tf^{a_2a_4b}\tf^{ba_3a_1},\\
    c\bigg(\usegraph{12}{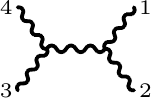}\bigg)&=c\bigg(\usegraph{13}{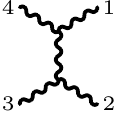}\bigg)-c\bigg(\!\usegraph{13}{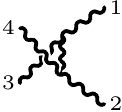}\bigg),\\
  \end{aligned}
\end{align}
and in the fundamental representation they are commutation relations,
\begin{align}\label{eq:jacobi2}
  \begin{aligned}
    T^b_{i_1\bar\imath_2}\tf^{ba_3a_4}
    &=T^{a_3}_{i_1\bar\jmath}T^{a_4}_{j\bar\imath_2}-T^{a_4}_{i_1\bar\jmath}T^{a_3}_{j\bar\imath_2} =[T^{a_3},T^{a_4}]_{i_1\bar\imath_2},\\
    c\bigg(\usegraph{12}{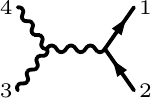}\bigg)&=c\bigg(\usegraph{13}{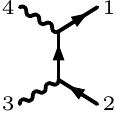}\bigg)-c\bigg(\!\usegraph{13}{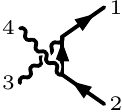}\bigg).
  \end{aligned}
\end{align}
The identities lead to relationships between the color factors of the form
\begin{equation}
  c_i = c_j - c_k.
\end{equation}
Furthermore, the flip of two legs at a pure-adjoint vertex leads to an overall sign flip, $\tf^{abc}=-\tf^{acb}$.  The same behavior can be implemented when swapping fundamental with anti-fundamental legs by defining new generators $T^a_{\bar\imath j}=-T^a_{j\bar\imath}$.

Kinematic numerators $n_i$ satisfying color-kinematics duality --- we shall refer to these as color-dual --- obey the same linear relations as their respective color factors.  We demand that
\begin{equation}
  \begin{aligned}
    n_i = n_j - n_k & \quad \Leftrightarrow \quad c_i = c_j - c_k, \\
    n_i \rightarrow -n_i & \quad \Leftrightarrow \quad c_i \rightarrow -c_i.
  \end{aligned}
\end{equation}
The second identity incorporates a sign change for a flip of two legs at a given cubic vertex. 

The existence of a set of color-dual numerators is generally non-trivial. At tree level, their existence has been proven in $D$-dimensional pure (super-)Yang-Mills theories using an inversion formula between numerators and color-ordered amplitudes~\cite{Kiermaier,BjerrumBohr:2010hn}.  This relies on the partial amplitudes satisfying BCJ relations~\cite{Bern:2008qj,BjerrumBohr:2009rd,Stieberger:2009hq,Feng:2010my}. For (super-)Yang-Mills theories with arbitrary fundamental matter, where less is known, the corresponding BCJ relations~\cite{Johansson:2014zca} have been proven for QCD~\cite{delaCruz:2015dpa}.

When including matter multiplets it is also useful to consider the two-term identity
\begin{align}
  \label{eq:twoTerm}
  c\bigg(\usegraph{12}{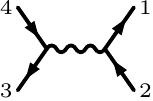}\bigg) \stackrel{?}{=} c\bigg(\usegraph{13}{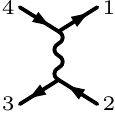}\bigg)
\end{align}
for indistinguishable matter multiplets.  Depending on the gauge-group representation, this may or may not hold.  Although fundamental representations of generic gauge groups do not obey this relation, the generator of U(1) does, as well as generators of particular tensor representations of U($N_c$).  It turns out that it is possible to find color-dual numerators that fulfill these two-term identities.  However, the two-term identities are not necessary for the double-copy prescription~\cite{Bern:2010ue} that we will now discuss.

\subsection{Pure supergravities from the double-copy prescription}\label{sec:doublecopyreview}

Our main motivation for finding color-dual numerators of gauge-theory amplitudes is the associated double-copy prescription: this offers a simple way to obtain gravity amplitudes in a variety of (supersymmetric) theories.  The inclusion of fundamental-matter multiplets widens the class of gravity amplitudes obtainable in this way; in particular, it enables the double-copy construction of pure $\cN<4$ supergravity amplitudes, including $\cN=0$ Einstein gravity~\cite{Johansson:2014zca}.  In this paper we will mostly be interested in pure $\cN=4$ (half-maximal) supergravity at two loops; however, the approach that we now review also applies more generally.

The basic setup is as follows.  Take a pair of $\cN,\cM\leq2$ SYM multiplets, $\cV_\cN$ and $\cV'_{\cM}$, both of which are non-chiral --- for instance, the former should include $2^\cN$ chiral and $2^\cN$ anti-chiral on-shell states.  Then analyze their double copy: one recovers a \emph{factorizable} graviton multiplet of the form
\begin{equation}
\label{FactM}
  \cV_\cN \otimes \cV'_\cM \equiv \mathcal{H}_{\cN + \cM} \oplus X_{\cN + \cM} \oplus \widebar{X}_{\cN + \cM},
\end{equation}
where $\mathcal{H}_{\cN+\cM}$ is a (non-chiral) pure-graviton multiplet of interest.  The unwanted gravitational matter and antimatter multiplets, $X$ and $\widebar{X}$, can be canceled by a double copy between complex matter multiplets $\Phi_\cN$ and $\Phi'_\cM$:
\begin{equation}
  \begin{aligned}
  \label{ghostsM}
    X_{\cN + \cM} &\equiv \Phi_\cN \otimes \widebar{\Phi}'_\cM,\\
    \widebar{X}_{\cN + \cM} &\equiv \widebar{\Phi}_\cN \otimes \Phi'_\cM.
  \end{aligned}
\end{equation}
In this paper $\cN=\cM=2$; the gauge-theory matter multiplets are hypermultiplets (hypers). The complex matter $\Phi_\cN$ (and the conjugate $\widebar{\Phi}_\cN$) belong to the fundamental (antifundamental) representation of the gauge group $G$.  In order to cancel the matter in \eqn{FactM} one assigns opposite statistics to $\Phi_\cN$ (but not to $\Phi'_\cM$), which in turn means that $X$, $\widebar{X}$ in \eqn{ghostsM} are ghosts.

The usual numerator double copy is modified when concealing unwanted matter states. Starting from a color-dual presentation of the SYM amplitude~(\ref{eq:colorKinDualAmp}) one replaces the color factors~$c_i$ by numerators of the second copy as usual.  However, one also now includes an appropriate sign that enforces the opposite statistics of internal hyper loops, and the numerator is conjugated (reversing matter arrows),
\begin{equation}
  c_i \rightarrow (-1)^{\lvert i\rvert}\bar{n}'_i\,,
\end{equation}
where $\lvert i \rvert$ is the number of matter loops in the given diagram. It is useful to generalize the above replacement by promoting $(-1)$ be a tunable parameter, $N_f$, which counts the number of fundamental matter multiplets in the first SYM copy (setting $N_f=1$ in the second copy),
\begin{equation}
  c_i \rightarrow (N_f)^{\lvert i\rvert}\bar{n}'_i\,.
\end{equation}
In the supergravity theory, the number of complexified matter multiplets is then $N_X=(1+N_f)$ where, depending on the amount of supersymmetry, $X$ is either a complexified vector, hyper, chiral multiplet or complex scalar. For instance, when $N_f=0$ the factorizable double copy is recovered (we take $0^0=1$).

After the double copy is performed, we can assemble terms on common denominators and thus consider (super-)gravity numerators given by
\begin{equation}
  N_i = \sum_{D_i=D_j} (N_f)^{\lvert j \rvert}n_j \bar{n}'_j\,,
\end{equation}
where all numerators sharing the same propagator $D_i$ are summed over, i.e. all possible routings of internal vector and complex matter multiplets.  The bar on top of the second numerator implies a reversal of hyper loop directions, i.e. conjugating the hypermultiplets.  It is sufficient that only one of the two numerator copies, $n_i$ or $\bar{n}'_i$, is color-dual.  We are left with complete (super-)gravity amplitudes:
\begin{align}\label{eq:fulldcamp}
  \cM_m^{(L)}=
  i^{L-1}\left(\frac{\kappa}{2}\right)^{m+2L-2}\sum_{\text{cubic graphs }\Gamma_i}
  \int\!\frac{\d^{LD}\ell}{(2\pi)^{LD}}\frac{(N_f)^{\lvert i\rvert}}{S_i}\frac{n_i\bar{n}'_i}{D_i},
\end{align}
where $\kappa$ is the gravitational coupling.

\section{Trees and cuts in \texorpdfstring{$\cN=2$}{N=2} SQCD}
\label{sec:treesAndCuts}

In this section we introduce $\cN=2$ super-QCD (SQCD) in $D=4$ dimensions.  This theory consists of $\cN=2$ SYM coupled to supersymmetric matter multiplets (hypermultiplets) in the fundamental representation of a generic gauge group $G$. For simplicity, we consider the limit of massless hypers.\footnote{Amplitudes with massive hypers can be recovered from the massless case by considering the $D=6$ version of the theory and reinterpreting the extra-dimensional momenta of the hypers as a mass.} $\cN=2$ SQCD's particle content is similar to that of $\cN=4$ SYM, so we refrain from describing its full Lagrangian; instead, we project its tree amplitudes out of the well-known ones in $\cN=4$ SYM.  We then proceed to calculate generalized unitarity cuts.  As our intention is to evaluate cuts in $D=4-2\eps$ dimensions, we also consider six-dimensional trees.  These are carefully extracted from the $\cN=(1,1)$ SYM theory --- the six-dimensional uplift of four-dimensional $\cN=4$ SYM.

The on-shell vector multiplet of $\cN=4$ SYM contains $2^\cN=16$ states:
\begin{align}\label{eq:neq4multiplet}
  \cV_{\cN=4}(\eta^I)=A^++\eta^I\psi^+_I+\frac{1}{2}\eta^I\eta^J\varphi_{IJ}+\frac{1}{3!}\eps_{IJKL}\eta^I\eta^J\eta^K\psi_-^L+\eta^1\eta^2\eta^3\eta^4A_-,
\end{align}
where we have introduced chiral superspace coordinates $\eta^I$ with SU(4) R-symmetry indices $\{I,J,\ldots\}$.  The $\cN=4$ multiplet is CPT self-conjugate and thus it is not chiral.  The apparent chirality is an artifact of the notation; it can equally well be expressed in terms of anti-chiral superspace coordinates $\etab_I$:
\begin{align}
  \cV_{\cN=4}(\etab_I)=\etab_1\etab_2\etab_3\etab_4A^++\cdots+A_-.
\end{align}

The on-shell spectrum of $\cN=2$ SYM contains $2\times2^\cN=8$ states, which can be packaged into one chiral $V_{\cN=2}$ and one anti-chiral $\widebar{V}_{\cN=2}$ multiplet, each with four states.  As explained in ref.~\cite{Johansson:2014zca}, it is convenient to combine them into a single non-chiral multiplet $\cV_{\cN=2}$ using the already-introduced superspace coordinates:
\begin{subequations}\label{eq:neq2multiplet}
  \begin{align}
    \cV_{\cN=2}(\eta^I)&=V_{\cN=2}+\eta^3\eta^4\widebar{V}_{\cN=2},\label{eq:neq2nonchiral}\\
    V_{\cN=2}(\eta^I)&=A^++\eta^I\psi^+_I+\eta^1\eta^2\varphi_{12},\label{eq:neq2chiral}\\
    \widebar{V}_{\cN=2}(\eta^I)&=\varphi_{34}+\eps_{I34J}\eta^I\psi_-^J+\eta^1\eta^2A_-,\label{eq:neq2antichiral}
  \end{align}
\end{subequations}
where the SU(2) indices $I,J=1,2$ are inherited from SU(4).  As is obvious, this is a subset of the full $\cN=4$ multiplet:
\begin{align}\label{eq:neq4decomp}
  \cV_{\cN=4}=\cV_{\cN=2}+\Phi_{\cN=2}+\widebar{\Phi}_{\cN=2},
\end{align}
where we have introduced a hypermultiplet and its conjugate parametrized as
\begin{subequations}\label{eq:hypermultiplets}
  \begin{align}
    \Phi_{\cN=2}(\eta^I)&=
    (\psi^+_3+\eta^I\varphi_{I3}+\eta^1\eta^2\psi_-^4)\eta^3,\\
    \widebar{\Phi}_{\cN=2}(\eta^I)&=
    (\psi^+_4+\eta^I\varphi_{I4}-\eta^1\eta^2\psi_-^3)\eta^4.
  \end{align}
\end{subequations}
One can easily check that these elements make up the full $\cN=4$ spectrum~(\ref{eq:neq4multiplet}).

The particle content of $\cN=2$ SQCD is the same as that of ordinary $\cN=4$ SYM, except that the hypermultiplet $\Phi_{\cN=2}$ ($\widebar{\Phi}_{\cN=2}$) transforms in the fundamental (anti-fundamental) representation of the gauge group $G$.  We also generalize to $N_f=\delta_\alpha^\alpha$ flavors by attaching flavor indices $\{\alpha,\beta,\ldots\}$ to the hypers, i.e. $(\Phi_{\cN=2})^\alpha$, $(\widebar{\Phi}_{\cN=2})_\alpha$.  The different representation affects the color structure of the amplitudes and the flavor adds more structure; however, when considering color-ordered tree amplitudes with identically-flavored hypers the distinction between $\cN=2$ SQCD and $\cN=4$ SYM is insignificant. The former amplitudes are obtainable from the latter by decomposing according to the above $\cN=2$ multiplets, as we shall now demonstrate.  For multi-flavor $\cN=2$ SQCD tree amplitudes the procedure to extract them from the $\cN=4$ ones is more complicated, although some tricks have been developed for this purpose~\cite{DelDuca:1999rs,Melia:2013epa}.

\subsection{Four dimensions}

To calculate four-dimensional tree amplitudes we start from the color-stripped Parke-Taylor formula for maximally-helicity-violating (MHV) scattering of $n$ states in $\cN=4$ SYM~\cite{Parke:1986gb,Nair:1988bq}:\footnote{Here we adopt the usual spinor-helicity notation, see for example~\cite{Dixon:1996wi,Elvang:2015rqa}.}
\begin{align}
  \langle\cV_{\cN=4}\cV_{\cN=4}\cdots\cV_{\cN=4}\rangle^\text{\tiny MHV}=
  i\frac{\delta^8(Q)}{\braket{12}\braket{23}\cdots\braket{n1}},
\end{align}
where for $\cN$ supersymmetries the supersymmetric delta function is
\begin{align}
  \delta^{2\cN}(Q)=\prod_{I=1}^\cN\sum_{i<j}^n\braket{i\,j}\eta_i^I\eta_j^I.
\end{align}
By inspection of the hypers~(\ref{eq:hypermultiplets}), $\Phi_{\cN=2}$ states are identified by a single $\eta^3$ factor and $\widebar{\Phi}_{\cN=2}$ states carry $\eta^4$.  In the full $\cN=2$ multiplet $\cV_{\cN=2}$~(\ref{eq:neq2multiplet}) a factor of $\eta^3\eta^4$ identifies a state belonging to $\widebar{V}_{\cN=2}$; anything else belongs to $V_{\cN=2}$.  This enables us to project out the relevant MHV amplitudes in $\cN=2$ SQCD, giving color-ordered amplitudes
\begin{subequations}\label{eq:4dtrees}
  \begin{align}
    \langle\cV_1\cV_2\cdots\cV_n\rangle^\text{\tiny MHV}
    &\!=\!i\frac{\delta^4(Q)}{\braket{12}\braket{23}\cdots\braket{n1}}
    \sum_{i<j}^n\braket{ij}^2\eta_i^3\eta_j^3 \eta_i^4\eta_j^4\,,\\
    \langle\cV_1 \cdots \Phi_i^\alpha \cdots \widebar{\Phi}_{j,\beta} \cdots \cV_n\rangle^\text{\tiny MHV}
    &\!=\!i\delta^\alpha_\beta
    \frac{\delta^4(Q)\,\eta_i^3\eta_j^4}{\braket{12}\braket{23}\cdots\braket{n1}}
    \sum_{k\neq i,j}^n\braket{ik}\braket{jk}\eta_k^4 \eta_k^3\,,\\
    \langle \cdots \Phi^\alpha_i \cdots \widebar{\Phi}_{j,\beta} \cdots
    \widebar{\Phi}_{k,\gamma}\cdots  \Phi^\delta_l \cdots \rangle^\text{\tiny MHV}
    &\!=\!i\delta^\alpha_\beta \delta^\delta_\gamma
    \frac{\delta^4(Q)\, \braket{il} \braket{jk} \eta_i^3  \eta_l^3 \eta_j^4 \eta_k^4 }{\braket{12}\braket{23}\cdots\braket{n1}}\,,
  \end{align}
\end{subequations}
where the ellipsis represents insertions of $\cV_{\cN=2}$.  We will only use tree amplitudes with up to five external legs, so MHV and $\overline{\rm MHV}$ are sufficient.

\subsection{Six dimensions}\label{sec:6dtheory}

The six-dimensional uplift of four-dimensional $\cN=4$ SYM is $\cN=(1,1)$ SYM; an on-shell superspace for this theory was introduced by Dennen, Huang and Siegel (DHS)~\cite{Dennen:2009vk}.  The DHS formalism uses a pair of Grassmann variables, $\eta_a$ and $\etat^{\dot{a}}$, carrying little-group indices in the two factors of $\text{SU}(2)\times\text{SU}(2)\sim\text{SO}(4)$.  The full non-chiral on-shell multiplet is
\begin{align}\label{eq:neq2d6multiplet}
  \cV_{\cN=(1,1)}(\eta_a,\etat^{\dot{a}})&=
  \phi+\chi^a\eta_a+\tilde{\chi}_{\dot{a}}\etat^{\dot{a}}+\phi'(\eta)^2+
           {g^a}_{\dot{a}}\eta_a\etat^{\dot{a}}+\bar{\phi}'(\etat)^2 \nn\\
           &\qquad+\bar{\tilde{\chi}}_{\dot{a}}(\eta)^2\etat^{\dot{a}}+\bar{\chi}^a\eta_a(\etat)^2+
           \bar{\phi}(\eta)^2(\etat)^2,
\end{align}
where $(\eta)^2=\frac{1}{2}\eta^a\eta_a$, $(\etat)^2=\frac{1}{2}\etat_{\dot{a}}\etat^{\dot{a}}$.  Note that in six dimensions CPT conjugation does not change the handedness of (anti-)chiral spinors $\chi^a$ ($\tilde{\chi}_{\dot{a}}$).

\subsubsection{Four-dimensional correspondence}

\begin{figure}[t]
  \centering
  \begin{subfigure}[b]{0.45\textwidth}
    \centering
      \includegraphics[keepaspectratio=true,clip=true]{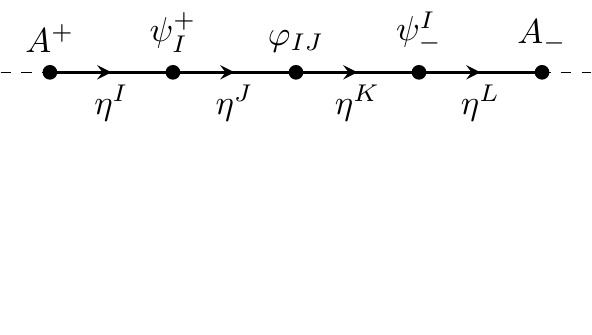}
    \caption{$D=4$, $\cN=4$}
  \end{subfigure}
  \begin{subfigure}[b]{0.45\textwidth}
    \centering
      \includegraphics[keepaspectratio=true,clip=true]{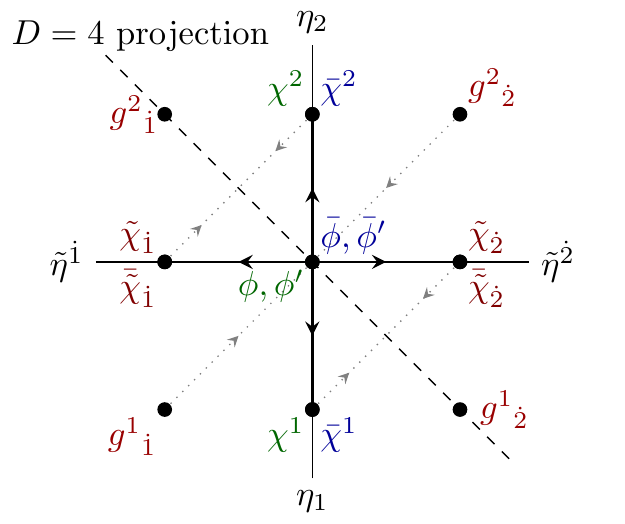}
    \caption{$D=6$, $\cN=(1,1)$}
  \end{subfigure}
  \caption{\small Weight-space diagrams for states in the (a) $D=4$, $\cN=4$ and (b) $D=6$, $\cN=(1,1)$ vector multiplets transforming under the little groups U(1) and $\text{SU}(2)\times\text{SU}(2)$ respectively.  The Grassmann variables~$\eta$ and~$\tilde{\eta}$ act as raising and lowering operators in the weight space~\cite{Boels:2009bv} as denoted by solid black arrows. Red-colored states belong to the vector multiplet; green- and blue-colored states belong to the two parts of the $D=6$, $\cN=(1,0)$ hypermultiplet. The dotted gray arrows show how 6D states are projected to the four-dimensional weight space.}
  \label{fig:weightSpace}
\end{figure}

The correspondence between four-dimensional $\cN=4$ SYM and six-dimensional $\cN=(1,1)$ SYM is most easily seen using a non-chiral superspace construction of the $\cN=4$ multiplet~\cite{Huang:2011um}.  Two chiral and two anti-chiral superspace coordinates are used to capture $\cN=4$ SYM's 16 on-shell states: we choose $\eta^1$, $\etab_2$, $\etab_3$ and $\eta^4$.  Switching between the superspaces is done with half-Fourier transforms:
\begin{align}\label{eq:etatransform}
  \Omega_\text{chiral}(\eta^I)&=
  \int\!\d\etab_2\d\etab_3\,e^{\etab_2\eta^2+\etab_3\eta^3}\Omega_\text{non-chiral}(\eta^{I=1,4},\etab_{I=2,3}).
\end{align}
In terms of the non-chiral superspace, a valid mapping to six dimensions is~\cite{Elvang:2011fx}
\begin{align}\label{eq:4d6dstatemapping}
  \eta_a\leftrightarrow(\eta^1,\bar{\eta}_2), && \etat^{\dot{a}}\leftrightarrow(\bar{\eta}_3,\eta^4).
\end{align}
This allows us to identify on-shell states between the four-dimensional $\cN=4$ and six-dimensional $\cN=(1,1)$ multiplets.  We do this for the $\cN=2$ vector~(\ref{eq:neq2multiplet}) and matter~(\ref{eq:hypermultiplets}) multiplets separately; the results are illustrated in fig.~\ref{fig:weightSpace}.

Starting with the $\cN=2$ vector multiplet~(\ref{eq:neq2multiplet}), we switch to the four-dimensional non-chiral superspace and then use the coordinate mapping~(\ref{eq:4d6dstatemapping}):
\begin{align}\label{eq:neq2transform}
  &\cV_{\cN=2}(\eta_a,\etat^{\dot{a}})
  =\int\!\d\eta^3\d\eta^2e^{\eta^2\etab_2+\eta^3\etab_3}\cV_{\cN=2}(\eta^I)\Big|_{(\eta^1,\bar{\eta}_2)\to\eta_a,(\bar{\eta}_3,\eta^4)\to\etat^{\dot{a}}}\\
  &\qquad=\etat^{\dot{1}}(\psi^+_2-\eta_1\varphi_{12}+\eta_2A^+-\eta_1\eta_2\psi^+_1)
  +\etat^{\dot{2}}(-\psi_-^1-\eta_1A_-+\eta_2\varphi_{34}-\eta_1\eta_2\psi_-^2).\nn
\end{align}
This clearly forms part of $\cV_{\cN=(1,1)}$: for instance, $\psi^+_2=\tilde{\chi}_{\dot{1}}$ and $\varphi_{12}={g^1}_{\dot{1}}$.  As it contains a six-dimensional vector and a pair of anti-chiral spinors, we identify it as the on-shell multiplet of $\cN=(1,0)$ SYM~\cite{deWit:2002vz}:
\begin{align}
  \cV_{\cN=(1,0)}(\eta_a,\etat^{\dot{a}})=
  \tilde{\chi}_{\dot{a}}\etat^{\dot{a}}+{g^a}_{\dot{a}}\eta_a\etat^{\dot{a}}+
  \bar{\tilde{\chi}}_{\dot{a}}(\eta)^2\etat^{\dot{a}}.
\end{align}
Hypermultiplets are transformed by the same procedure:
\begin{subequations}\label{eq:6dhypers}
  \begin{align}
    \Phi_{\cN=2}(\eta_a,\etat^{\dot{a}})&=
    \varphi_{23}+\eta_1\psi_-^4-\eta_2\psi^+_3-\eta_1\eta_2\varphi_{13},\\
    \widebar{\Phi}_{\cN=2}(\eta_a,\etat^{\dot{a}})&=
    \etat^{\dot{1}}\etat^{\dot{2}}(\varphi_{24}-\eta_1\psi_-^3-\eta_2\psi^+_4-\eta_1\eta_2\varphi_{14}).
  \end{align}
\end{subequations}
Each of these contains a chiral spinor and a pair of scalars, so we recognize them as the two parts of the $\cN=(1,0)$ hypermultiplet:
\begin{subequations}
  \begin{align}
    \Phi_{\cN=(1,0)}(\eta_a,\etat^{\dot{a}})&=\phi+\chi^a\eta_a+\phi'(\eta)^2,\\
    \widebar{\Phi}_{\cN=(1,0)}(\eta_a,\etat^{\dot{a}})&=
    \bar{\phi}'(\etat)^2+\bar{\chi}^a\eta_a(\etat)^2+\bar{\phi}(\eta)^2(\etat)^2.
  \end{align}
\end{subequations}
The three multiplets make up $\cV_{\cN=(1,1)}$~(\ref{eq:neq2d6multiplet}) as expected.  Notice that the above procedure was not unique: we could instead have chosen to identify $\cV_{\cN=2}$ with $\cV_{\cN=(0,1)}$, which carries the chiral spinors $\chi^a$ and $\bar{\chi}^a$.  The choice was made when we picked the specific mapping between four- and six-dimensional states~(\ref{eq:4d6dstatemapping}).\footnote{For instance, in ref.~\cite{Huang:2011um} the states are instead identified as $\eta_a\leftrightarrow(\eta^1,\etab_4)$, $\etat^{\dot{a}}\leftrightarrow(\etab_3,\eta^2)$.}

\subsubsection{Six-dimensional amplitudes}

Amplitudes in this theory are given in terms of the six-dimensional spinor-helicity formalism~\cite{Cheung:2009dc}.  For each particle $i$ they depend on the chiral and anti-chiral spinors, $\lambda_i$ and $\lambdat_i$, and the supermomenta,
\begin{align}
  \mathfrak{q}^A_i=\lambda_i^{A,a}\eta_{i,a}, &&
  \tilde{\mathfrak{q}}_{i,A}=\lambdat_{i,A,\dot{a}}\etat_i^{\dot{a}}\,. ~~~~~~  (\text{no sum over}~i)
\end{align}
The indices $\{A,B,\ldots\}$ belong to SU*(4) --- the spin group of SO(1,5) Lorentz symmetry.  For notational convenience we define the two four-brackets:
\begin{subequations}
  \begin{align}
    \braket{i^a,j^b,k^c,l^d}&\equiv
    \eps_{ABCD}\lambda_i^{A,a}\lambda_j^{B,b}\lambda_k^{C,c}\lambda_l^{D,d}\,,\\
        [i_{\dot{a}},j_{\dot{b}},k_{\dot{c}},l_{\dot{d}}]&\equiv
        \eps^{ABCD}\lambdat_{i,A,\dot{a}}\lambdat_{j,B,\dot{b}}\lambdat_{k,C,\dot{c}}\lambdat_{l,D,\dot{d}}\,.
  \end{align}
\end{subequations}
The four-point color-ordered tree amplitude of $\cN=(1,1)$ SYM is~\cite{Dennen:2009vk}
\begin{align}\label{eq:6dtree}
  \langle\cV_{\cN=(1,1)}\cV_{\cN=(1,1)}\cV_{\cN=(1,1)}\cV_{\cN=(1,1)}\rangle^{D=6}=
  -i\,\frac{\delta^4(\sum_i\mathfrak{q}_i)\delta^4(\sum_i\tilde{\mathfrak{q}}_i)}{st},
\end{align}
where $s=(p_1+p_2)^2$ and $t=(p_2+p_3)^2$; the fermionic delta functions are
\begin{subequations}
\begin{align}
  \delta^4\Big(\sum_i\mathfrak{q}_i\Big)&=
  \frac{1}{4!}\sum_{i,j,k,l}\braket{i^a,j^b,k^c,l^d}\eta_{i,a}\eta_{j,b}\eta_{k,c}\eta_{l,d},\\
  \delta^4\Big(\sum_i\tilde{\mathfrak{q}}_i\Big)&=
  \frac{1}{4!}\sum_{i,j,k,l}\left[i_{\dot{a}},j_{\dot{b}},k_{\dot{c}},l_{\dot{d}}\right]
  \etat_i^{\dot{a}}\etat_j^{\dot{b}}\etat_k^{\dot{c}}\etat_l^{\dot{d}}.
\end{align}
\end{subequations}
This amplitude is invariant under $\mathfrak{q}_i\leftrightarrow\tilde{\mathfrak{q}}_i$ --- a reflection of the non-chiral nature of $\cN=(1,1)$ SYM.  In six dimensions there is no notion of helicity sectors (like N$^k$MHV sectors in four dimensions) because rotations of the little group $\text{SO(4)}\sim\text{SU(2)}\times\text{SU(2)}$ mix them.  The above superamplitude is therefore a single expression that unifies all the four-dimensional helicity sectors.

We can now write down six-dimensional equivalents of the four-dimensional $\cN=2$ SQCD trees~(\ref{eq:4dtrees}).  States are identified using the $\etat^{\dot{a}}$ variables: we project these out of $\delta^4(\sum_i\tilde{\mathfrak{q}}_i)$ in the four-point amplitude~(\ref{eq:6dtree}).  With four external legs the color-ordered amplitudes are
\begin{subequations}\label{eq:6dtrees}
  \begin{align}
    \langle\cV_1\cV_2\cV_3\cV_4\rangle^{D=6}&=
    -i\,\frac{\delta^4(\sum_i\mathfrak{q}_i)}{st}[1_{\dot{a}},2_{\dot{b}},3_{\dot{c}},4_{\dot{d}}]
    \etat_1^{\dot{a}}\etat_2^{\dot{b}}\etat_3^{\dot{c}}\etat_4^{\dot{d}},\\
    \langle \Phi^\alpha_1 \, \widebar{\Phi}_{2,\beta}\cV_3\cV_4\rangle^{D=6}&=
    -i\delta^\alpha_\beta\frac{\delta^4(\sum_i\mathfrak{q}_i)}{st}[2_{\dot{1}},2_{\dot{2}},3_{\dot{c}},4_{\dot{d}}]
    \etat_2^{\dot{1}}\etat_2^{\dot{2}}\etat_3^{\dot{c}}\etat_4^{\dot{d}},\label{eq:6dtrees2}\\
    \langle \Phi_1^\alpha \, \widebar{\Phi}_{2,\beta} \, \widebar{\Phi}_{3,\gamma} \, \Phi_4^\delta\rangle^{D=6}&=
    -i\delta^\alpha_\beta\delta^\delta_\gamma\frac{\delta^4(\sum_i\mathfrak{q}_i)}{st}
    [2_{\dot{1}},2_{\dot{2}},3_{\dot{1}},3_{\dot{2}}]
    \etat_2^{\dot{1}}\etat_2^{\dot{2}}\etat_3^{\dot{1}}\etat_3^{\dot{2}},
    %
  \end{align}
\end{subequations}
where $\{\alpha,\beta,\ldots\}$ are flavor indices.  These amplitudes are not invariant under $\mathfrak{q}_i\leftrightarrow\tilde{\mathfrak{q}}_i$: our use of the $\cV_{\cN=(1,0)}$ multiplet implies that $\tilde{\mathfrak{q}}_i$ encodes information about the external states.  In principle, we will also need five-point amplitudes when computing three-particle cuts of the two-loop amplitude.  However, the five-point $\cN=(1,1)$ SYM tree amplitude~\cite{Dennen:2009vk} is less compact and, as we will discuss in section~\ref{sec:symmetriesandcuts}, it turns out that these cuts do not provide any new physical information compared to the four-dimensional ones given the constraints on our ans\"{a}tze.

When the momenta are taken in a four-dimensional subspace the six-dimensional trees should reproduce their four-dimensional counterparts~(\ref{eq:4dtrees}).  This is easily confirmed: write the six-dimensional spinors in terms of four-dimensional ones~\cite{Cheung:2009dc},
\begin{align}\label{eq:6dspinors4d}
  {\lambda^A}_a=
  \left(\begin{matrix}
    0 & \lambda_\alpha \\
    \lambdat^{\dot{\alpha}} & 0
  \end{matrix}\right), &&
  \lambdat_{A,\dot{a}}=
  \left(\begin{matrix}
    0 & \lambda^\alpha \\
    -\lambdat_{\dot{\alpha}} & 0
  \end{matrix}\right),
\end{align}
where here the Greek letters denote SL(2,$\mathbb{C}$) spinor indices (not to be confused with the previously-introduced flavor indices, the distinction should be clear from the context).  Then substitute into the six-dimensional tree amplitudes~(\ref{eq:6dtrees}).  Finally, switch to four-dimensional fermionic coordinates using the half-Fourier transform~(\ref{eq:etatransform}) and the state mapping~(\ref{eq:4d6dstatemapping}) on each of the external legs.  A helpful intermediate relationship is~\cite{Huang:2011um}
\begin{align}
  \int\!\bigg(\prod_{i=1}^4\d\bar{\eta}_{i,2}\bigg)\text{exp}\bigg(\sum_{i=1}^4\etab_{i,2}\eta^2_i\bigg)\left[\delta^4\Big(\sum_i\mathfrak{q}_i\Big)\bigg|_{
  	\substack{\eta_a\to(\eta^1,\bar{\eta}_2) \\ \etat^{\dot{a}}\to(\bar{\eta}_3,\eta^4)}
  }\right]=
  \frac{[12]}{\braket{34}}\delta^4(Q),
\end{align}
which exchanges the fermionic delta functions.  This procedure gives all external helicity sectors in four dimensions, but at four points only the MHV is non-zero.

\subsection{Cuts}

In the next section we will use ans\"{a}tze to construct color-dual numerators for $\cN=2$ SQCD.  The physical input for these will come from unitarity cuts.  In four dimensions the cuts are calculable using the MHV tree amplitudes~(\ref{eq:4dtrees}) --- the techniques involved are fairly standard and we will not repeat them here.  In six dimensions we follow ref.~\cite{Bern:2010qa}; however, the tree amplitudes~(\ref{eq:6dtrees}) create some new subtleties that we will now address.\footnote{For a good review of both four- and six-dimensional unitarity techniques see ref.~\cite{Bern:2011qt}.}

As we always take external states in a four-dimensional subspace, the key to this approach is finding six-dimensional spinor solutions for the on-shell loop momenta in terms of four-dimensional external spinors.  We write the six-dimensional massless on-shell condition as a four-dimensional massive condition:
\begin{align}
  \ell^2={\bar\ell \,}^2-(\ell_4)^2-(\ell_5)^2={\bar\ell\,}^2-m\mt=0,
\end{align}
where $\bar{\ell}$ is the part of $\ell$ living in the four-dimensional subspace.  The complex masses are related to the fifth and sixth components of $\ell$:
\begin{align}
\label{mmtilde}
  m=\ell_4-i\,\ell_5, && \mt=\ell_4+i\,\ell_5.
\end{align}
Now considering $\bar{\ell}$ as a massive on-shell four-momentum we write it in terms of two pairs of four-dimensional spinors, $(\lambda,\tilde{\lambda})$ and $(\rho,\tilde{\rho})$, as:
\begin{align}
  \bar{\ell}_{\alpha\dot{\alpha}}=
  \lambda_\alpha\lambdat_{\dot{\alpha}}+\frac{m\mt}{2\lambda\cdot\rho}\rho_\alpha\tilde{\rho}_{\dot{\alpha}}.
\end{align}
Appropriate six-dimensional spinors are then~\cite{Boels:2009bv}
\begin{align}\label{eq:6dspinorrep}
  {\lambda^A}_a=
  \left(\begin{matrix}
    \frac{i \mt}{\braket{\lambda\rho}}\rho_\alpha & \lambda_\alpha \\
    \lambdat^{\dot{\alpha}} & \frac{im}{[\rho\lambda]}\tilde{\rho}^{\dot{\alpha}}
  \end{matrix}\right), &&
  \lambdat_{A,\dot{a}}=
  \left(\begin{matrix}
    \frac{im}{\braket{\lambda\rho}}\rho^\alpha & \lambda^\alpha \\
    -\lambdat_{\dot{\alpha}} & \frac{i\mt}{[\lambda\rho]}\tilde{\rho}_{\dot{\alpha}}
  \end{matrix}\right).
\end{align}
By taking $\ell_4=\ell_5=0$, and therefore $m=\mt=0$, we recover the four-dimensional projections~(\ref{eq:6dspinors4d}).

\subsubsection{One-loop example}\label{sec:oneloopcut}

\begin{figure}[t]
  \centering
  \includegraphics[scale=.4,keepaspectratio=true,clip=true,trim = 0 0 30 0]{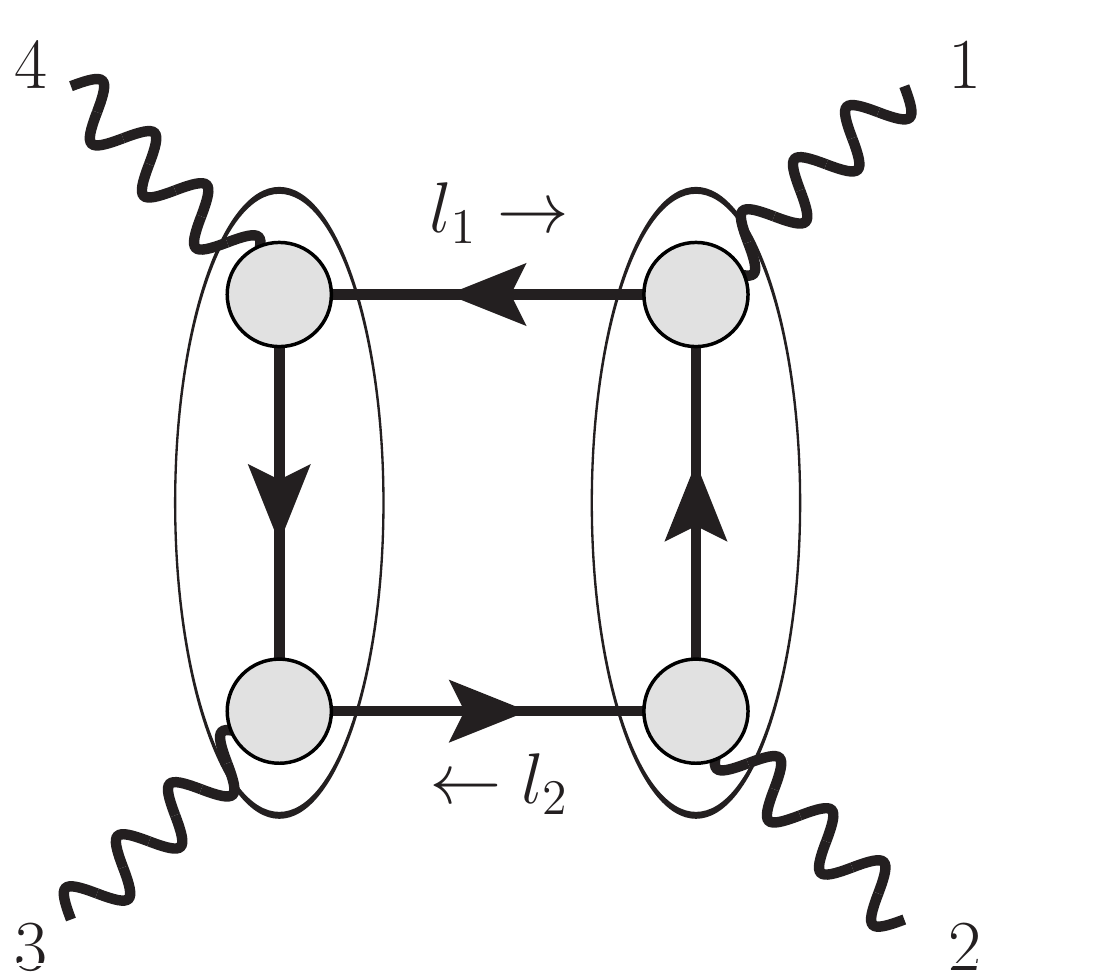}
  \caption{\small A one-loop quadruple cut with circulating hypermultiplets --- all four exposed lines are taken on shell.  To compute it, we first write it as a two-particle cut involving a pair of four-point tree amplitudes --- as indicated by the transparent blobs --- and then we multiply by the inverse poles exposed inside these blobs.  The quadruple cut limit is then finite.
    \label{fig:hyperboxcut}}
\end{figure}

We now illustrate the cutting method by calculating the one-loop quadruple cut in fig.~\ref{fig:hyperboxcut}. Without loss of generality we set $N_f=1$; powers of $N_f$ can be restored at any point by counting the number of hyper loops. The quadruple cut integrand is then
\begin{align}
  \text{Cut}\bigg(\usegraph{9}{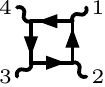}\bigg)=
  \int\bigg(\prod_{i=1}^2\d^2\eta_{l_i}\d^2\etat_{l_i}\bigg)
  (l_2-p_3)^2A^{(0)}_\text{L}\times(l_1-p_1)^2A^{(0)}_\text{R}.
\end{align}
Instead of starting from the (rather complicated) three-point six-dimensional amplitudes we study factorization limits of the simpler four-point amplitudes.  Using~(\ref{eq:6dtrees2}),
\begin{subequations}   
  \begin{align}
    A^{(0)}_\text{L}=-i\frac{\delta^4(\sum_{i\in L}\mathfrak{q}_i)}{s(l_2-p_3)^2}
    [(l_1)_{\dot{1}},(l_1)_{\dot{2}},3_{\dot{c}},4_{\dot{d}}]
    \etat_{l_1}^{\dot{1}}\etat_{l_1}^{\dot{2}}\etat_3^{\dot{c}}\etat_4^{\dot{d}},\\
    A^{(0)}_\text{R}=-i\frac{\delta^4(\sum_{i\in R}\mathfrak{q}_i)}{s(l_1-p_1)^2}
    [(l_2)_{\dot{1}},(l_2)_{\dot{2}},1_{\dot{a}},2_{\dot{b}}]
    \etat_{l_2}^{\dot{1}}\etat_{l_2}^{\dot{2}}\etat_1^{\dot{a}}\etat_2^{\dot{b}},
  \end{align}
\end{subequations}
and, with $\ell=l_1=p_1+p_2+l_2$, the on-shell conditions are $\ell^2=(\ell-p_1)^2=(\ell-p_1-p_2)^2=(\ell+p_4)^2=0$.  Putting the pieces together, and summing over the supersymmetric states, we obtain the cut for internal hypers:
\begin{align}\label{eq:hyperboxcut}
  \text{Cut}\bigg(\usegraph{9}{nBoxBar}\bigg)
  &=-\frac{\delta^4(\sum_i\mathfrak{q}_i)}{s}[(l_2)_{\dot{1}},(l_2)_{\dot{2}},1_{\dot{a}},2_{\dot{b}}]
       [(l_1)_{\dot{1}},(l_1)_{\dot{2}},3_{\dot{c}},4_{\dot{d}}]
       \etat_1^{\dot{a}}\etat_2^{\dot{b}}\etat_3^{\dot{c}}\etat_4^{\dot{d}}.
\end{align}
This expression can now be converted to external spinors and loop momenta using the previously-given expressions. The full details on how to work this out may be found in ref.~\cite{Bern:2010qa}; we simply quote the final results below.

The external states are purely four-dimensional, and thus through the half-Fourier transform~(\ref{eq:etatransform}) we may split up the cut into the contributions from sectors of different powers of the four-dimensional Grassmann variables $\eta_i^I$.  Unlike the maximally-supersymmetric case, the $\cN=2$ SYM loop amplitudes will not have automatically-vanishing integrands when all external states are picked from the chiral vector multiplet $V_{\cN=2}$ --- thus, besides the MHV sector we will also have an ``all-chiral'' helicity sector (as well as an ``all-anti-chiral'' sector that is trivially obtained by CPT conjugation). A further sector would be the one-chiral-vector case (and its CPT conjugate), but we will not consider it here.  All sectors are consistent with $\cN=2$ supersymmetry, but the all-chiral integrand will only give rise to ${\cal O}(\epsilon)$ contributions in $D=4-2\epsilon$ dimensions. Nevertheless, the integrand of the all-chiral sector is important as it gives non-vanishing supergravity amplitudes through the double copy.

We will come back to the all-chiral case shortly, but for now consider the MHV sector. The cut (\ref{eq:hyperboxcut}) then becomes
\begin{align}
  \text{Cut}_\text{MHV}\bigg(\usegraph{9}{nBoxBar}\bigg)&=
  \,m\mt\left(\frac{\kappa_{12}+\kappa_{34}}{s}+\frac{\kappa_{23}+\kappa_{14}}{t}+
  \frac{\kappa_{13}+\kappa_{24}}{u}\right)\\
  &\qquad\qquad\qquad
  +2i\eps(1,2,3,\bar{\ell})\frac{\kappa_{13}-\kappa_{24}}{u^2}+
  \frac{st}{2u^2}(\kappa_{13}+\kappa_{24}),\nn
\end{align}
where we have introduced the variables
\begin{align}\label{eq:kappa}
  \kappa_{ij}=\frac{[12][34]}{\braket{12}\braket{34}}\delta^4(Q)\braket{i\,j}^2\eta_i^3\eta_j^3\eta_i^4\eta_j^4
\end{align}
to identify contributions from external legs $i$, $j$ belonging to the anti-chiral vector multiplet $\widebar{V}_{\cN=2}$.

The six-dimensional cut contains all the information to be formally re-interpreted as a $D$-dimensional cut. Strictly speaking, it is only valid for $D\le6$ since the theory does not exist beyond six dimensions, but we may write it in a dimensionally-agnostic form. Parametrizing the dimension as $D=4-2\epsilon$, the loop momentum is conveniently split into the four- and extra-dimensional parts as $\ell=\bar{\ell}+\mu$.  As external momenta are $p_i$ are purely four-dimensional, we expect that $\mu$  appears in the Lorentz-invariant combination $\mu^2=-\mu\cdot\mu>0$. Indeed, we may eliminate $m$ and $\mt$ in favor of $\mu^2=\mt m$.  The cut then agrees with the $D$-dimensional fundamental box numerator found in ref.~\cite{Johansson:2014zca}.

Now consider the all-chiral sector, which is identified by all external states belonging to the chiral $V_{\cN=2}$ multiplet. In this case the cut (\ref{eq:hyperboxcut}) becomes
\begin{align}\label{eq:hyperboxcutplus}
  \text{Cut}_\text{all-chiral}\bigg(\usegraph{9}{nBoxBar}\bigg)=
  -\frac{[12][34]}{\braket{12}\braket{34}}m^2\delta^4(Q).
\end{align}
Interestingly, rotational symmetry in the extra-dimensional direction has been broken. Naively, we cannot eliminate the complex variable $m^2$ in favor of the rotational invariant $\mu^2$ (as we did in the MHV sector).  This is because the external states probe the extra-dimensional space and pick out a chiral direction. Indeed, the component amplitudes in the all-chiral sector always have either scalars or fermions (of the vector multiplet) on the external legs, and these particles are secretly six dimensional (e.g. the scalars are the extra-dimensional gluons). There is no all-gluon amplitude in the all-chiral sector, and thus there is no component amplitude in this sector that involves only four-dimensional external states.

We can avoid the issue with complex momenta by instead choosing to work with a five-dimensional embedding corresponding to $m=\mt$.  This should be sufficient to parametrize the physical content of the dimensionally-regulated theory.  In five dimensions the spinors belong to USp(2,2) $\sim$ SO(1,4), which means that they can be made to satisfy a reality condition (see e.g. recent work~\cite{Badger:2017gta}):
\begin{align}
  \lambdat_{A,\dot{a}}=\Sigma^5_{AB}\lambda^B_a,
\end{align}
where $\Sigma^5=i\,\sigma_0\otimes\sigma_2$ is an element of the six-dimensional Clifford algebra.  This effectively identifies the little-group SU(2) labels $a$ and $\dot{a}$.  One can then write $\mu^2=m^2=\mt^2$, which is the desired result.

\subsubsection{Two-loop cuts}\label{sec:2lcuts}

\begin{figure}[t]
  \centering
  \begin{subfigure}{0.25\textwidth}
    \centering
    \includegraphics[width=0.8\textwidth]{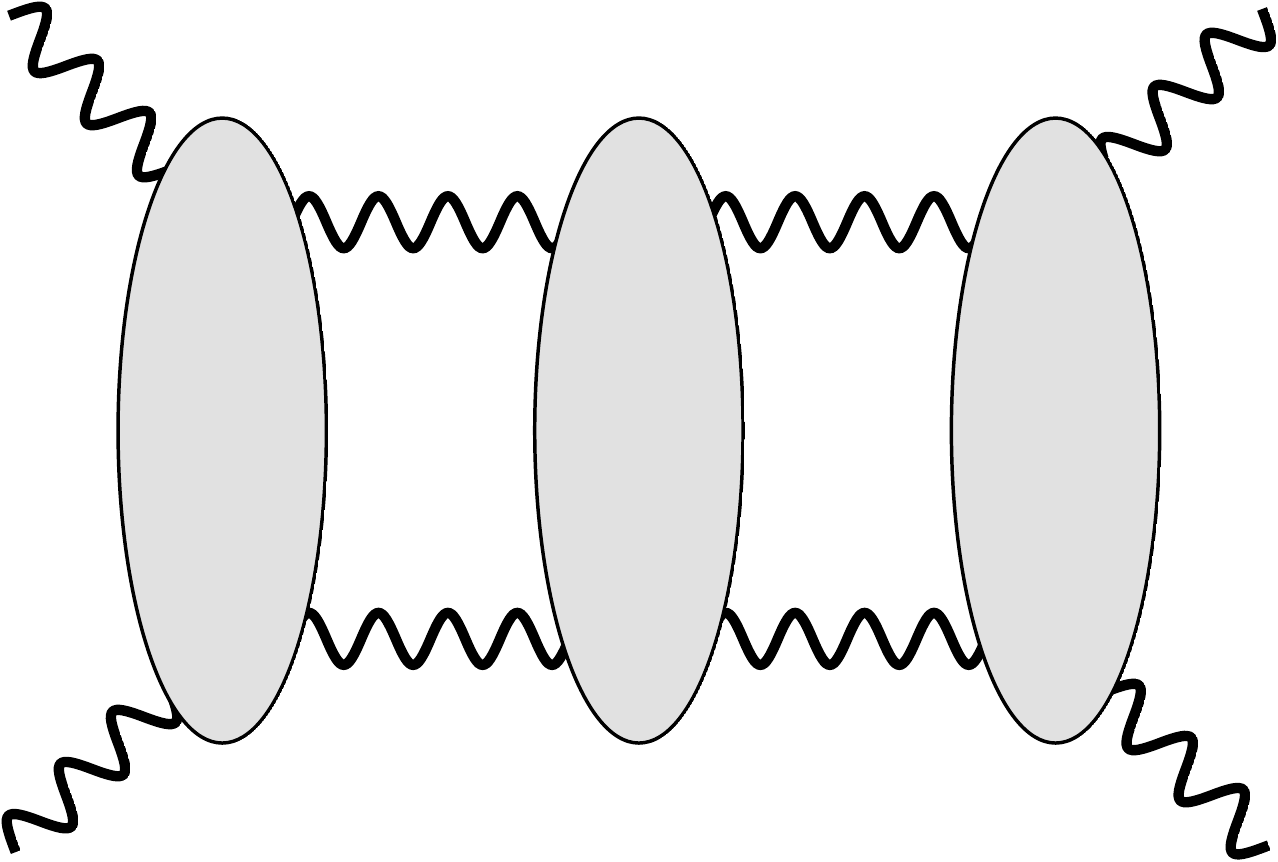}
    \caption[a]{}
  \end{subfigure}
  \begin{subfigure}{0.25\textwidth}
    \centering
    \includegraphics[width=0.8\textwidth]{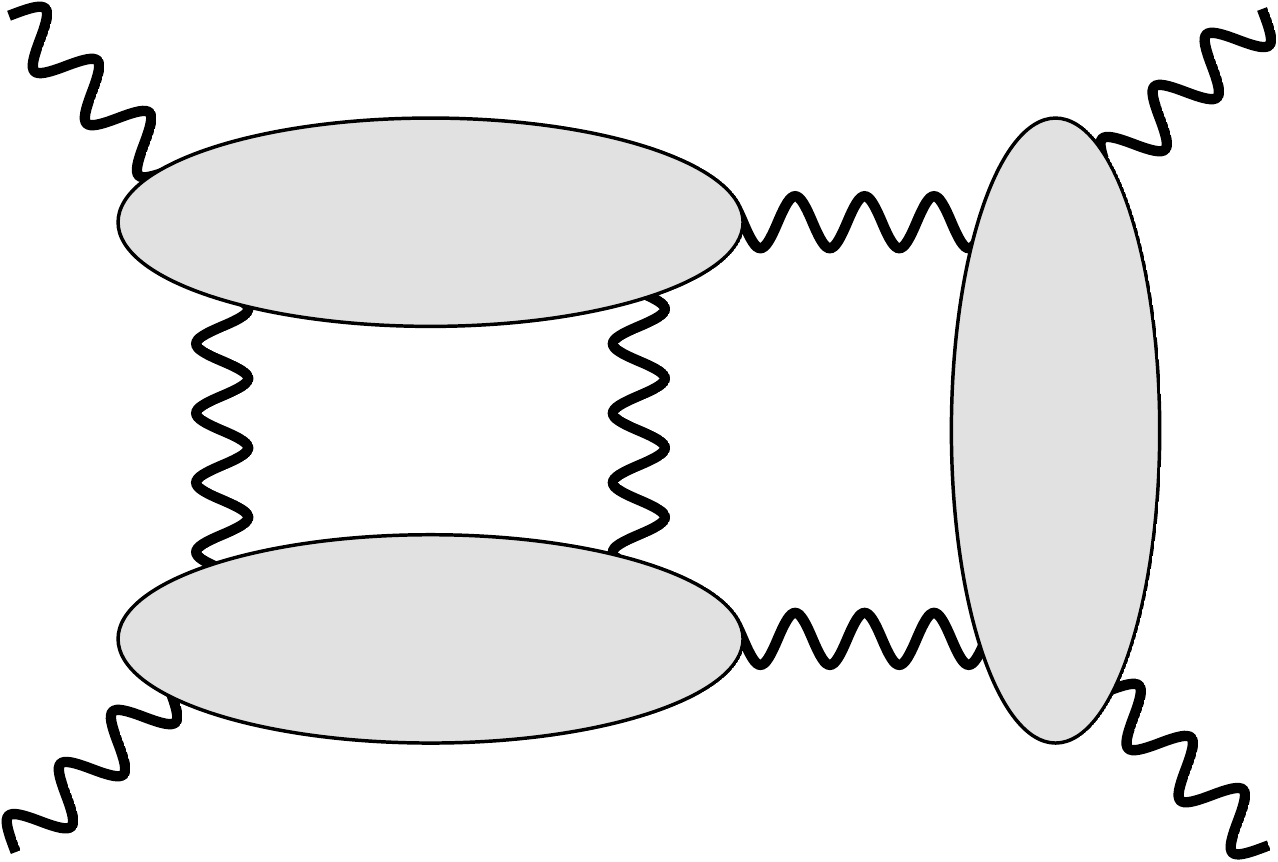}
    \caption[a]{}
  \end{subfigure}
  \begin{subfigure}{0.25\textwidth}
    \centering
    \includegraphics[width=0.8\textwidth]{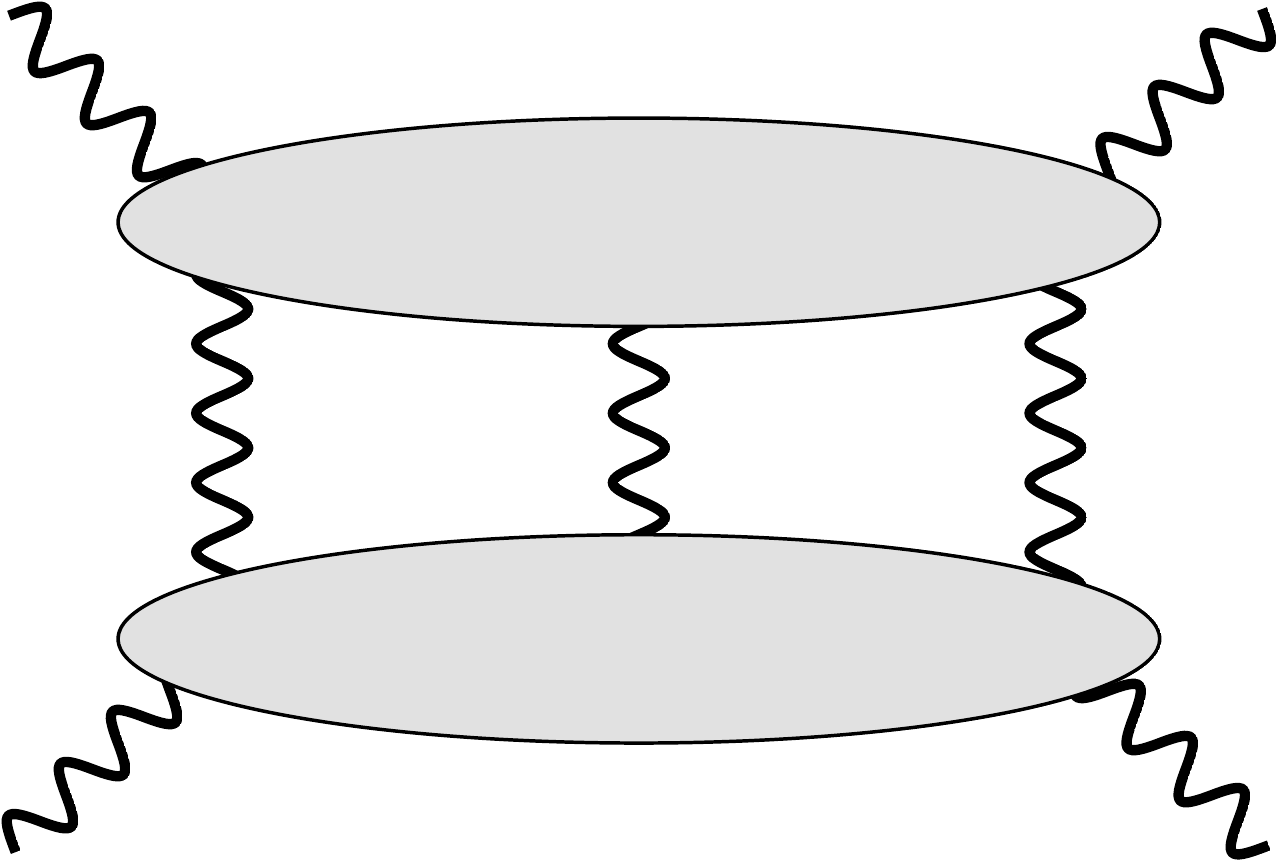}
    \caption[a]{}
  \end{subfigure}
  \caption{\small The three two-loop spanning cuts involving only pure $\cN=2$ states.  All exposed propagators are understood as being taken on shell.
    \label{fig:purecuts}}
\end{figure} 

\begin{figure}[t]
  \centering
  \begin{subfigure}{0.25\textwidth}
    \centering
    \includegraphics[width=0.8\textwidth]{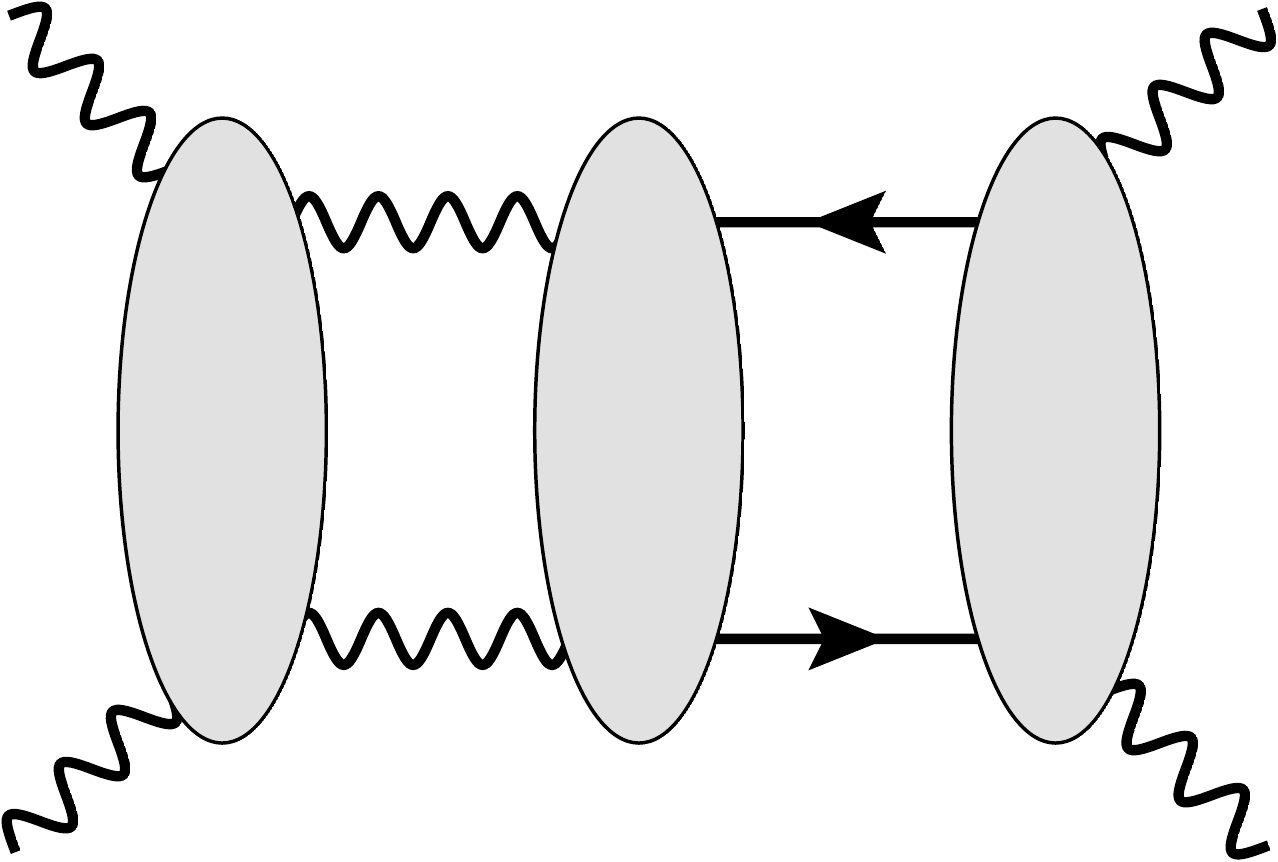}
    \caption[a]{}
    \label{fig:hypercut1}
  \end{subfigure}
  \begin{subfigure}{0.25\textwidth}
    \centering
    \includegraphics[width=0.8\textwidth]{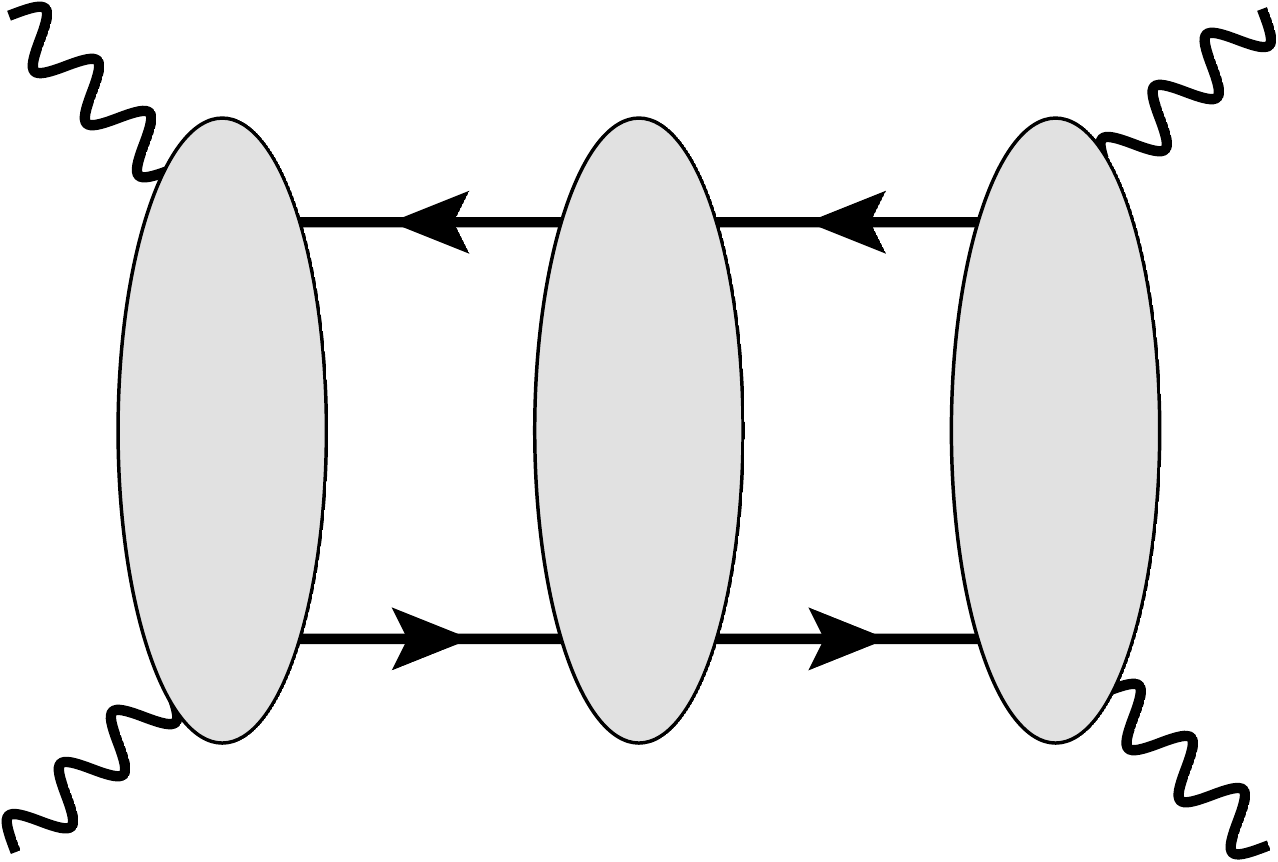}
    \caption[a]{}
    \label{fig:hypercut2}
  \end{subfigure}
  \begin{subfigure}{0.25\textwidth}
    \centering
    \includegraphics[width=0.8\textwidth]{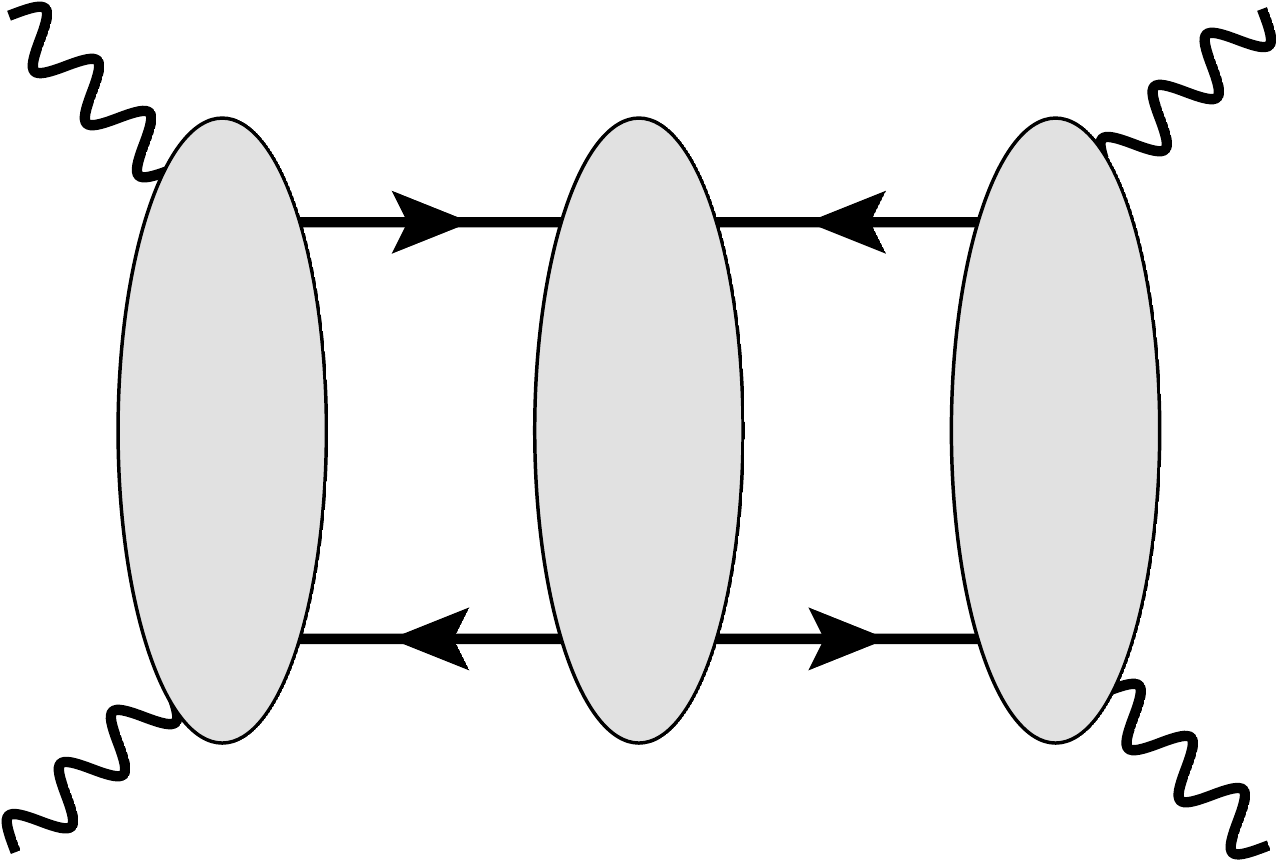}
    \caption[a]{}
    \label{fig:hypercut3}
  \end{subfigure}
  \begin{subfigure}{0.25\textwidth}
    \centering
    \includegraphics[width=0.8\textwidth]{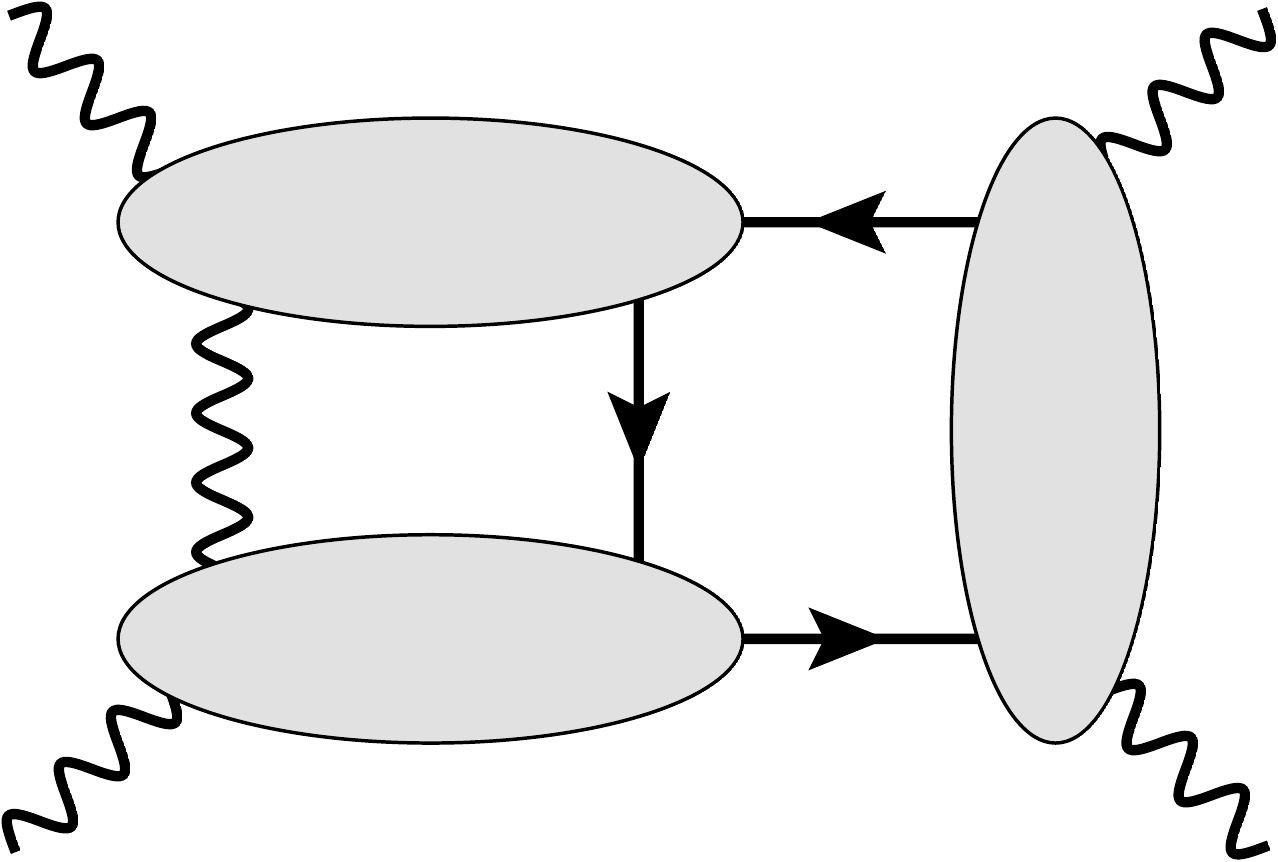}
    \caption[a]{}
  \end{subfigure}
  \begin{subfigure}{0.25\textwidth}
    \centering
    \includegraphics[width=0.8\textwidth]{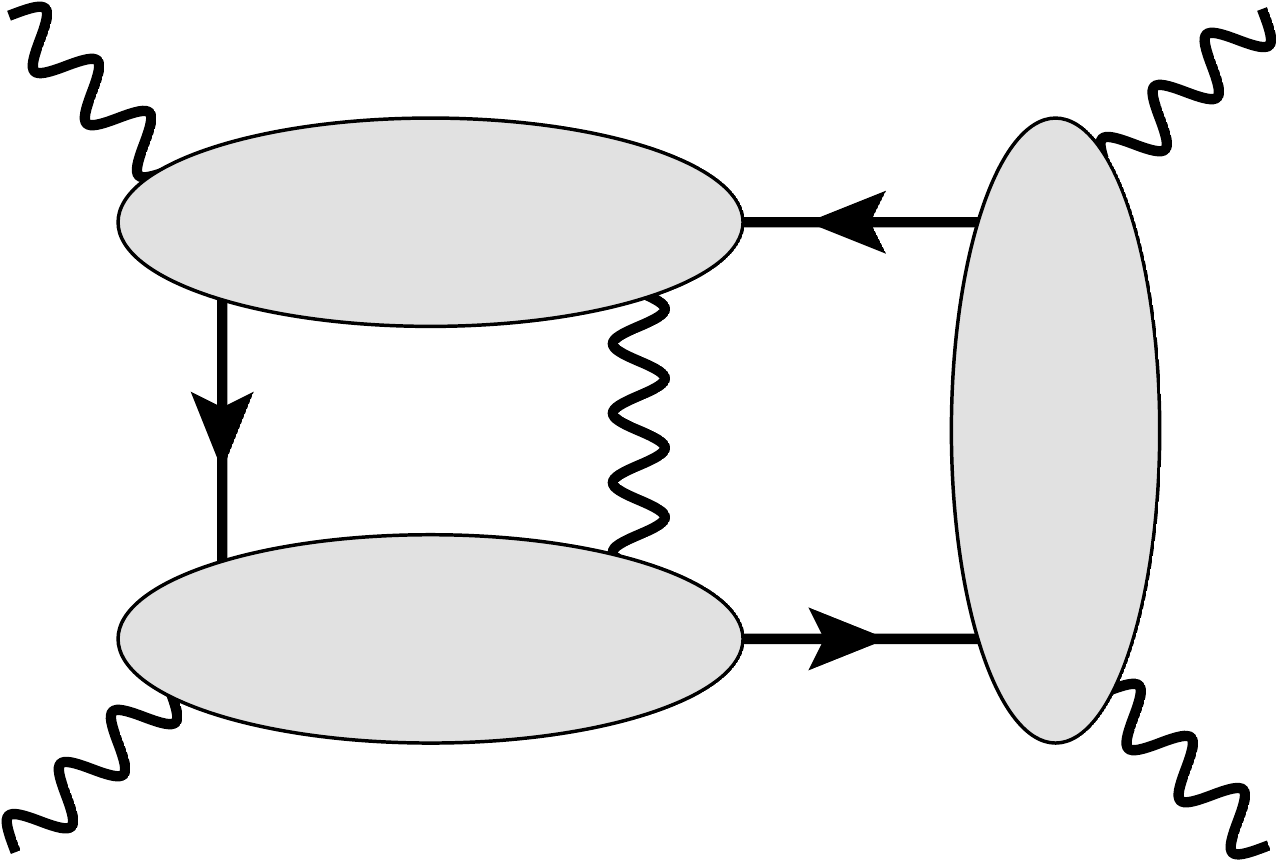}
    \caption[a]{}
  \end{subfigure}
  \begin{subfigure}{0.25\textwidth}
    \centering
    \includegraphics[width=0.8\textwidth]{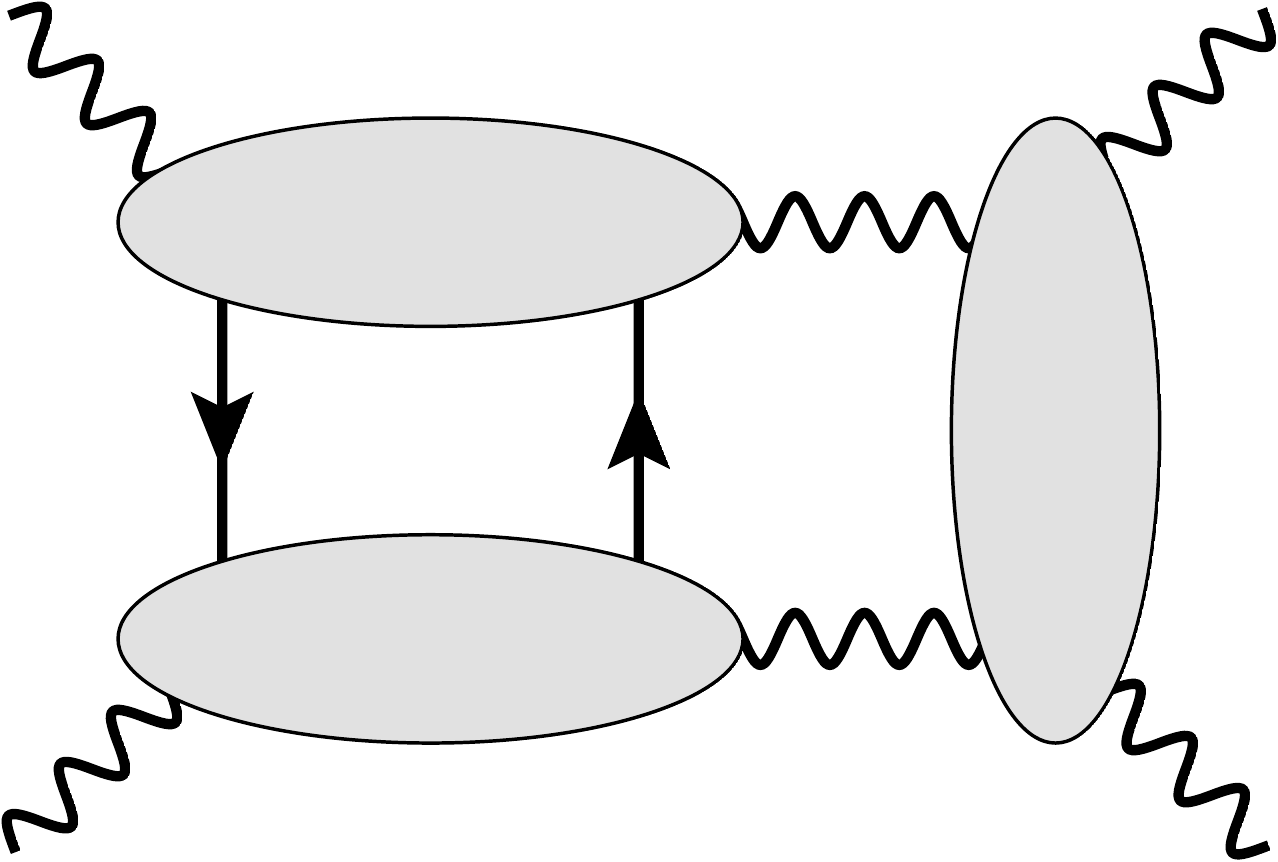}
    \caption[a]{}
  \end{subfigure}
  \begin{subfigure}{0.25\textwidth}
    \centering
    \includegraphics[width=0.8\textwidth]{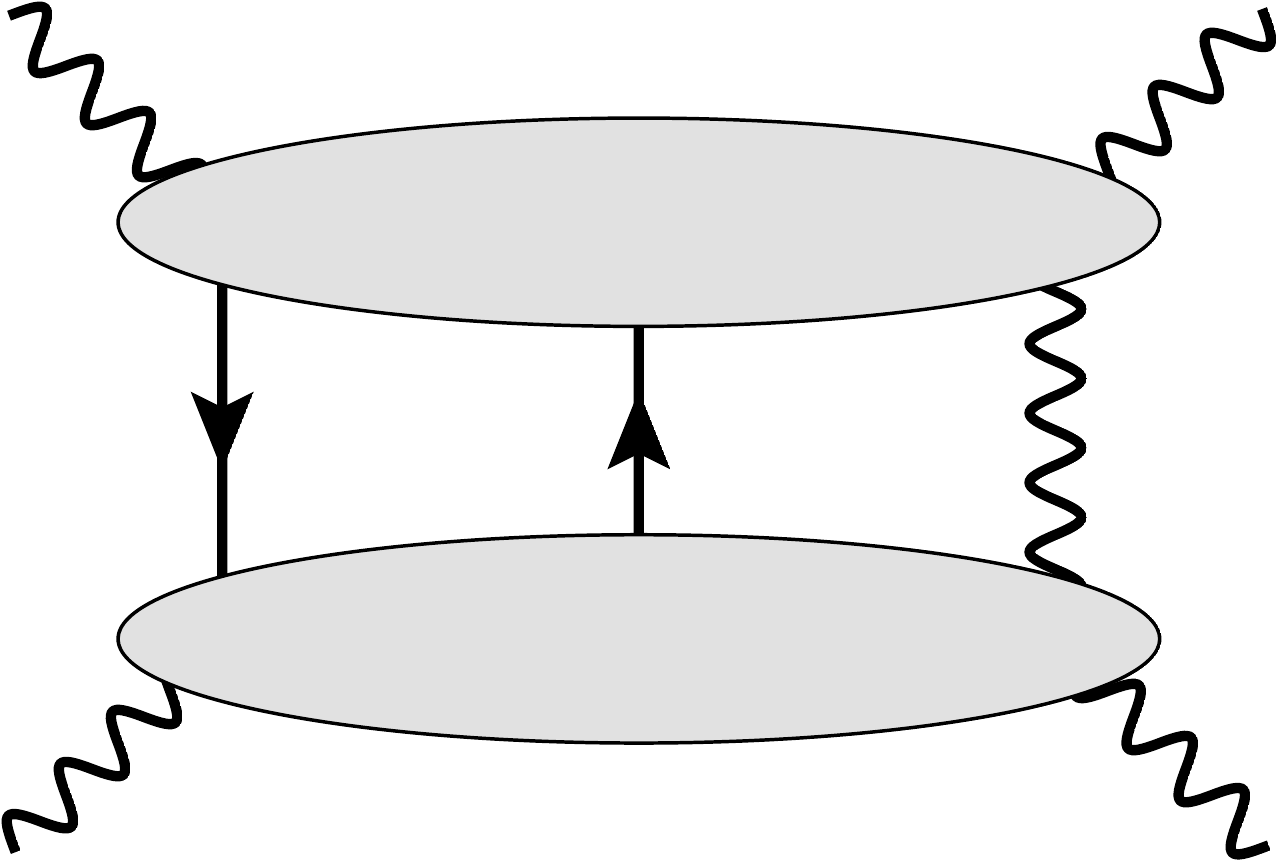}
    \caption[a]{}
  \end{subfigure}
  \begin{subfigure}{0.25\textwidth}
    \centering
    \includegraphics[width=0.8\textwidth]{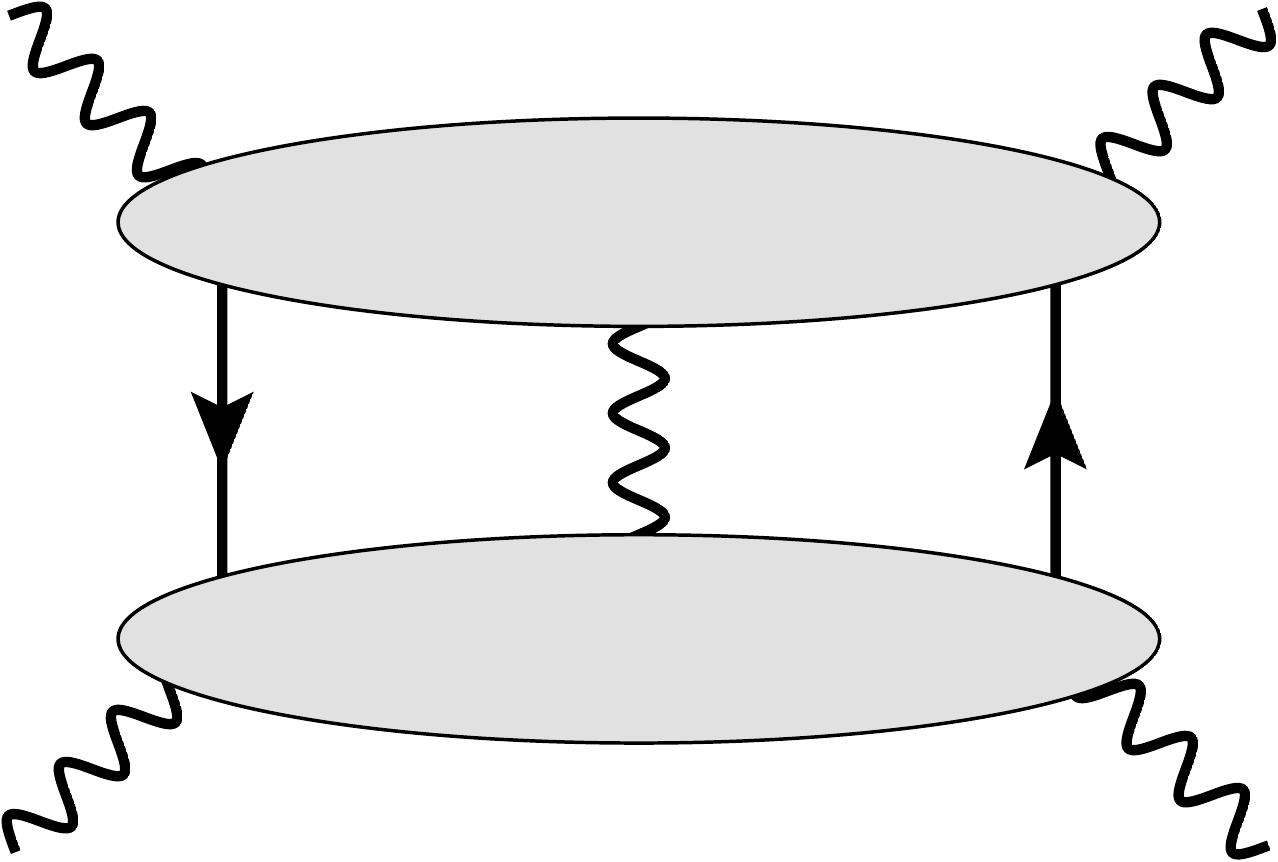}
    \caption[a]{}
  \end{subfigure}
  \caption{\small The eight two-loop spanning cuts involving hypermultiplets required at four points.  All exposed propagators are understood as being taken on shell.  Notice that cuts (b) and (c) differ by the routing of fundamental hypermultiplets on the right-hand side.
    \label{fig:hypercuts}}
\end{figure} 

The two-loop cuts that we wish to calculate are displayed in figs.~\ref{fig:purecuts} and \ref{fig:hypercuts}.  As the one-loop feature of complex extra-dimensional momenta related to chiral states remains, at two loops we choose to keep all the degrees of freedom of the six-dimensional momenta.  The integrands presented will therefore be unavoidably chiral.  We will continue to use the complex $m$ and $\mt$ variables, which we generalize to $m_i$ and $\mt_i$ for different loop momenta $\ell_i=\bar{\ell}_i+\mu_i$.  These are related to the extra-dimensional parts of the loop momenta by
\begin{align}
\mu_{ij}=\mt_{(i}m_{j)}, && \eps(\mu_i,\mu_j)=i\,\mt_{[i}m_{j]},
\end{align}
where $\mu_{ij}=-\mu_i\cdot\mu_j>0$.  Again, for the four-point amplitude the issue of complex momenta is only relevant to the all-chiral sector, while the MHV sector will also have a chiral dependence introduced through the antisymmetric combination $\eps(\mu_1,\mu_2)$.  As we shall see, a non-chiral $D$-dimensional amplitude will emerge in half-maximal supergravity when the double copy is performed between the amplitudes of a chiral and an anti-chiral gauge theory.

The procedure for calculating the two-loop cuts is exactly the same as we have outlined at one loop. A minor distinction is that we avoid excessive analytical manipulations of the cuts: for the purposes of fitting an ans\"{a}tze in the next section we found it sufficient to compute the cuts numerically on various phase-space points.

\section{Calculation of \texorpdfstring{$\cN=2$}{N=2} SQCD numerators}
\label{sec:SQCDcalc}

\renewcommand\thesubfigure{\arabic{subfigure}}
\begin{figure}[t]
  \centering
  \begin{subfigure}{0.19\textwidth}
    \centering
    \includegraphics[keepaspectratio=false,scale=1,clip=true]{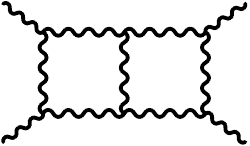}
    \caption{}
  \end{subfigure}
  \begin{subfigure}{0.19\textwidth}
    \centering
    \includegraphics[keepaspectratio=true,scale=1,clip=true]{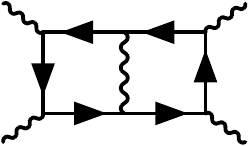}
    \caption{}
  \end{subfigure}
  \begin{subfigure}{0.19\textwidth}
    \centering
    \includegraphics[keepaspectratio=true,scale=1,clip=true]{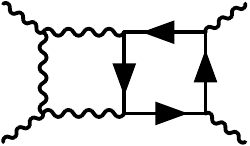}
    \caption{}
  \end{subfigure}
  \begin{subfigure}{0.19\textwidth}
    \centering
    \includegraphics[keepaspectratio=false,scale=1,clip=true]{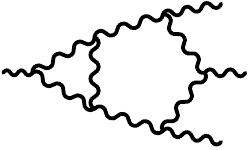}
    \caption{}
  \end{subfigure}
  \begin{subfigure}{0.19\textwidth}
    \centering
    \includegraphics[keepaspectratio=true,scale=1,clip=true]{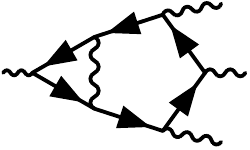}
    \caption{}
  \end{subfigure}
  \begin{subfigure}{0.19\textwidth}
    \centering
    \includegraphics[keepaspectratio=true,scale=1,clip=true]{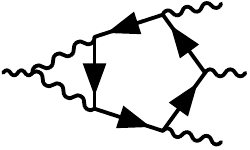}
    \caption{}
  \end{subfigure}
  \begin{subfigure}{0.19\textwidth}
    \centering
    \includegraphics[keepaspectratio=true,scale=1,clip=true]{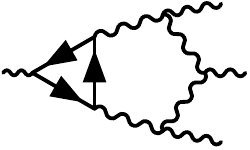}
    \caption{}
  \end{subfigure}
  \begin{subfigure}{0.19\textwidth}
    \centering
    \includegraphics[keepaspectratio=false,scale=1,clip=true]{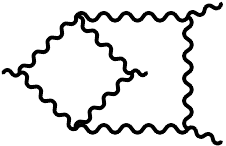}
    \caption{}
  \end{subfigure}
  \begin{subfigure}{0.19\textwidth}
    \centering
    \includegraphics[keepaspectratio=false,scale=1,clip=true]{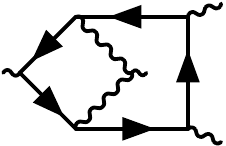}
    \caption{}
  \end{subfigure}
  \begin{subfigure}{0.19\textwidth}
    \centering
    \includegraphics[keepaspectratio=false,scale=1,clip=true]{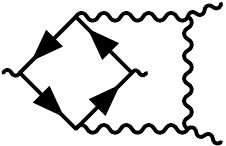}
    \caption{}
  \end{subfigure}
  \begin{subfigure}{0.19\textwidth}
    \centering
    \includegraphics[keepaspectratio=false,scale=1,clip=true]{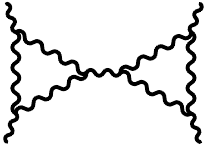}
    \caption{}
  \end{subfigure}
  \begin{subfigure}{0.19\textwidth}
    \centering
    \includegraphics[keepaspectratio=true,scale=1,clip=true]{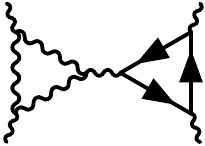}
    \caption{}
  \end{subfigure}
  \begin{subfigure}{0.19\textwidth}
    \centering
    \includegraphics[keepaspectratio=true,scale=1,clip=true]{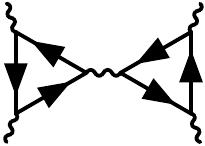}
    \caption{}
  \end{subfigure}
  \begin{subfigure}{0.19\textwidth}
    \centering
    \includegraphics[keepaspectratio=false,scale=1,clip=true]{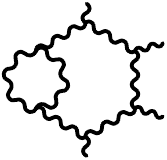}
    \caption{}
  \end{subfigure}
 \begin{subfigure}{0.19\textwidth}
    \centering
    \includegraphics[keepaspectratio=true,scale=1,clip=true]{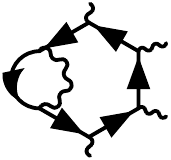}
    \caption{}
  \end{subfigure}
  \begin{subfigure}{0.19\textwidth}
    \centering
    \includegraphics[keepaspectratio=true,scale=1,clip=true]{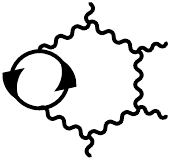}
    \caption{}
  \end{subfigure}
  \begin{subfigure}{0.19\textwidth}
    \centering
    \includegraphics[keepaspectratio=false,scale=1,clip=true]{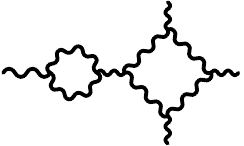}
    \caption{}
  \end{subfigure}
  \begin{subfigure}{0.19\textwidth}
    \centering
    \includegraphics[keepaspectratio=true,scale=1,clip=true]{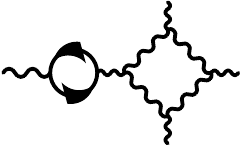}
    \caption{}
  \end{subfigure}
  \begin{subfigure}{0.19\textwidth}
    \centering
    \includegraphics[keepaspectratio=true,scale=1,clip=true]{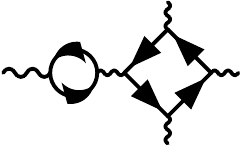}
    \caption{}
  \end{subfigure}
  \begin{subfigure}{0.19\textwidth}
    \centering
    \includegraphics[keepaspectratio=false,scale=1,clip=true]{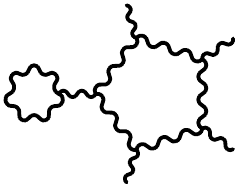}
    \caption{}
  \end{subfigure}
  \begin{subfigure}{0.23\textwidth}
    \centering
    \includegraphics[keepaspectratio=false,scale=.85,clip=true]{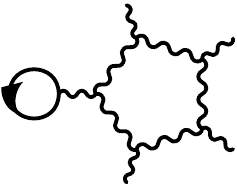}
    \caption{}
  \end{subfigure}
  \begin{subfigure}{0.23\textwidth}
    \centering
    \includegraphics[keepaspectratio=true,scale=.85,clip=true]{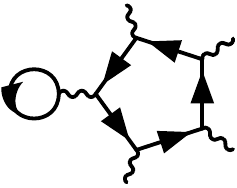}
    \caption{}
  \end{subfigure}
  \begin{subfigure}{0.23\textwidth}
    \centering
    \includegraphics[keepaspectratio=false,scale=.85,clip=true]{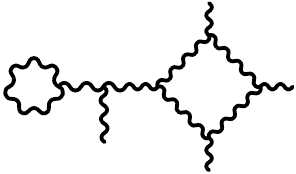}
    \caption{}
  \end{subfigure}
  \begin{subfigure}{0.23\textwidth}
    \centering
    \includegraphics[keepaspectratio=false,scale=.85,clip=true]{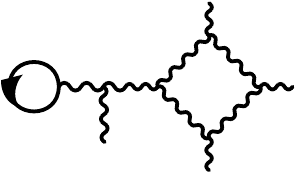}
    \caption{}
  \end{subfigure}
  \caption{\small A complete list of non-vanishing graphs and graphs corresponding to master numerators. The eight master graphs that we choose to work with are~(1)--(5), (13), (19) and (22). While tadpoles and external bubbles are dropped from the final amplitudes it is useful to consider them at intermediate steps of the calculation.
    \label{fig:masterdiags}}
\end{figure}

In this section we describe the computation of color-dual $\cN=2$ SQCD numerators.  All external states are taken from the vector multiplet $\cV_{\cN=2}$~(\ref{eq:neq2multiplet}). Without loss of generality, we consider color-dual numerators with a single hypermultiplet flavor ($N_f=1$); powers of $N_f$ are straightforward to restore at the end by counting the number of hyper loops in a given diagram. We compute both the all-chiral-vector and MHV external-helicity sectors, and as the discussion for the two sectors is mostly the same, we present them alongside each other.  The full list of diagrams involved in our all-chiral and MHV solutions is given in fig.~\ref{fig:masterdiags}.  Expressions for their numerators are presented in section~\ref{sec:allplussolns} and appendix~\ref{sec:MHV_sol} respectively.  The solutions are also provided in ancillary files attached to the \texttt{arXiv} version of this paper.

In the MHV sector we provide two alternative solutions.  The first of these includes non-zero numerators corresponding to bubble-on-external-leg and tadpole diagrams --- diagrams (17)--(24) in fig.~\ref{fig:masterdiags}.  All of these diagrams have propagators of the form $1/p^2 \sim 1/0$ that are ill-defined in the on-shell limit (unless amputated away).  Their appearance in the solution follow from certain desirable but auxiliary relations that we choose to impose on the color-dual numerators, making them easier to find through an ansatz.  As we shall see, contributions from these diagrams either vanish upon integration or can be dropped because the physical unitarity cuts are insensitive to them.  However, as they potentially can give non-vanishing contributions to the ultraviolet (UV) divergences in $\cN=2$ SQCD, we also provide an alternative color-dual solution in which such terms are manifestly absent.  The two solutions may be found in separate ancillary files.

To aid the discussion we generally represent numerators pictorially (as we have already done for color factors and cuts).  For example, the double-box numerator is represented as
\begin{align}
  n_1(1234;\ell_1,\ell_2)=n\bigg(\usegraph{9}{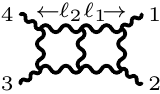}\bigg).
\end{align}
Similar correspondences are made for the other diagrams in fig.~\ref{fig:masterdiags}.

\subsection{Master numerators}

We begin by identifying a basis of master numerators for two-loop four-point diagrams and for external vector multiplets.  All other numerators, called descendants, are expressed as linear combinations of the masters using Jacobi identities.  For instance, double-triangle numerators are given as differences of double boxes:
\begin{align}\label{eq:doubletrianglejacobi}
  \begin{aligned}
    n_{11}(1234;\ell_1,\ell_2)&=
    n_1(1234;\ell_1,\ell_2)-n_1(1243;\ell_1,p_3+p_4-\ell_2),\\
    n\bigg(\usegraph{9}{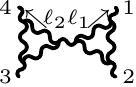}\bigg)&=
    n\bigg(\usegraph{9}{nBoxBox}\bigg)-n\bigg(\usegraph{10}{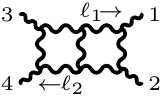}\bigg).
  \end{aligned}
\end{align}
Considering the combined equation system of all all Jacobi (\ref{eq:jacobi1}) and commutation (\ref{eq:jacobi2}) relations, it is straightforward to find eight masters: two with no internal matter content, three with a single matter loop and three with two matter loops.  The same choice is made for both the all-chiral and MHV sectors: these are specified in fig.~\ref{fig:masterdiags}.  Our task is to calculate expressions for the numerators of these diagrams.

The reduction of a given numerator to a linear combination of masters is generally not unique.  Additional numerator relations therefore give a first set of consistency constraints for the masters.  Further constraints come from (i) the requirement that all numerators satisfy the symmetries of their color factors and (ii) that the overall integrand matches the unitarity cuts given in figs.~\ref{fig:purecuts} and \ref{fig:hypercuts}.  In the MHV sector, we satisfy these requirements by finding suitable ans\"{a}tze for the master numerators and solving for constraints on the relevant coefficients.  In the all-chiral sector this is unnecessary: the system is sufficiently simple that the solution can simply be postulated by knowledge of the maximal cuts.

\subsection{Ansatz construction}

Our procedure for constructing ans\"{a}tze for the eight MHV-sector master numerators is a straightforward two-loop extension of the discussion in ref.~\cite{Johansson:2014zca}.  The ansatz for each master numerator is of the same form.  The basic building blocks include the Mandelstam invariants $s_{ij}=(p_i+p_j)^2$ together with contractions of loop momenta~$\ell_i$ $(i=1,2)$ and external momenta~$p_i$ $(i=1, \dots, 4)$.  As in section~\ref{sec:oneloopcut} we decompose $D$-dimensional loop momenta into their four- and extra-dimensional parts as $\ell_i=\bar{\ell}_i+\mu_i$.  The following set of kinematic variables is then sufficient:
\begin{equation}
  M=\left\{s_{ij},\ell_i^s,\ell_i^t,\ell_i^u,\ell_i\cdot\ell_j,\mu_{ij}\right\},
  \label{SetM}
\end{equation}
where
\begin{equation}
  \begin{aligned}
    \ell_i^s &\equiv 2 \ell_i \cdot(p_1 + p_2), & \ell_i^t &\equiv 2 \ell_i \cdot(p_2 + p_3), & \ell_i^u &\equiv 2 \ell_i \cdot(p_3 + p_1), \\
    & & \mu_{ij} &= - \mu_i \cdot \mu_j.
  \end{aligned}
\end{equation}
Note that $M$ contains 14 independent objects after accounting for $u=-s-t$, where $s=s_{12}$, $t=s_{23}$ and $u=s_{13}$.  We also include contractions with the four-dimensional Levi-Civita tensor:
\begin{equation}
  \Xi = \left\{\eps(1, 2, 3, \bar{\ell}_1), \eps(1, 2, 3, \bar{\ell}_2), \eps(1, 2, \bar{\ell}_1, \bar{\ell}_2), \eps(2, 3, \bar{\ell}_1, \bar{\ell}_2), \eps(3, 1, \bar{\ell}_1, \bar{\ell}_2)\right\},
\end{equation}
which acts only on the four-dimensional parts of the loop momenta. To be specific about normalizations, given some momenta $k_{i}$, we define $\eps(k_1,k_2,k_3,k_4)\equiv{\rm Det}(k_{i \mu})$.

The last ingredient is the antisymmetric combination $\eps(\mu_1,\mu_2)$ --- we saw earlier how it arises from the chiral nature of the $\cN=(1,0)$ theory (see section~\ref{sec:2lcuts}).  It is an extra-dimensional echo of the six-dimensional Levi-Civita tensor:
\begin{equation}
  \eps(\mu_1,\mu_2) =\frac{\eps^{(6)}(v_1,v_2,v_3,v_4,\ell_1,\ell_2)}{\eps(v_1,v_2,v_3,v_4)}={\rm Det}(\mu_1, \mu_2) ,
\end{equation}
where $v_i$ are some four-dimensional vectors, such as external momenta and polarizations (their specific form is not important since they cancel out in the above ratio).  It satisfies $\eps(\mu_1,\mu_2)^2=\mu_{11}\mu_{22}-\mu_{12}^2$.

We label external states using four-dimensional variables: in section~\ref{sec:treesAndCuts} we introduced variables~$\kappa_{ij}$ that carry the correct helicity weight for all MHV helicity configurations~(\ref{eq:kappa}).  The indices $i$ and $j$ label the legs belonging to the anti-chiral part of the vector multiplet $\widebar{V}_{\cN=2}$, i.e. those containing negative-helicity gluons.  With these building blocks, sufficient ans\"{a}tze for the master numerators are
\begin{equation}\label{eq:mhvansatz}
  n_i(1234;\ell_1,\ell_2)\!=\!
  \sum_{j<k}\frac{\kappa_{jk}}{s_{jk}^2}\!\left(\sum_mc_{i,jk}^m M_m^{(3)}\!+\!
  i\sum_{m,n}d_{i,jk}^{mn} \Xi_m M_n^{(1)}\!+\!
  i\sum_m e_{i,jk}^m \eps(\mu_1,\mu_2)M_m^{(2)}\!\right).
\end{equation}
The objects $c_{i,jk}^m$, $d_{i,jk}^{mn}$ and $e_{i,jk}^m$ are free parameters to be solved for; $M^{(N)}$ denotes the set of monomials of engineering dimension $2N$ built from the set $M$ in \eqn{SetM}:
\begin{equation}
  M^{(N)}=\left\{\prod_{i=1}^N z_i ~\bigg|~ z_i \in M\right\}.
\end{equation}
Note that there exist non-linear relations between the monomials in $M^{(N)}$ and the Levi-Civita objects in $\Xi$.  In principle, we could use these to slightly reduce the size of the ansatz; however, the benefit in doing so is marginal. Furthermore, working with an over-determined set of building blocks is helpful when searching for a compact form of the integrand.

Finally, we should comment on the fact that the kinematic numerators we choose to work with have poles of the form $\kappa_{jk}/s_{jk}^2$. Because we choose to work with gauge-invariant building blocks (as opposed to polarization-dependent numerators) the price to pay is some non-locality in the numerators.\footnote{One reason for this is that the number of gauge-invariant local functions one can write down is strongly dependent on the engineering dimension. If one wants to have access to sufficiently many of these while keeping the overall dimension small one is forced to compensate with momentum factors in the denominator.} While the specific form of the non-locality is not a priori obvious we find that the observation of ref.~\cite{Johansson:2014zca} carries over to two loops: the MHV numerators have at worst $\kappa_{jk}/s_{jk}^{4-\cN}$ poles, where $\cN$ is the number of supersymmetries. 

\subsection{Symmetries and unitarity cuts}\label{sec:symmetriesandcuts}

Graph symmetries, or automorphisms, can be used to obtain relations between numerators with different external-leg orderings and differently parametrized loop momenta.  For instance, the pure-adjoint double-box numerator has two symmetries: mirroring through the horizontal and vertical axes,
\begin{align}
  n\bigg(\usegraph{9}{nBoxBox}\bigg)=
  n\bigg(\usegraph{10}{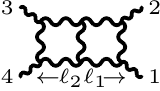}\bigg)=
  n\bigg(\usegraph{9}{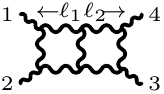}\bigg).
\end{align}
Non-uniqueness of the numerators in terms of the masters leaves further graph isomorphisms to solve for.  In the MHV sector we implement the graph isomorphisms of all graphs as constraints on the ans\"{a}tze of the eight master numerators.

Further physical information comes from demanding that the color-dual form of the integrand matches the required spanning cuts: three of these are pure-adjoint (fig.~\ref{fig:purecuts}) and the remaining eight contain hypermultiplets in the loops (fig.~\ref{fig:hypercuts}) with $N_f=1$.  In the latter case there are six four-particle cuts and three three-particle cuts.  However, we find that the three-particle cuts do not provide any new information: once an ansatz solution satisfying the four-particle cuts and automorphisms has been found, evaluating the resulting integrand on the three-particle cuts forces all remaining freedom to cancel.  For this reason, we found it unnecessary to evaluate these cuts in six dimensions (in four dimensions these cuts provided a useful check).

\subsection{Additional constraints}

Merely enforcing symmetries and unitarity cuts does not provide unique all-chiral or MHV color-dual numerators.  We therefore seek additional  auxiliary constraints to impose on our numerators.  Here we discuss three in particular: these constraints, while having no effect on the double copy, helpfully clean up the remaining parameter space while revealing important physical properties.

\subsubsection{Matter-reversal symmetry}
\label{sec:matterReversalSym}

We impose invariance under direction reversal of hypermultiplets.  For instance,
\begin{align}
  n\bigg(\usegraph{9}{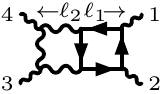}\bigg)=n\bigg(\usegraph{9}{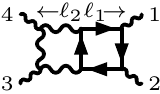}\bigg).
\end{align}
This constraint was already used in ref.~\cite{Johansson:2014zca}: it is made possible by the $\cN=2$ fundamental and antifundamental hypermultiplets~(\ref{eq:hypermultiplets}) having the same particle content up to the R-symmetry index.

\subsubsection{Two-term identities}
\label{sec:twoTermIdentities}
The two-term identities were first introduced in eq.~(\ref{eq:twoTerm}); they are consistent with all unitarity cuts given that $N_f=1$.  They effectively reduce the number of master numerators to five due to these three relations:
\begin{subequations}\label{eq:twoTerm_explicit}
  \begin{align}
    n\bigg(\usegraph{9}{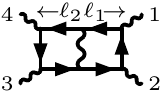}\bigg) &= n\bigg(\usegraph{9}{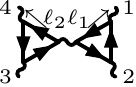}\bigg),\label{eq:db2term}\\
    n\bigg(\!\usegraph{9}{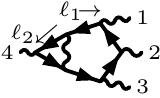}\bigg) &= n\bigg(\!\usegraph{9}{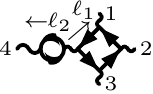}\bigg),\\
    n\bigg(\!\usegraph{9}{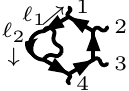}\!\bigg) &=  n\bigg(\!\!\usegraph{12}{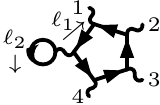}\bigg),
  \end{align}
\end{subequations}
where each numerator on the right-hand side is a master.  Numerators of diagrams containing two hypermultiplet loops are completely determined in terms of those containing only a single hypermultiplet loop.

\subsubsection{\texorpdfstring{Matching with the $\cN=4$}{N=4} limit}\label{sec:neq4idents}

The observation that $\cN=2$ SQCD's field content is contained by that of $\cN=4$ SYM has already proved useful: we used it in section~\ref{sec:treesAndCuts} to obtain four-dimensional tree amplitudes by projecting out of those in $\cN=4$ SYM.  The connection between these theories offers another constraint on our color-dual numerators: that summing $\cN=2$ SQCD numerators over internal multiplets corresponding to the field content of $\cN=4$ SYM should reproduce those color-dual numerators.  Again taking the double-box numerator as an example, we demand that
\begin{align}
\label{Neq4relation}
  &n^{[\cN=4\text{ SYM}]}\bigg(\usegraph{9}{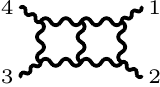}\bigg)\\
  &\qquad=n\bigg(\usegraph{9}{nBoxBox}\bigg)+ 2 n\bigg(\usegraph{9}{nBoxBarBox}\bigg) + 2 n\bigg(\usegraph{9}{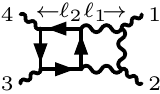}\bigg) + 2 n\bigg(\usegraph{9}{nBoxBoxBar}\bigg) ,\nn
\end{align}
where the factors of 2 come from our implementation of matter-reversal symmetry.  Similar identities should hold for all other topologies.

In the all-chiral sector this is easily implemented as the corresponding $\cN=4$ SYM numerators are all zero.  In the MHV sector there are two non-zero numerators~\cite{Bern:2008qj}:
\begin{subequations}
  \begin{align}
    n^{[\cN=4\text{ SYM}]}\bigg(\usegraph{10}{nBoxBoxi}\bigg)&=
    s(\kappa_{12}+\kappa_{13}+\kappa_{14}+\kappa_{23}+\kappa_{24}+\kappa_{34}),\\
    n^{[\cN=4\text{ SYM}]}\bigg(\!\!\usegraph{8}{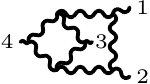}\bigg)&=
    s(\kappa_{12}+\kappa_{13}+\kappa_{14}+\kappa_{23}+\kappa_{24}+\kappa_{34}),
  \end{align}
\end{subequations}
where we have expanded out the components of the $\cN=4$ supersymmetric delta function $\delta^8(Q)$ into pieces that have $\cN=2$ supersymmetry.

This matching to the $\cN=4$ limit goes together naturally with the two-term identities.  To illustrate this, consider forming a double-triangle numerator using the Jacobi identity~(\ref{eq:doubletrianglejacobi}):
\begin{align}
  \begin{aligned}
    &n^{[\cN=4\text{ SYM}]}\bigg(\usegraph{9}{nTriTri}\bigg)=
    n^{[\cN=4\text{ SYM}]}\bigg(\usegraph{9}{nBoxBox}\bigg)-
    n^{[\cN=4\text{ SYM}]}\bigg(\usegraph{10}{nBoxBox1243}\bigg)\\
    &\qquad=n\bigg(\usegraph{9}{nTriTri}\bigg) + 2 n\bigg(\usegraph{9}{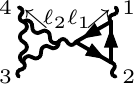}\bigg)+2n\bigg(\usegraph{9}{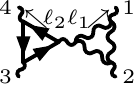}\bigg) + 4 n\bigg(\usegraph{9}{nTriBarTriBar}\bigg).
  \end{aligned}
\end{align}
In the second line we have decomposed the double-triangle numerator into its $\cN=2$ SQCD constituents like we did for the double box.  The above statements are naturally consistent with each other when the two-term identity between the double-box and double-triangle numerators~(\ref{eq:db2term}) is enforced.

Another beneficial aspect of matching to $\cN=4$ SYM is that it further reduces the number of masters required.  All of the off-shell identities identities contain a single SQCD numerator with pure-adjoint particle content.  Therefore, the identities can be used to uniquely specify pure-adjoint $\cN=2$ numerators in terms of those containing hypermultiplets.  Taken together with the two-term identities, this means that we need only solve for numerators whose diagrams contain a single hypermultiplet loop: in our case we need only the three masters (2), (3) and (5) in fig.~\ref{fig:masterdiags}.

\subsection{All-chiral solutions}\label{sec:allplussolns}

The all-chiral-vector sector of $\cN=2$ SQCD was not discussed in ref.~\cite{Johansson:2014zca}; we include the one-loop solution here for completeness. It will integrate to zero, but it is needed for the subsequent supergravity construction.  There are two non-zero numerators, both of them boxes:
\begin{subequations}\label{eq:1lplusnumerators}
  \begin{align}
    n\bigg(\usegraph{10}{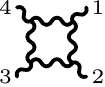}\bigg)&=
    2\cT\delta^4(Q)m^2,\\
    n\bigg(\usegraph{10}{nBoxBar}\bigg)&=
    -\cT\delta^4(Q)m^2,
  \end{align}
\end{subequations}
where $m$ is defined in \eqn{mmtilde} and we have introduced the permutation-invariant prefactor
\begin{align}\label{eq:cT}
  \cT=\frac{[12][34]}{\braket{12}\braket{34}}.
\end{align}
We recall that working with tree amplitudes in six-dimensional $\cN = (1,0)$~SYM leads to all-chiral numerators expressed in terms of purely-four dimensional objects together with the $(6-4)$-dimensional variables~$m_i$.  The latter numerator satisfies the box cut~(\ref{eq:hyperboxcutplus}) and a similar check can easily be done on the former.

There are nine non-vanishing two-loop numerators.  Due to the simplicity of the cuts for the all-chiral helicity configuration it is possible to postulate and check the numerators' expressions directly; the two-term and $\cN = 4$ identities require that the solution is unique.  Six of them have the topology of a planar or non-planar double box:\begin{subequations}\label{eq:2lplussoln}
  \begin{align}
    n\bigg(\usegraph{9}{nBoxBox}\bigg)=n\bigg(\!\!\usegraph{9}{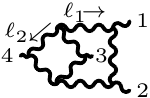}\bigg)&=
    2\cT\delta^4(Q)s(m_3^2-m_1m_2),\\
    n\bigg(\usegraph{9}{nBoxBoxBar}\bigg)=n\bigg(\!\!\usegraph{9}{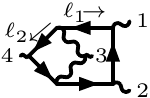}\bigg)&=
    \cT\delta^4(Q)sm_1m_2,\\
    n\bigg(\usegraph{9}{nBoxBarBox}\bigg)&=-\cT\delta^4(Q)sm_1m_3,\\
    n\bigg(\!\usegraph{9}{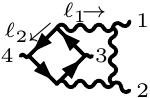}\bigg)&=-\cT\delta^4(Q)sm_2m_3,
  \end{align}
where $m_3=m_1+m_2$.  The other three non-zero numerators correspond to double-triangle diagrams:
  \begin{align}
    n\bigg(\usegraph{9}{nTriTri}\bigg)&=4\cT\delta^4(Q)sm_1m_2,\\
    n\bigg(\usegraph{9}{nTriBarTri}\bigg)&=-2\cT\delta^4(Q)sm_1m_2,\\
    n\bigg(\usegraph{9}{nTriBarTriBar}\bigg)&=\cT\delta^4(Q)sm_1m_2.
  \end{align}
\end{subequations}
We encourage the reader to check that these nine numerators match to $\cN=4$ SYM and satisfy matter-reversal and two-term identities; also that they satisfy Jacobi identities, commutation relations and graph symmetries.  That all of these numerators integrate to zero follows straightforwardly from averaging over global rotations and parity of the extra-dimensional momentum space.

\subsection{MHV solutions, bubbles and tadpoles}\label{sec:bubblestadpoles}

Our first two-loop MHV solution was found by matching to $\cN=4$ SYM and demanding matter-reversal and two-term identities (as described above).  We also imposed the vanishing of a pentagon-triangle master numerator --- diagram (5) in fig.~\ref{fig:masterdiags} --- which fixed all remaining freedom.  The resulting solution is particularly compact: it contains the nineteen non-zero numerators listed in appendix~\ref{sec:MHV_sol}.  Eight of these are pure-adjoint and the rest contain matter hypermultiplets.  The solution is also given in the ancillary files attached to the \texttt{arXiv} version of this paper. 

This MHV solution also includes numerators corresponding to bubble-on-external-leg and tadpole diagrams that are singular because of propagators being automatically on shell.  We remarked earlier that these diagrams either vanish after integration (if carefully regulated) and/or can be dropped since the physical unitarity cuts are insensitive to them. Let us explain this in more detail.  If we focus on only the relevant integrals (and not on the external propagators), the integrals are scaleless, and thus on general grounds are expected to vanish in dimensional regularization. However, because of the singular external propagators the details are slightly more intricate. 

Consider two typical integrals we encounter with numerators listed in appendix~\ref{sec:MHV_sol}:
\begin{subequations}\label{eq:bubbleTadBehav}
  \begin{align}
    n\bigg(\!\usegraph{9}{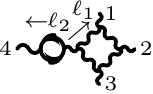}\!\bigg) &\propto 2\, \ell_2\cdot(p_4 - \ell_2),\label{eq:bubBehav}\\
    n\bigg(\!\!\usegraph{12}{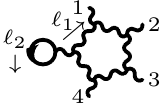}\bigg) &\propto \ell_1\cdot\ell_2\,\label{eq:tad1Behav} .
  \end{align}
\end{subequations}
The tadpole integral (\ref{eq:tad1Behav}) is trivially zero since the integrand is odd in $\ell_2$.  For the bubble (\ref{eq:bubBehav}) we use off-shell external momenta $p_4^2\neq 0$ to regulate the singular denominator.  We can rewrite the numerator as $p_4^2-\ell_2^2-(p_4-\ell_2)^2$, where the last two terms correspond to scaleless tadpole integrals that can be dropped. After the first term is canceled against the $1/p_4^2$ pole the integrand becomes non-singular and we can safely take the $p_4^2 \rightarrow 0$ limit. The remaining bubble diagram is then also scaleless, thus all terms are zero in dimensional regularization.  

In $\cN=2$ SQCD the above integrals can be dropped, but in half-maximal supergravity the same diagrams will appear with squared numerators through the double copy.  Will the bubble-on-external-leg and tadpole integrals also vanish in that case? Yes they will. The bubble integral~(\ref{eq:bubBehav}) behaves as
\begin{align}
  \frac{1}{p_4^2}\int\!\frac{\d^D\ell_2}{(2\pi)^D}\frac{\big[2\, \ell_2\cdot(p_4-\ell_2)\big]^2}{\ell_2^2 (\ell_2-p_4)^2}
  \propto(-p_4^2)^{D/2} ~ \mathop{\longrightarrow}_{p_4^2 \rightarrow 0} ~ 0\,,
\end{align}
where again $p_4$ is kept off shell as a regulator until after integration. 

Similarly, the gravitational tadpole integral~(\ref{eq:tad1Behav})  can be argued to vanish by promoting the soft line to have a tiny amount of off-shell momentum~$k^2 \neq 0$ (for example, by considering a weakly-curved background that slightly violates energy conservation). Injecting momentum $k$ into the vertex nearest the tadpole integral~(\ref{eq:tad1Behav}) still gives a scaleless integral:
\begin{align}
  \frac{1}{k^2}\int\! \frac{\d^D\ell_2}{(2\pi)^D} \frac{\big[\ell_1\cdot\ell_2\big]^2}{\ell_2^2}  =0.
\end{align}
Alternatively, injecting it by creating a new vertex on the tadpole $\ell_2$ line takes us back to the above bubble integral.

We also consider a second MHV solution. If we relax the two-term identities (\ref{eq:twoTerm}) and the requirement of a simple relation to the $\cN=4$ SYM numerators (\ref{Neq4relation}), this allows for an alternative presentation that has manifestly-vanishing bubble-on-external-leg and tadpole diagrams (it also satisfies matter-reversal symmetry).  It contains six pure-adjoint numerators and twelve with hypermultiplets. While the expressions for each numerator in this solution are slightly less compact, it has somewhat better UV power counting for each individual diagram and loop momenta. The main reason for this is the absence of the bubble-on-external-leg and tadpole diagrams, which are the worst-behaved diagrams by superficial power counting.  Given the absence of these diagrams, and the good power counting, it  should be useful for studies of the UV behavior of supergravities that can be obtained as a double copy involving $\cN=2$ SQCD. The solution is given in appendix~\ref{sec:MHV_sol}, as well as in the ancillary files of the \texttt{arXiv} version of this paper.

\section{Half-maximal supergravity amplitudes}
\label{sec:HMSG}

We will now use the double copy to construct four-point $\cN=4$ (half-maximal) supergravity amplitudes coupled to an arbitrary number $N_V$ of $\cV_{\cN=4}$ vector multiplets. As discussed in the previous section, we can drop all bubble-on-external-leg and tadpole integrals. Then, in both helicity configurations currently being considered, the one-loop amplitude is presented in the form
\begin{align}
  \cM^{(1),[\cN=4\text{ SG}]}_4
  &=\left(\frac{\kappa}{2}\right)^4\sum_{\sigma \in S_4}
  \bigg\{\frac{1}{8}\mathcal{I}^D\bigg[N^{[\cN=4\text{ SG}]}\bigg(\usegraph{9}{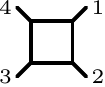}\bigg)\bigg]\\
  & \qquad ~~
  +\frac{1}{4}\mathcal{I}^D\bigg[N^{[\cN=4\text{ SG}]}\bigg(\usegraph{9}{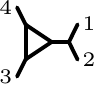}\bigg)\bigg]
  +\frac{1}{16}\mathcal{I}^D\bigg[N^{[\cN=4\text{ SG}]}\bigg(\!\usegraph{5}{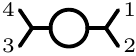}\!\bigg)\bigg]\bigg\},\nn
\end{align}
and the two-loop amplitude is
\begin{align}\label{eq:sgamp}
  \cM^{(2),[\cN=4\text{ SG}]}_4
  &=i\left(\frac{\kappa}{2}\right)^6 \sum_{\sigma \in S_4}
    \bigg\{\frac{1}{4}\mathcal{I}^D\bigg[N^{[\cN=4\text{ SG}]}\bigg(\usegraph{10}{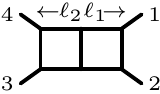}\bigg)\bigg]\\
    & \qquad ~~
    +\frac{1}{2}\mathcal{I}^D\bigg[N^{[\cN=4\text{ SG}]}\bigg(\!\usegraph{9}{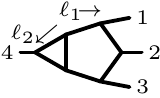}\bigg)\bigg] 
    +\frac{1}{4}\mathcal{I}^D\bigg[N^{[\cN=4\text{ SG}]}\bigg(\!\!\usegraph{9}{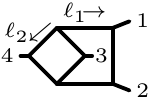}\bigg)\bigg]\nn\\
    &\qquad ~~
    +\frac{1}{8}\mathcal{I}^D\bigg[N^{[\cN=4\text{ SG}]}\bigg(\usegraph{9}{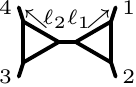}\bigg)\bigg]
    +\frac{1}{4}\mathcal{I}^D\bigg[N^{[\cN=4\text{ SG}]}\bigg(\!\usegraph{9}{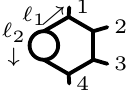}\bigg)\bigg] \bigg\},\nn
\end{align}
where $S_4$ is the four-object permutation group of the external legs.  

The $D$-dimensional $L$-loop integration operator $\mathcal{I}^D$ acts on gravity numerators $N_i$ as
\begin{align}
  \mathcal{I}^D[N_i]\equiv\int\!\frac{\d^{LD}\ell}{(2\pi)^{LD}}\frac{N_i}{D_i},
\end{align}
where $D_i$ are the propagator denominators of graph $i$.  The denominators include poles from all $(1+3L)$ lines in the cubic graphs, not counting the amputated external four legs. The gravity numerators $N_i$ will be assembled from sums of double-copied gauge-theory numerators; the solid lines in the above diagrams indicate that the relevant sum over contributions has been carried out.
When specializing to the all-chiral-helicity sector, to be described below, only the box diagram is non-zero at one loop; the first three diagrams are non-zero at two loops.

Before introducing the $N_i$, let us describe the action of the double copy on the asymptotic states. The particle content of pure $\cN=4$ supergravity is described by a pair of CPT-conjugate multiplets, one chiral and the other anti-chiral, which we combine into a single non-chiral on-shell multiplet~\cite{Carrasco:2013ypa}:\footnote{Note, the gravitini $\psi_I^+$, $\psi^I_-$ are distinct from the similarly-named states in the gauge theory~(\ref{eq:neq4multiplet}).}
\begin{align}\label{eq:neq4sgmultiplet}
  \begin{aligned}
    \cH_{\cN=4}&=H_{\cN=4}+\eta^5\eta^6\eta^7\eta^8\widebar{H}_{\cN=4},\\
    H_{\cN=4}&=h^{++}+\eta^I\psi^+_I+\frac{1}{2}\eta^I\eta^JA^+_{IJ}+\frac{1}{3!}\eps_{IJKL}\eta^I\eta^J\eta^K\chi_+^L+\eta^1\eta^2\eta^3\eta^4\,\bar{t},\\
    \widebar{H}_{\cN=4}&=t+\eta^I\chi^-_I+\frac{1}{2}\eta^I\eta^JA^-_{IJ}+\frac{1}{3!}\eps_{IJKL}\eta^I\eta^J\eta^K\psi_-^L+\eta^1\eta^2\eta^3\eta^4\,h_{--}.
  \end{aligned}
\end{align}
Additional vector multiplets $\cV_{\cN=4}$ are described by eq.~(\ref{eq:neq4multiplet}).  External states in our amplitudes always belong to $\cH_{\cN=4}$: we classify their helicities by whether they belong to $H_{\cN=4}$ or $\widebar{H}_{\cN=4}$.  The all-chiral sector has all of them in $H_{\cN=4}$; the MHV sector has two external states belonging to $\widebar{H}_{\cN=4}$ and the rest to $H_{\cN=4}$.

Half-maximal supergravity amplitudes are particularly accessible because, in addition to being obtainable using the $\cN=2$ SQCD numerators derived above, they are also obtainable using a double copy of $\cN=4$ SYM with pure Yang Mills.  The latter double copy route has already been used to perform several studies on the ultraviolet structure of these amplitudes up to four loops~\cite{Bern:2011rj,BoucherVeronneau:2011qv,Bern:2012cd,Bern:2012gh,Bern:2013qca,Bern:2013uka,Bern:2014lha,Bern:2017lpv}.  Therefore, as a check on the results of the last section, we will compare the results of the two double copies.  In the all-chiral sector we will do this directly at the level of individual numerators; in the MHV sector we will consider the two-loop ultraviolet (UV) structure in $D=5-2\eps$ dimensions.

\subsection{\texorpdfstring{$\cN=0\otimes\cN=4$}{N=0xN=4} construction}

As this double-copy route has been repeatedly considered in the literature, what follows will mostly be a review.  The half-maximal theory is a ``factorizable'' supergravity, meaning that the amplitudes and states may be described as a double copy involving only vector multiplets:
\begin{align}
  \cH_{\cN=4}=\cV_{\cN=0}\otimes\cV_{\cN=4},
\end{align}
where $\cV_{\cN=0}$ is the on-shell vector ``multiplet'' of pure Yang-Mills theory,
\begin{align}
  \cV_{\cN=0}=A^++\eta^1\eta^2\eta^3\eta^4A_-.
\end{align}
This is a projection of the full $\cV_{\cN=4}$ multiplet (\ref{eq:neq4multiplet}).  The $H_{\cN=4}$ and $\widebar{H}_{\cN=4}$ multiplets are each associated with one of the two gluon helicity states:
\begin{align}
  H_{\cN=4}=A^+\otimes\cV_{\cN=4}, && \widebar{H}_{\cN=4}=A_-\otimes\cV_{\cN=4}.
\end{align}
Therefore, the all-chiral and MHV sectors of $\cN=4$ supergravity can be traced back to the double copy of the $\cN=4$ SYM MHV sector with the all-plus and MHV sectors of pure Yang-Mills theory, respectively.  

The extension to include $N_V$ vector multiplets is described in terms of the on-shell states as~\cite{Carrasco:2013ypa}
\begin{align}\label{eq:neq0neq4copy}
  \cH_{\cN=4}\oplus(N_V\, \cV_{\cN=4})=(\cV_{\cN=0}\oplus N_V\, \varphi)\otimes\cV_{\cN=4},
\end{align}
where we now include $N_V$ real scalars $\varphi$ on the non-supersymmetric Yang-Mills side.

\subsubsection{One loop}

The four-point one-loop amplitude for both $\cN=4$ SYM and all-plus sector of Yang-Mills can be described by box diagrams~\cite{Green:1982sw,Bern:1995db}:
\begin{subequations}
  \begin{align}
    n^{[\cN=4\text{ SYM}]}\bigg(\usegraph{10}{nBox}\bigg)&=\cT\delta^8(Q),\\
    n^{[\cN=0\text{ YM}]}\bigg(\usegraph{10}{nBox}\bigg)&=\cT(D_s-2)\mu^4.
  \end{align}
\end{subequations}
The permutation-invariant helicity prefactor $\cT$ was introduced in eq.~(\ref{eq:cT}); the $S_4$ symmetry of these numerators ensures that they are manifestly color dual. The double copy for the $\cN=4$ supergravity amplitude in the all-chiral sector is given by the product of these two gauge-theory numerators
\begin{align}\label{eq:neq4sgbox}
  N^{[\cN=4\text{ SG}]}\bigg(\usegraph{10}{nBoxs}\bigg)=\cT^2(D_s-2)\mu^4\delta^8(Q).
\end{align}
The non-vanishing of this amplitude after integration can be attributed to an anomaly of the R-symmetry U(1) factor of $\cN=4$ supergravity~\cite{Carrasco:2013ypa} (see also ref.~\cite{Freedman:2017zgq}). We will reproduce this amplitude using the one-loop all-chiral $\cN=2$ SQCD numerators.

Let us remark on a useful detail before proceeding to two loops.  Pure Yang-Mills amplitudes, such as the one above, are often written in terms of a state parameter $D_s$ which counts the effective dimension of the gluon states. It may be taken to be independent of the momentum dimension $D$.  Common four-dimensional regularization schemes include the 't~Hooft-Veltman scheme, $D_s=4-2\eps$, and the four-dimensional helicity scheme, $D_s=4$~\cite{Bern:2002zk}.  Via dimensional reduction one may include $N_V=D_s-4$ real scalars $\varphi$ by considering them as additional states of the gluon; by application of the double copy~(\ref{eq:neq0neq4copy}) these become the desired $N_V$ vector multiplets in the supergravity theory.

\subsubsection{Two loops}

The four-point one-loop amplitude of $\cN=4$ SYM can be written in terms of two non-zero color-dual numerators~\cite{Bern:2008qj}.  These were already given in section~\ref{sec:neq4idents}; they can be written more compactly using the full on-shell supersymmetry
\begin{align}
  n^{[\cN=4\text{ SYM}]}\bigg(\usegraph{10}{nBoxBoxi}\bigg)=
  n^{[\cN=4\text{ SYM}]}\bigg(\!\!\usegraph{8}{nBoxBoxNPi}\bigg)=s\,\cT\delta^8(Q).
\end{align}
The corresponding all-plus Yang-Mills numerators were first given in ref.~\cite{Bern:2013yya}. We write the relevant ones on a more compact form found by O'Connell and one of the present authors~\cite{Mogull:2015adi}. We will need the following two
\begin{subequations}
  \begin{align}
    n^{[\cN=0\text{ YM}]}\bigg(\usegraph{10}{nBoxBox}\bigg)&=
    \cT\left(\!s\,F_1(\mu_1,\mu_2)\!+\!\frac{1}{2}(D_s-2)^2\mu_{11}\mu_{22}(\ell_1+\ell_2)^2\right)\!,\\
    n^{[\cN=0\text{ YM}]}\bigg(\!\!\usegraph{8}{nBoxBoxNP}\bigg)&=
    s\,\cT F_1(\mu_1,\mu_2),
  \end{align}
\end{subequations}
and the remaining numerators can be ignored since they do not contribute to the double copy with the $\cN=4$ SYM numerators. We have introduced the following function of extra-dimensional components~\cite{Badger:2013gxa}:
\begin{align}\label{eq:F1}
  F_1(\mu_1,\mu_2)=16(\mu_{12}^2-\mu_{11}\mu_{22})+(D_s-2)(\mu_{11}\mu_{22}+\mu_{11}\mu_{33}+\mu_{22}\mu_{33}),
\end{align}
with $\mu_{33}=\mu_{11}+\mu_{22}+2\mu_{12}$.  The double copy gives three non-zero all-chiral $\cN=4$ supergravity contributions:
\begin{subequations}\label{eq:2lneq4sgnums}
  \begin{align}
    N^{[\cN=4\text{ SG}]}\bigg(\usegraph{10}{nBoxBoxs}\bigg)&=
    \cT^2s^2\,\delta^8(Q)F_1(\mu_1,\mu_2),
    \label{eq:neq4sgBoxBox}\\
    N^{[\cN=4\text{ SG}]}\bigg(\!\!\usegraph{8}{nBoxBoxNPs}\bigg)&=
    \cT^2s^2\,\delta^8(Q)F_1(\mu_1,\mu_2),
    \label{eq:neq4sgBoxBoxNP}\\
    N^{[\cN=4\text{ SG}]}\bigg(\usegraph{9}{nTriTris}\bigg)&=
    (D_s-2)^2\cT^2s^2\,\delta^8(Q)\mu_{11}\mu_{22}.
    \label{eq:neq4sgTriTri}
  \end{align}
\end{subequations}
A slight rearrangement of the presentation has been performed; we moved the reducible part of the double-box numerator, carrying an overall factor of $(D_s-2)^2$, into a separate diagram, the double triangle. (The factor of $1/2$ was absorbed into the symmetry factor.)

\subsection{\texorpdfstring{$\cN=2\otimes\cN=2$}{N=2xN=2} construction}

Following our discussion in section~\ref{sec:doublecopyreview}, the double copy of four-dimensional $\cN=2$ SYM with itself gives, in addition to the desired $\cH_{\cN=4}$ graviton multiplet, two $\cV_{\cN=4}$ vector multiplets~\cite{Carrasco:2012ca,Chiodaroli:2013upa,Ochirov:2013xba}:
\begin{align}\label{eq:n2n2doublecopy}
  \cH_{\cN=4}\oplus2\cV_{\cN=4}=\cV_{\cN=2}\otimes\cV_{\cN=2}.
\end{align}
These extra $\cV_{\cN=4}$ multiplets can be removed by including fundamental matter hypermultiplets $\Phi_{\cN=2}$ and $\widebar{\Phi}_{\cN=2}$:
\begin{align}
  \cV_{\cN=4}=\Phi_{\cN=2}\otimes\widebar{\Phi}_{\cN=2}=\widebar{\Phi}_{\cN=2}\otimes\Phi_{\cN=2},
\end{align}
and promoting them to ghosts.  More generally, by using $N_f$ hypermultiplet flavors in one of the gauge-theory copies (and $N_f=1$ for the other copy), i.e. considering $\cN=2$ SQCD, this construction is generalizable to $N_V=2(1+N_f)$ vector multiplets.

Dimensional regularization requires us to construct the loop integrands in $D>4$ dimensions. It is therefore convenient to consider a six-dimensional uplift of the four-dimensional double copy~(\ref{eq:n2n2doublecopy}):
\begin{align}
  \cH_{\cN=(1,1)}=\cV_{\cN=(1,0)}\otimes\cV_{\cN=(0,1)},
\end{align}
where $\cH_{\cN=(1,1)}$ is the multiplet of $\cN=(1,1)$ supergravity (its on-shell particle content is described in ref.~\cite{deWit:2002vz}).  The precise form of $\cH_{\cN=(1,1)}$ will not concern us here; it is enough for us to know that, upon dimensional reduction to four dimensions, it gives the desired $\cH_{\cN=4}$ and $\cV_{\cN=4}$ multiplets.  A similar six-dimensional interpretation works for the double copy of hypermultiplets:
\begin{align}
  \cV_{\cN=(1,1)}
  =\Phi_{\cN=(1,0)}\otimes\widebar{\Phi}_{\cN=(0,1)}
  =\widebar{\Phi}_{\cN=(1,0)}\otimes\Phi_{\cN=(0,1)},
\end{align}
where we gave the details of $\Phi_{\cN=(1,0)}$ and $\widebar{\Phi}_{\cN=(1,0)}$ in \eqn{eq:6dhypers}.

It is well known that the uplift of $\cN=4$ supergravity to six dimensions is not unique. Besides the $\cN=(1,1)$ theory, there exists the chiral $\cN=(2,0)$ supergravity  (see ref.~\cite{deWit:2002vz} for further details), which has the following state content in the factorizable double copy:
\begin{align}
  \cH_{\cN=(2,0)}\oplus \cT_{\cN=(2,0)}=\cV_{\cN=(1,0)}\otimes\cV_{\cN=(1,0)},
\end{align}
The extra self-dual tensor multiplet $\cT_{\cN=(2,0)}$ can be removed (or more can be added) by exploiting that the same multiplet appears in the double copy of the matter multiplets:
\begin{align}
  \cT_{\cN=(2,0)} =\Phi_{\cN=(1,0)}\otimes\Phi_{\cN=(1,0)}.
\end{align}
By removing the bar on the second factor we emphasize that the half-hypers should transform in a pseudo-real representation in the gauge theories.  This avoids over-counting the number of tensor multiplets that are obtained in the double copy. 

In terms of the six-dimensional gauge-theory numerators, double-copying to obtain pure $\cN=(1,1)$-type supergravities is simple: multiply the $\cN=(1,0)$ SQCD numerators by those corresponding to the chiral-conjugate theory, obtained by swapping $m_i\leftrightarrow\mt_i$ (which reverses the sign of the Levi-Civita invariant $\epsilon(\mu_i, \mu_j)$).  For the $\cN=(2,0)$-type supergravities simply square the $\cN=(1,0)$ SQCD numerators.  For five-dimensional loop momenta, the two alternative double copies become identical as $m_i=\mt_i$ and $\epsilon(\mu_i, \mu_j)=0$. 

Since the external states of our numerators carry four-dimensional momenta, they can be properly identified using the four-dimensional double copy~(\ref{eq:n2n2doublecopy}).  The chiral and anti-chiral supergravity multiplets, $H_{\cN=4}$ and $\widebar{H}_{\cN=4}$, are associated with the chiral and anti-chiral $\cN=2$ multiplets, $V_{\cN=2}$ and $\widebar{V}_{\cN=2}$, respectively:
\begin{align}
  H_{\cN=4}=V_{\cN=2}\otimes V_{\cN=2}, &&
  \widebar{H}_{\cN=4}=\widebar{V}_{\cN=2}\otimes\widebar{V}_{\cN=2}.
\end{align}
The cross terms between external states $V_{\cN=2}$ and $\widebar{V}_{\cN=2}$ give the extra two $\cV_{\cN=4}$ multiplets that should be manually truncated away in the pure four-dimensional theory.  

In terms of the helicity sectors of SQCD the double copy can be performed in each sector separately since the cross-terms between the sectors should integrate to zero. This vanishing is necessary in order for the gravitational R-symmetry SU(4) to emerge out of the R-symmetry of the two gauge-theory factors SU(2)$\times$SU(2). Thus the all-chiral and MHV sectors of $\cN=4$ supergravity can be isolated by considering color-dual numerators of $\cN=2$ SQCD belonging to the all-chiral and MHV sectors respectively.  At two loops, these are the sets of numerators provided in the previous section.

\subsubsection{One loop}

In the all-chiral sector the only non-zero one-loop numerators are boxes.  The relevant $\cN=(1,1)$-type double copy gives the following half-maximal supergravity numerator in generic dimensions:
\begin{align}
  N^{[\cN=4\text{ SG}]}\bigg(\usegraph{9}{nBoxs}\bigg)
  =\left|n\bigg(\usegraph{9}{nBox}\bigg)\right|^2+(D_s-6)\left|n\bigg(\usegraph{9}{nBoxBar}\bigg)\right|^2,
  \label{allchiralbox}
\end{align}
where we have replaced $N_f$ using  $N_V=2(1+N_f)$ and $N_V=D_s-4$.  The modulus-square notation of the numerators indicates multiplication between the chiral and chiral-conjugate numerators: these are related by $m_i\leftrightarrow\mt_i$ and $\epsilon(\mu_i, \mu_j) \rightarrow -\epsilon(\mu_i, \mu_j)$ (for real external states chiral conjugation is the same as complex conjugation). Pure theories in $D=4,5,6$ dimensions can be obtained by setting $D_s=D$.  After plugging in the $\cN=2$ SQCD numerators given earlier~(\ref{eq:1lplusnumerators}) into   \eqn{allchiralbox}, and using $\mu^2=\mt m$, one recovers the box numerator~(\ref{eq:neq4sgbox}) of the $(\cN=0)\otimes(\cN=4)$ construction.  

The $\cN=(2,0)$-type double copy gives the following $\cN=(2,0)$ supergravity box numerator in six dimensions:\footnote{The Grassmann-odd parameters $\eta_i^I$ should not be literally squared in \eqn{allchiralbox2} or \eqn{allchiralbox}; instead the R-symmetry index should be shifted $\eta_i^I \rightarrow \eta_{i}^{I+4}$ in one of the copies.}
\begin{align}
  N^{[\cN=(2,0)\text{ SG}]}\bigg(\usegraph{9}{nBoxs}\bigg)
  =\left(n\bigg(\usegraph{9}{nBox}\bigg)\right)^2+(N_T-1)\left(n\bigg(\usegraph{9}{nBoxBar}\bigg)\right)^2,
  \label{allchiralbox2}
\end{align}
where the modulus square is replaced by an ordinary square and the parameter $N_T=1+2N_f$ counts the number of self-dual $\cN=(2,0)$ tensor multiplets in the six-dimensional theory (with $N_T=0$ being the pure theory). In general, for any diagram and at any loop order, we can obtain the $\cN=(2,0)$ supergravity numerators from the $\cN=(1,1)$ ones by replacing the modulus square with an ordinary square and replacing $(D_s-6)$ with $(N_T-1)$, as exemplified above.

In the MHV sector the one-loop amplitude of half-maximal gravity obtained from SQCD has already been considered by Ochirov and one of the present authors~\cite{Johansson:2014zca}.  In this case the triangle and bubble diagrams will also contribute to the amplitude.

\subsubsection{Two loops}

At two loops there are more non-zero numerators to consider (see ancillary files for the assembled supergravity numerators). For instance, the double copy of the double-box numerators gives 
\begin{align}
  &N^{[\cN=4\text{ SG}]}\bigg(\usegraph{9}{nBoxBoxs}\bigg)=\\
  &\left|n\bigg(\usegraph{9}{nBoxBox}\bigg)\!\right|^2\!\!+
  (D_s-6)\left(\left|n\bigg(\usegraph{9}{nBoxBoxBar}\bigg)\!\right|^2\!\!+
  \left|n\bigg(\usegraph{9}{nBoxBarBox}\bigg)\!\right|^2\!\!+
  \left|n\bigg(\usegraph{9}{nBoxBarBox3412}\bigg)\!\right|^2\right),\nn
\end{align}
where the last two terms are related by relabelling and reparametrization of the states and momenta. 
It is a straightforward exercise to show that, in the all-chiral sector, this matches the double-box numerator~(\ref{eq:neq4sgBoxBox}) found by the $(\cN=0)\otimes(\cN=4)$ construction: one assembles the combinations of $m_i$ and $\mt_i$ into the extra-dimensional function $F_1(\mu_1,\mu_2)$ given earlier~(\ref{eq:F1}) using $\mu_{ij}=\mt_{(i}m_{j)}$.

In the all-chiral sector, there are only two more non-zero supergravity numerators. They are found using similar double copies:
\begin{align}
  &N^{[\cN=4\text{ SG}]}\bigg(\!\!\usegraph{9}{nBoxBoxNPs}\bigg)=\\
  &\,\,\left|n\bigg(\!\!\usegraph{9}{nBoxBoxNP}\bigg)\right|^2\!\!+
  (D_s-6)\left(\left|n\bigg(\!\!\usegraph{9}{nBoxBoxBarNP}\bigg)\right|^2\!\!+
  \left|n\bigg(\!\!\usegraph{9}{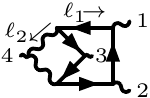}\bigg)\right|^2\!\!+
  \left|n\bigg(\!\!\usegraph{9}{nBoxBarBoxNP2}\bigg)\right|^2\right),\nn\\
  &N^{[\cN=4\text{ SG}]}\bigg(\usegraph{9}{nTriTris}\bigg)=\,\,\left|n\bigg(\usegraph{9}{nTriTri}\bigg)\!\right|^2\\
  &\qquad+(D_s-6)\left(\left|n\bigg(\usegraph{9}{nTriBarTri}\bigg)\!\right|^2\!\!+
  \left|n\bigg(\usegraph{9}{nTriBarTri3412}\bigg)\!\right|^2\right)\!+
  (D_s-6)^2\left|n\bigg(\usegraph{9}{nTriBarTriBar}\bigg)\!\right|^2,\nn
\end{align}
where, again, some of the terms are related by relabelling and reparametrization. 
In the all-chiral sector, these expressions perfectly match the $(\cN=0)\otimes(\cN=4)$ double copy given in eqs.~(\ref{eq:neq4sgBoxBoxNP}) and~(\ref{eq:neq4sgTriTri}).  

Identical formulas apply for the MHV-sector supergravity numerators, of which there are two more:
\begin{align} 
 &N^{[\cN=4\text{ SG}]}\bigg(\!\usegraph{9}{nPentaTris}\bigg)=\\
   &\left|n\bigg(\!\usegraph{9}{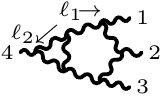}\bigg)\right|^2
   \!\!+\!(D_s-6)\left(\left|n\bigg(\!\usegraph{9}{nPentaTriBar}\bigg)\right|^2\!\!+\!
   \left|n\bigg(\!\usegraph{9}{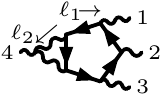}\bigg)\right|^2\!\!+\!
   \left|n\bigg(\!\usegraph{9}{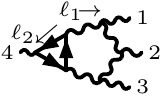}\bigg)\right|^2\right),\nn\\
   &N^{[\cN=4\text{ SG}]}\bigg(\!\usegraph{9}{nHexBubs}\bigg)=
   \left|n\bigg(\!\usegraph{9}{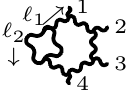}\bigg)\right|^2\\
   &\qquad+(D_s-6)\left(\left|n\bigg(\!\usegraph{9}{nHexBubBar}\bigg)\right|^2
   +\left|n\bigg(\!\usegraph{9}{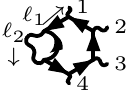}\bigg)\right|^2
   +\left|n\bigg(\!\usegraph{9}{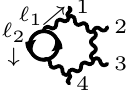}\bigg)\right|^2\right),\nn
 \end{align}
As already explained, we drop the remaining bubble-on-external-leg and tadpole diagrams since they integrate to zero in dimensional regularization (see section~\ref{sec:bubblestadpoles}). 

Pure half-maximal supergravity numerators (including $\cN=(1,1)$ but not $\cN=(2,0)$) are obtained by setting $D_s=D$.  The $\cN=(2,0)$ supergravity two-loop numerators can be obtained from the above formulas by replacing the modulus square with an ordinary square and $(D_s-6)$ with $(N_T-1)$.

\subsection{Enhanced ultraviolet cancellations in \texorpdfstring{$D=5-2\eps$}{D=5-2e} dimensions}

In $D=5-2\eps$ the two-loop pure half-maximal supergravity amplitude is known to be UV-finite for all external helicity configurations.  This was demonstrated in ref.~\cite{Bern:2012gh} as an example of an enhanced cancellation. While a potentially valid counterterm seems to exist, recent arguments confirm that half-maximal $D=5$ supergravity is finite at two loops~\cite{Bossard:2011tq,Bern:2012gh,Bossard:2012xs,Bossard:2013rza}.  Using the obtained pure supergravity amplitude~(\ref{eq:sgamp}) we confirm this enhanced UV cancelation.  In the all-chiral sector, the only UV-divergent integrals are the planar and non-planar double boxes, divergences for which may be found in ref.~\cite{Bern:2012gh}.  It is a simple exercise to show that these contributions, when substituted into the two-loop supergravity amplitude~(\ref{eq:sgamp}), give an overall cancellation.

In the MHV sector, where there is a wider range of non-trivial integrals involved, the UV calculation provides a useful check on our results.  Following the procedure outlined in ref.~\cite{Bern:2010tq}, we consider the limit of small external momenta with respect to the loop momenta $|p_i|\ll|\ell_j|$ in the integrand~(\ref{eq:sgamp}).  This is formally achieved by taking $p_i\to\delta p_i$ for a small parameter $\delta$, keeping only the leading term.  The resulting integrand is then reduced to a sum of vacuum-like integrands of the form~\cite{Grozin:2003ak}
\begin{align}
  \int\frac{\d^D\ell_1\d^D\ell_2}{\big(i\pi^{\frac{D}{2}}\big)^2}
  \frac{1}{(-\ell_1^2)^{\nu_1}(-\ell_2^2)^{\nu_2}(-(\ell_1+\ell_2+k)^2)^{\nu_3}}=
  H^D(\nu_1,\nu_2,\nu_3)(-k^2)^{D-\nu_1-\nu_2-\nu_3},
  \label{UVBubbleIntegral}
\end{align}
where we have introduced a scale $k^2$ to regulate intermediate IR divergences and
\begin{align}
  H^D(\nu_1,\nu_2,\nu_3)\equiv
  \frac{\Gamma\left(\frac{D}{2}-\nu_1\right)\Gamma\left(\frac{D}{2}-\nu_2\right)\Gamma\left(\frac{D}{2}-\nu_3\right)}{\Gamma\left(\nu_1\right)\Gamma(\nu_2)\Gamma(\nu_3)}\frac{\Gamma(\nu_1+\nu_2+\nu_3-D)}{\Gamma\left(\frac{3D}{2}-\nu_1-\nu_2-\nu_3\right)}.
\end{align}
Coefficients of these integrals depend only on the external momenta.  When $D=5-2\eps$ there are no subdivergences to subtract so the leading UV behavior in each integral is at most $\cO(\eps^{-1})$.

This requires us to eliminate all loop-momentum dependence in the numerators.  First we remove $\eps(\mu_1,\mu_2)$ using $\eps(\mu_1,\mu_2)^2=\mu_{11}\mu_{22}-\mu_{12}^2$; any odd powers integrate to zero.  All remaining loop-momentum dependence can then be converted to contractions of the form $\ell_i\cdot\ell_j$ using\footnote{This prescription also works for extra-dimensional components.  We write $\mu_{ij}=\tilde{\eta}_{\mu\nu}\ell_i^\mu\ell_j^\nu$ where $\tilde{\eta}_{\mu\nu}$ is the extra-dimensional part of the metric, $\eta_{\mu\nu}=\bar{\eta}_{\mu\nu}-\tilde{\eta}_{\mu\nu}$.}
\begin{align}
  \ell_i^{\mu}\ell_j^{\nu}&\to\eta^{\mu\nu}\frac{\ell_i\cdot\ell_j}{D}.
\end{align}
Contractions $\ell_i\cdot\ell_j$ can then be converted to inverse propagators, which shifts the $\nu_i$ indices in \eqn{UVBubbleIntegral}.  Similar reductions hold for higher tensor ranks; the relevant identities follow from Lorentz invariance. 

We present the MHV-sector UV divergences diagram by diagram. Diagrams containing tadpoles or bubbles on external legs are dropped since they vanish for dimensional reasons in gravity (the integrals evaluate to positive powers of the momentum or mass used to regulate the infrared singularities, implying that no logarithmically-dependent UV poles survive). For the remaining diagrams, considering the numerators of the first MHV solution (see appendix \ref{sec:MHV_sol1}), we obtain the following UV divergences:
\begingroup
\allowdisplaybreaks
\begin{subequations}
  \begin{align}
    &\mathcal{I}^{5-2\eps} \bigg[N^{[\cN=4\text{ SG}]}\bigg(\usegraph{9}{nBoxBoxs}\bigg)\bigg]=\\
    &\,\,\frac{1}{(4\pi)^5}\left(\frac{(2-D_s)\pi}{70\eps}(\kappa_{12}^2+\kappa_{34}^2)-
    \frac{(29D_s-142)\pi}{210\eps}(\kappa_{13}^2+\kappa_{14}^2+\kappa_{23}^2+\kappa_{24}^2)\right)+\cO(\eps^0),\nn\\
     &\mathcal{I}^{5-2\eps}\bigg[N^{[\cN=4\text{ SG}]}\bigg(\!\usegraph{9}{nPentaTris}\bigg)\bigg]=\\
    &\,\,\frac{1}{(4\pi)^5}\frac{(D_s-2)\pi}{5\eps}\left(\kappa_{12}^2+\kappa_{13}^2+\kappa_{14}^2+\kappa_{23}^2+\kappa_{24}^2+\kappa_{34}^2\right)+\cO(\eps^0),\nn\\
    &\mathcal{I}^{5-2\eps}\bigg[N^{[\cN=4\text{ SG}]}\bigg(\!\!\usegraph{9}{nBoxBoxNPs}\bigg)\bigg]=\\
    &\,\,\frac{1}{(4\pi)^5}\left(\frac{(18-23D_s)\pi}{105\eps}(\kappa_{12}^2+\kappa_{34}^2)-
    \frac{2(4+5D_s)\pi}{105\eps}(\kappa_{13}^2+\kappa_{14}^2+\kappa_{23}^2+\kappa_{24}^2)\right)+\cO(\eps^0),\nn\\
    &\mathcal{I}^{5-2\eps}\bigg[N^{[\cN=4\text{ SG}]}\bigg(\usegraph{9}{nTriTris}\bigg)\bigg]=\cO(\eps^0), \\
    &\mathcal{I}^{5-2\eps}\bigg[N^{[\cN=4\text{ SG}]}\bigg(\!\usegraph{9}{nHexBubs}\bigg)\bigg]= \nn \\
    &\,\,-\frac{1}{(4\pi)^5}\frac{(D_s-2)\pi}{6\eps}\left(\kappa_{12}^2+\kappa_{13}^2+\kappa_{14}^2+\kappa_{23}^2+\kappa_{24}^2+\kappa_{34}^2\right)+\cO(\eps^0).
  \end{align}
\end{subequations}
\endgroup
These integrals cancel among themselves when substituted into the assembled amplitude~(\ref{eq:sgamp}), and after summing over permutations of external legs.  Using the numerators of the second MHV solution (see appendix \ref{sec:MHV_sol2}), we arrive at the same vanishing result. This confirms that the UV divergence is absent in $D=5$ dimensions. 

\section{Conclusions and outlook}
\label{sec:conclusions}
 
In this paper we have computed the two-loop integrand of four-vector scattering in $\cN=2$ SQCD in a color-dual form. This provides the first example of color-dual numerators in a two-loop amplitude containing $N_f$ fundamental matter multiplets.

The calculation builds on and extends the previous one-loop work by Ochirov and one of the present authors~\cite{Johansson:2014zca}. Through the double-copy construction, the color-dual $\cN=2$ SQCD numerators can be recycled into the construction of pure and matter-coupled $\cN=4$ supergravity amplitudes.  In particular, we have considered numerators belonging to the MHV and all-chiral sectors of $\cN=2$ SQCD, where, in the latter, all external states belong to the chiral $V_{\cN=2}$ multiplet~(\ref{eq:neq2chiral}). The latter are non-zero before (and zero after) integration, but they are needed as the double copy gives non-vanishing gravity amplitudes in the all-chiral sector. Indeed, this peculiar behavior of the all-chiral sector is needed in order to give correct U(1)-anomalous amplitudes in $\cN=4$ supergravity~\cite{Carrasco:2013ypa}, while not introducing any corresponding anomalies in the $\cN=2$ gauge theory~\cite{Freedman:2017zgq}. 

We found the two-loop numerators by fitting ans\"{a}tze to physical data from generalized unitarity cuts and utilizing kinematic Jacobi relations and graph symmetries.  Furthermore, we used the fact that there is a close connection between the states of $\cN=2$ SQCD and $\cN=4$ SYM after accounting for simple differences in gauge-group representation and flavor.  For the color-dual numerators of $\cN=2$ SQCD we identified the separate contributions from the vector multiplets and the fundamental hypermultiplets.  We then constrained them by demanding that appropriate linear combinations of the numerators sum up to their known $\cN=4$ SYM counterparts.  For this to work seamlessly, we amended the kinematic numerator relations with a specific (auxiliary) two-term identity~(\ref{eq:twoTerm}) valid in the limit $N_f\rightarrow1$.  This enabled us to find unique color-dual numerators in both the MHV and all-chiral helicity sectors.

Somewhat surprisingly, the two-loop MHV solution found using the above rules includes non-zero numerators corresponding to bubble-on-external-leg and tadpole diagrams. Heuristic expectations from power counting of individual diagrams suggest that these diagrams should be absent~\cite{Bern:2012uf, Boels:2012ew, Nohle:2013bfa,Johansson:2014zca,Yang:2016ear}; however, this expectation seems not to be compatible with the two-term identity. The appearance of cubic tadpoles and external bubbles is potentially troublesome since they have singular propagators; nevertheless, these diagrams vanish after integration in the massless theories and can thus be dropped. 

Since general expectations from $\cN=2$ supersymmetry suggest that color-dual numerators should exist where the singular bubble-on-external-leg and tadpole diagrams are absent~\cite{Johansson:2014zca}, we searched and found such a solution that differs only slightly from the first one. This second solution does not obey the auxiliary two-term identity (which is optional from the point of view of color-kinematics duality) but it has somewhat better diagram-by-diagram and loop-by-loop UV power counting than the first solution. This potentially makes it better suited for studies of UV behavior in supergravity theories constructed out of double copies involving $\cN=2$ SQCD.

The need to use dimensional regularization for the amplitudes required us to define the gauge and gravity theories in $D=4-2\epsilon$ dimensions. Specifically, it is necessary to precisely know the integrand in $D>4$ dimensions: this led us to consider the six-dimensional spinor-helicity formalism~\cite{Cheung:2009dc} and the corresponding on-shell superspace~\cite{Dennen:2009vk}. In six dimensions the $\cN=2$ SQCD theory lifts to a chiral $\cN=(1,0)$ SYM theory with fundamental hypers. The chirality has important consequences: in the all-chiral-vector sector we encounter numerators depending on complexified extra-dimensional momenta, and in the MHV sector it was necessary to introduce $\eps(\mu_1,\mu_2)$ to account for a dependence on the six-dimensional Levi-Civita tensor.

However, this chiral dependence canceled upon double-coping chiral and anti-chiral sets of numerators corresponding to an $\cN=(1,0) \, \otimes \, \cN=(0,1)$ construction: we successfully obtained $D\le6$-dimensional $\cN=4$ supergravity numerators from the $\cN=(1,1)$ supergravity theory.  Alternatively, chiral $\cN=(2,0)$ supergravity amplitudes in six dimensions could be obtained from our numerators using the $\cN=(1,0)\,\otimes \, \cN=(1,0)$ double copy. In five dimensions, both constructions reduce to the same expressions corresponding to amplitudes of $D=5$ half-maximal supergravity. We explicitly calculated the UV divergences of the two-loop supergravity integrals in $D=5-2\eps$ dimensions, and showed that the $1/\epsilon$ poles non-trivially cancel out in the full amplitude.

\begin{table}
  \centering
  \begin{tabular}{l c c c}
    \toprule
    6D $\text{SYM}\otimes\text{SYM}$ &  $\cV\otimes\cV$ & \multicolumn{1}{c@{~$\oplus$}}{$(\Phi\otimes\widebar{\Phi})$}  & $(\widebar{\Phi}\otimes\Phi)$ \\
    \midrule
    \multirow{2}{*}{$\cN = (0,0)\otimes(0,0)$} & \multirow{2}{*}{${h^{ab}}_{\dot{a}\dot{b}}\oplus B^{ab}\oplus B_{\dot{a}\dot{b}} \oplus \phi$} & \multicolumn{1}{c@{~$\oplus$}}{${A^a}_{\dot{a}}$} & ${A^a}_{\dot{a}}$ \\ & & \multicolumn{1}{c@{~$\oplus$}}{$B^{ab}\oplus\phi$} & $B_{\dot{a}\dot{b}}\oplus\phi$ \\
    \midrule
    \multirow{2}{*}{$\cN = (1,0)\otimes(0,0)$} & \multirow{2}{*}{$\cH_{\cN = (1,0)}\oplus\cT_{\cN=(1,0)}$} & $\cT_{\cN=(1,0)}$ & -- \\ & & \multicolumn{1}{c@{~$\oplus$}}{$\cV_{\cN=(1,0)}$} & $\cV_{\cN=(1,0)}$ \\
    \midrule
    $\cN = (1,1)\otimes(0,0)$ & $\cH_{\cN=(1,1)}$ & \multicolumn{2}{c}{--} \\
    \midrule
    $\cN = (1,0)\otimes(0,1)$ & $\cH_{\cN=(1,1)}$ & \multicolumn{1}{c@{~$\oplus$}}{$\cV_{\cN=(1,1)}$} & $\cV_{\cN=(1,1)}$ \\
    \midrule
    $\cN = (1,0)\otimes(1,0)$ & $\cH_{\cN=(2,0)}\oplus\cT_{\cN=(2,0)}$ & $\cT_{\cN=(2,0)}$ & -- \\
    \midrule
    $\cN = (1,1)\otimes(1,0)$ & $\cH_{\cN=(2,1)}$ & \multicolumn{2}{c}{--}\\
    \midrule
    $\cN = (1,1)\otimes(1,1)$ & $\cH_{\cN=(2,2)}$ & \multicolumn{2}{c}{--}\\
    \bottomrule
  \end{tabular}  
  \caption{\small Summary of the double copy of on-shell vector and matter multiplets relevant for obtaining pure supergravities in six dimensions for various amounts of supersymmetry. The multiplets are: graviton~$\cH$, tensor~$\cT$, vector~$\cV$, and hyper/chiral $\Phi$. In the non-supersymmetric case there are two natural choices for the matter double copy, leading to either vectors or tensors and scalars; however, neither perfectly matches the matter content in the $\cV\otimes\cV$ product. For the supersymmetric cases it is always possible to find matching matter states in the $\cV\otimes\cV$ and $\Phi\otimes\widebar{\Phi}$ products, implying the latter can be used as ghosts. For the (1,0) and (2,0) supergravities, the matter double copy can be done using a half-hypermultiplet in a pseudo-real representation~\cite{Chiodaroli:2015wal} (instead of fundamental); this gives a self-dual tensor multiplet.}
  \label{tab:sixD}
\end{table}

An important aspect of the double copy when moving between dimensions is that the precise details for constructing the pure supergravities is sensitive to the dimension $D$ and chirality of the theory.  In four dimensions, we have to subtract two unwanted internal matter multiplets to arrive at pure supergravity, whereas in six dimensions the pure $\cN=(1,1)$ theory is factorizable (i.e. the double copy of $\cN=(1,0)\otimes\cN=(0,1)$ SYM exactly produces the pure theory). In contrast, pure $\cN=(2,0)$ supergravity is non-factorizable, and thus to obtain it one needs to subtract out exactly one self-dual $\cN=(2,0)$ tensor multiplet from the $\cN=(1,0) \otimes \cN=(1,0)$ double copy. This is straightforwardly done for our two-loop numerators using ghosts of type $\Phi \otimes \overline{\Phi}$ (and dropping the contributions of states $\overline{\Phi} \otimes \Phi$ to avoid overcount). Similarly, for pure $\cN=(1,0)$ supergravity one needs to subtract out a single $\cN=(1,0)$  tensor multiplet from the $\cN=(1,0) \otimes \cN=(0,0)$ double copy. Table~\ref{tab:sixD} lists the multiplet decomposition of different six-dimensional double copies. As can be seen from this table, except for the non-supersymmetric case, the strategy for obtaining pure six-dimensional supergravities is a direct generalization of the four-dimensional situation~\cite{Johansson:2014zca}: the states in the matter-antimatter tensor product can be used as ghosts to remove unwanted states in the vector-vector tensor product. 

Several extensions of this work are possible. Using the obtained integrands a next step is to complete the integration of the two-loop amplitudes in $\cN=2$ SQCD and half-maximal supergravities in various dimensions. A particularly interesting case to study is the super-conformal boundary point of SQCD, where color and flavor are balanced $N_f=2N_c$. Both leading and subleading $N_c$ contributions can be obtained from the current integrand, since all non-planar diagrams are included. Previous four-point one- and two-loop results in this theory include the works of refs.~\cite{Glover:2008tu,Dixon2008talk, Andree:2010na,Leoni:2014fja, Leoni:2015zxa}. 

Finding color-dual numerators with further-reduced supersymmetry is an important task--- this would enable calculations of two-loop pure $\cN<4$ supergravity amplitudes.  As these supergravities are not factorizable, a conventional double-copy approach involving only adjoint-representation particles is not possible. The $\cN=4$ SYM multiplet and tree amplitudes could be further decomposed into lower-supersymmetric pieces in order to provide input for the calculation (see related work~\cite{EngelundJohanssonKalin}).  An interesting challenge would be to obtain two-loop pure Einstein gravity amplitudes using the double copy, removing the unwanted dilaton and axion trough the ghost prescription of ref.~\cite{Johansson:2014zca}, in order to recalculate the UV divergence first found by Goroff and Sagnotti~\cite{Goroff:1985sz,Goroff:1985th, vandeVen:1991gw}. Recently, the role of evanescent operators for the two-loop UV divergence was clarified in ref.~\cite{Bern:2015xsa} (see also refs.~\cite{Bern:2017puu, Bern:2017tuc}); it would be interesting to examine how such contributions enter into a double-copy construction of the two-loop amplitude. 

Finally, a potentially challenging but rewarding future task is to consider the three-loop amplitudes in $\cN=2$ SQCD using the same building blocks for the numerators as used in the current work, and imposing color-kinematics duality. This calculation would be a stepping stone towards obtaining the pure $\cN<4$ supergravity amplitudes at three loops, and a natural direction to take if one wants to elucidate the UV behavior of these theories. 

\begin{acknowledgments}
  The authors would like thank Simon Badger, Marco Chiodaroli, Oluf Engelund, Alexander Ochirov, Donal O'Connell and Tiziano Peraro for useful discussions and for collaborations on related work.  The research is supported by the Swedish Research Council under grant 621-2014-5722, the Knut and 
Alice Wallenberg Foundation under grant KAW 2013.0235, and the Ragnar S\"{o}derberg Foundation under grant S1/16. 
The research of G.M. is also supported by an STFC Studentship ST/K501980/1.
\end{acknowledgments}

\appendix
\section{Two-loop MHV solutions}
\label{sec:MHV_sol}
We present two distinct versions of the color-dual four-point two-loop $\cN=2$ SQCD amplitude in the MHV sector, manifesting different properties and auxiliary constraints on the numerators. The following short-hand notations are adopted:
\begin{equation}
  \begin{aligned}
    \ell_i^s &\equiv 2 \ell_i \cdot(p_1 + p_2)\,, & \ell_i^t &\equiv 2 \ell_i \cdot(p_2 + p_3)\,, & \ell_i^u &\equiv 2 \ell_i \cdot(p_3 + p_1)\,,\\
    \ell_i &= \bar{\ell}_i + \mu_i\,, & \mu_{ij} &= - \mu_i\cdot\mu_j\,, & \ell_3 &\equiv \ell_1 + \ell_2\,,
  \end{aligned}
\end{equation}
where $p_i$ are four-dimensional external momenta and $\ell_i$ are internal $D$-dimensional loop momenta. Without loss of generality we set $N_f = 1$ in the numerators; powers of $N_f$ can be restored by counting the number of hyper loops in each diagram.
The symmetry factors~$S_i$ used in eq.~(\ref{eq:colorKinDualAmp}) are given by the number of permutations of the internal lines that leaves the graph invariant. The symmetry factor is 2 for the graphs 14, 17, 20 and 23 in fig.~\ref{fig:masterdiags}; all other graphs have symmetry factor 1.

The following two subsections list the numerators for the two MHV solutions discussed in the paper; the same expressions, together with color and symmetry factors, are also given in machine-readable ancillary files submitted to \texttt{arXiv}. We also provide a similar file containing the two-loop all-chiral solution discussed in section~\ref{sec:allplussolns}:
\begin{itemize}
\item First two-loop  MHV solution: \texttt{solMHV1.txt},
\item Second two-loop MHV solution: \texttt{solMHV2.txt},
\item All-chiral two-loop solution: \texttt{solAllChiral.txt}.
\end{itemize}

\subsection{First solution}
\label{sec:MHV_sol1}
The first MHV solution incorporates: matter-reversal symmetry (sec.~\ref{sec:matterReversalSym}), two-term identities (sec.~\ref{sec:twoTermIdentities}) and matching with the $\cN=4$ limit (sec.~\ref{sec:neq4idents}).  There are 19 non-zero numerators labelled according to fig.~\ref{fig:masterdiags}:
\begingroup
\allowdisplaybreaks 
\begin{align*}
  &
  \begin{aligned}
    \MoveEqLeft[1] n_1 \equiv \, n\bigg(\usegraph{9}{nBoxBox}\bigg) = \\
    &\big(s-2(\mu_{13}+\mu_{22})\big)(\kappa_{12}+\kappa_{34})+2i\eps(\mu_1,\mu_2)(\kappa_{12}-\kappa_{34})\\
    &
    \begin{array}{l c l}
      +\bigg[ &
        \begin{multlined}[9cm]
          u\big((\ell_2^s)^2-(\ell_2^t)^2+(\ell_2^u)^2+\ell_1^s\ell_3^s-\ell_1^t\ell_3^t+\ell_1^u\ell_3^u-\ell_3^t\ell_2^u+\ell_2^t\ell_3^u\big) \\[-12pt]
          +2s\big(\ell_1^u\ell_3^u+(\ell_2^u)^2\big) + 2 su^2-4su(\ell_1{\cdot}\ell_3+\ell_2^2+\mu_{13}+\mu_{22})
        \end{multlined}
        & \bigg]\frac{\kappa_{13}+\kappa_{24}}{2u^2}\\
      +\big[ & t\leftrightarrow u & \big]\frac{\kappa_{14}+\kappa_{23}}{2t^2}
    \end{array}\\
    &+\!\big[(\ell_1^u + 2 p_3{\cdot}\ell_2)\eps(1,2,3,\bar{\ell}_1) + (\ell_2^u + 2 p_1{\cdot}\ell_1)\eps(1,2,3,\bar{\ell}_2) + s \, \eps(1,3,\bar{\ell}_1,\bar{\ell}_2) \big]\tfrac{4i(\kappa_{13} - \kappa_{24})}{u^2}\\
    &+\!\big[\ell_3^t\eps(1,2,3,\bar{\ell}_1) + \ell_2^t \eps(1,2,3,\bar{\ell}_2) + t\,\eps(1,2,\bar{\ell}_1,\bar{\ell}_2)\big]\tfrac{4i(\kappa_{14} - \kappa_{23})}{t^2},
  \end{aligned}\\
  &
  \begin{aligned}
    \MoveEqLeft[1] n_2 \equiv n\bigg(\usegraph{9}{nBoxBoxBar}\bigg) =\\
    &-\mu_{12}(\kappa_{12} + \kappa_{34}) +i \eps(\mu_1,\mu_2) (\kappa_{12} - \kappa_{34}) \\
    &
    \begin{array}{l @{{}\big[{}} c @{{}\big]{}} l}
      + & u \big(\ell_1^s\ell_2^s+(\ell_1^u-\ell_1^t)(\ell_2^t+\ell_2^u)\big) + 2 s \big(\ell_1^u\ell_2^u - 2 u (\ell_1 {\cdot} \ell_2 + \mu_{12})\big) & \frac{\kappa_{13} + \kappa_{24}}{4u^2}\\
      + & t \leftrightarrow u &\frac{\kappa_{14} + \kappa_{23}}{4 t^2}
    \end{array}\\
    &
    \begin{array}{l @{{}\big[{}} c @{{}\big]{}} l}
      + &\ell_2^u \eps(1,2,3,\bar{\ell}_1) + u \, \eps(1,2,\bar{\ell}_1,\bar{\ell}_2) & \frac{2i(\kappa_{13} - \kappa_{24})}{u^2} \\
      + & t \leftrightarrow u & \frac{2i(\kappa_{14} - \kappa_{23})}{t^2},
    \end{array}
  \end{aligned}\\
  &
  \begin{aligned}
    \MoveEqLeft[1] n_3 \equiv n\bigg(\usegraph{9}{nBoxBarBox}\bigg) =\\
    &\mu_{13}(\kappa_{12}+\kappa_{34}) -i\eps(\mu_1,\mu_2)(\kappa_{12} - \kappa_{34})\\
    &
    \begin{array}{l @{{}\big[{}} c @{{}\big]{}} l}
      - & u\big(\ell_1^s\ell_3^s+(\ell_1^u-\ell_1^t)(\ell_3^t+\ell_3^u)\big)+2s\big(\ell_1^u\ell_3^u - 2 u(\ell_1 {\cdot} \ell_3 + \mu_{13})\big) & \frac{\kappa_{13} + \kappa_{24}}{4u^2} \\
      - & t \leftrightarrow u & \frac{\kappa_{14} + \kappa_{23}}{4 t^2}\\
    \end{array}\\
    &
    \begin{array}{l @{{}\big[{}} c @{{}\big]{}} l}
      - & \ell_3^u\eps(1,2,3,\bar{\ell}_1) + u \, \eps(1,2,\bar{\ell}_1,\bar{\ell}_2) & \frac{2i(\kappa_{13}-\kappa_{24})}{u^2}\\
      - & t \leftrightarrow u & \frac{2i(\kappa_{14}-\kappa_{23})}{t^2},
    \end{array}
  \end{aligned}\\
  &
  \begin{aligned}
    \MoveEqLeft[1] n_4 \equiv n\bigg(\!\usegraph{9}{nPentaTri}\bigg) =\\
    &
    \begin{array}{r @{{}\big[{}} c @{{}\big]{}} l}
      & \ell_1^s(\ell_2^t+\ell_2^u) - (\ell_1^t + \ell_1^u)\ell_2^s + 4 s \ell_2{\cdot}\ell_3 & \frac{\kappa_{12}+\kappa_{34}}{2s}\\
      + & s\leftrightarrow u & \frac{\kappa_{13} + \kappa_{24}}{2u}\\
      + & s\leftrightarrow t & \frac{\kappa_{14} + \kappa_{23}}{2t}\\
    \end{array}\\
    &+4i\big(\eps(3,4,\bar{\ell}_1,\bar{\ell}_2)\tfrac{\kappa_{12}-\kappa_{34}}{s} + \eps(2,4,\bar{\ell}_1,\bar{\ell}_2)\tfrac{\kappa_{13}-\kappa_{24}}{u} - \eps(1,4,\bar{\ell}_1,\bar{\ell}_2)\tfrac{\kappa_{14}-\kappa_{23}}{t}\big)\\
    &+2i\eps(\mu_1,\mu_2)(\kappa_{12} - \kappa_{34} + \kappa_{13} - \kappa_{24} - \kappa_{14} + \kappa_{23}),
  \end{aligned}\\
  &
  \begin{aligned}
    \MoveEqLeft[1] n_7 \equiv  n\bigg(\!\usegraph{9}{nPentaTriRBar}\bigg) =\\
    &
    \begin{array}{r c l}
      - \big[ & \ell_2{\cdot}\ell_3 + \frac{1}{4s}\big(\ell_1^s(\ell_2^t+\ell_2^u) - (\ell_1^t + \ell_1^u)\ell_2^s\big) & \big](\kappa_{12} + \kappa_{34})\\
      - \big[ & s \leftrightarrow u & \big](\kappa_{13} + \kappa_{24})\\
      - \big[ & s \leftrightarrow t & \big](\kappa_{14} + \kappa_{23})\\
    \end{array}\\
    &-\!\!\big[\ell_1^s\eps(1,2,3,\bar{\ell}_2) - \ell_2^s\eps(1,2,3,\bar{\ell}_1) + s \big(\eps(1,2,\bar{\ell}_1,\bar{\ell}_2) + 2 \eps(1,3,\bar{\ell}_1,\bar{\ell}_2)\big)\big]\tfrac{2i(\kappa_{12}-\kappa_{34})}{s^2}\\
    &-\!\!\big[\ell_1^t\eps(1,2,3,\bar{\ell}_2) - \ell_2^t\eps(1,2,3,\bar{\ell}_1) + t \big(\eps(2,3,\bar{\ell}_1,\bar{\ell}_2) + 2 \eps(1,3,\bar{\ell}_1,\bar{\ell}_2)\big)\big]\tfrac{2i(\kappa_{14}-\kappa_{23})}{t^2}\\
    &- \eps(2,4,\bar{\ell}_1,\bar{\ell}_2)\tfrac{2i(\kappa_{13} - \kappa_{24})}{u}\\
    &-i \eps(\mu_1,\mu_2)(\kappa_{12} - \kappa_{34} + \kappa_{13} - \kappa_{24} - \kappa_{14} + \kappa_{23}),
  \end{aligned}\\
  &
  \begin{aligned}
    \MoveEqLeft[1] n_8 \equiv n\bigg(\!\!\usegraph{9}{nBoxBoxNP}\bigg) = \\
    & \!\!\big[2s^2 + (\ell_1^t + \ell_1^u)\ell_2^s - \ell_1^s (\ell_2^t + \ell_2^u) - 4 s(\ell_2{\cdot}\ell_3 + \mu_{13} + \mu_{22}) \big]\tfrac{\kappa_{12} + \kappa_{34}}{2s}\\
    & - \eps(3,4,\bar{\ell}_1,\bar{\ell}_2) \tfrac{4i(\kappa_{12} - \kappa_{34})}{s}\\
    &
    \begin{array}{l c l}
      + \bigg[ &
        \begin{multlined}[9cm]
          u\big((\ell_1^s)^2 - (\ell_1^t)^2 + (\ell_1^u)^2 + \ell_2^s \ell_3^s - \ell_2^t \ell_3^t + \ell_2^u \ell_3^u + \ell_1^s \ell_2^u - \ell_1^u \ell_2^s\big)  \\[-12pt]
          - 4 u^2 \ell_2{\cdot}\ell_3 + 2 s\big(u^2 + \ell_1^u\ell_3^u + (\ell_2^u)^2 - 2 u(\ell_1{\cdot}\ell_3 + \ell_2^2  + \mu_{13} + \mu_{22})\big)
        \end{multlined}
        & \bigg]\frac{\kappa_{13} + \kappa_{24}}{2 u^2}\\
      + \big[ & t \leftrightarrow u & \big]\frac{\kappa_{14} + \kappa_{23}}{2 t^2}
    \end{array}\\
    &+\!\! \bigg[
      \begin{multlined}[9cm]
        (\ell_1^u + 2 p_3{\cdot}\ell_2)\eps(1,2,3,\bar{\ell}_1) +(\ell_2^u + 2 p_1{\cdot}\ell_1)\eps(1,2,3,\bar{\ell}_2) \\
        + s\, \eps(1,3,\bar{\ell}_1,\bar{\ell}_2) - u\, \eps(2,4,\bar{\ell}_1,\bar{\ell}_2) 
      \end{multlined}
      \bigg] \tfrac{4i(\kappa_{13} - \kappa_{24})}{u^2}\\
    &+\!\!\big[\ell_1^t \eps(1,2,3,\bar{\ell}_1) + \ell_3^t \eps(1,2,3,\bar{\ell}_2) - t\, \eps(2,4,\bar{\ell}_1,\bar{\ell}_2) \big] \tfrac{4i(\kappa_{14} - \kappa_{23})}{t^2}\\
    &-2i\eps(\mu_1,\mu_2)(\kappa_{13} - \kappa_{24} - \kappa_{14} + \kappa_{23}),
  \end{aligned}\\
  & n_9 \equiv  n\bigg(\!\!\usegraph{9}{nBoxBoxBarNP}\bigg) = n\bigg(\usegraph{9}{nBoxBoxBar}\bigg)  \equiv n_2, \\
  &
  \begin{aligned}
    \MoveEqLeft[1] n_{10} \equiv  n\bigg(\!\!\usegraph{9}{nBoxBarBoxNP2}\bigg) =\\
    &- \big((\ell_1^t + \ell_1^u)\ell_2^s - \ell_1^s(\ell_2^t + \ell_2^u) - 4 s(\ell_2{\cdot}\ell_3 + \mu_{23})\big)\frac{\kappa_{12} + \kappa_{34}}{4s}\\
    &
    \begin{array}{l @{{}\big[{}} c @{{}\big]{}} l}
      -&
      u(\ell_2^s\ell_3^s - \ell_2^t\ell_3^t + \ell_1^s\ell_2^u - \ell_1^u\ell_2^s) + (s-t)\ell_2^u\ell_3^u + 4 u(t\, \ell_2{\cdot}\ell_3 - s\, \mu_{23}) & \frac{\kappa_{13}+\kappa_{24}}{4 u^2}\\
      - & t \leftrightarrow u & \frac{\kappa_{14} + \kappa_{23}}{4 t^2}\\
    \end{array}\\
    &-\!\!\big[
      \ell_1^s\eps(1,2,3,\bar{\ell}_2) - \ell_2^s\eps(1,2,3,\bar{\ell}_1) + s\big(\eps(1,2,\bar{\ell}_1,\bar{\ell}_2) - 2\eps(2,3,\bar{\ell}_1,\bar{\ell}_2)\big)\big] \tfrac{2i(\kappa_{12} - \kappa_{34})}{s^2}\\
    &-\!\!\left[\tfrac{2i}{u^2}\big(\ell_2^u\eps(1,2,3,\bar{\ell}_3) + u \,\eps(2,3,\bar{\ell}_1,\bar{\ell}_2)\big) -i \eps(\mu_1,\mu_2)\right](\kappa_{13} - \kappa_{24})\\
    &-\!\!\left[\tfrac{2i}{t^2}\big(\ell_3^t\eps(1,2,3,\bar{\ell}_2) - t \,\eps(2,4,\bar{\ell}_1,\bar{\ell}_2)\big) +i \eps(\mu_1,\mu_2)\right](\kappa_{14} - \kappa_{23}),
  \end{aligned}\\
  &
  \begin{aligned}
    \MoveEqLeft[1] n_{11} \equiv n\bigg(\usegraph{9}{nTriTri}\bigg) = \\
    & - 4 \mu_{12}(\kappa_{12} + \kappa_{34}) + 4i\eps(\mu_1,\mu_2)(\kappa_{12} - \kappa_{34})\\
    &
    \begin{array}{l @{{}\big[{}} c @{{}\big]{}} l}
      + & u \big(\ell_1^s \ell_2^s + (\ell_1^u - \ell_1^t)(\ell_2^t + \ell_2^u)\big) + 2 s \big(\ell_1^u \ell_2^u - 2 u (\ell_1{\cdot}\ell_2 + \mu_{12})\big) & \frac{\kappa_{13} + \kappa_{24}}{u^2}\\
      + & t \leftrightarrow u & \frac{\kappa_{13} + \kappa_{24}}{t^2}
    \end{array}\\
    & +\!\!\bigg[
      \begin{multlined}[9cm]
        (\ell_2^u + 2 p_3{\cdot}\ell_2)\eps(1,2,3,\bar{\ell}_1) + 2 p_1{\cdot}\ell_1 \eps(1,2,3,\bar{\ell}_2) \\
        + u\, \eps(1,2,\bar{\ell}_1,\bar{\ell}_2) + s\, \eps(1,3,\bar{\ell}_1,\bar{\ell}_2)
      \end{multlined}
      \bigg] \tfrac{4i(\kappa_{13} - \kappa_{24})}{u^2}\\
    & +\!\!\bigg[
      \begin{multlined}[9cm]
        (\ell_2^t + 2 p_3{\cdot}\ell_2)\eps(1,2,3,\bar{\ell}_1) + 2 p_2{\cdot}\ell_1 \eps(1,2,3,\bar{\ell}_2) \\
        + t\, \eps(1,2,\bar{\ell}_1,\bar{\ell}_2) - s\, \eps(2,3,\bar{\ell}_1,\bar{\ell}_2)
      \end{multlined}
      \bigg] \tfrac{4i(\kappa_{14} - \kappa_{23})}{t^2},
  \end{aligned}\\
  &
  \begin{aligned}
    \MoveEqLeft[1] n_{12} \equiv  n\bigg(\usegraph{9}{nTriBarTri}\bigg) =\\
    &2 \mu_{12}(\kappa_{12} + \kappa_{34}) - 2i \eps(\mu_1,\mu_2)(\kappa_{12} - \kappa_{34})\\
    &
    \begin{array}{l @{{}\big[{}} c @{{}\big]{}} l}
      - & u\big(\ell_1^s \ell_2^s + (\ell_1^u - \ell_1^t)(\ell_2^u + \ell_2^t)\big) +2s\big(\ell_1^u\ell_2^u - 2 u(\ell_1{\cdot}\ell_2 + \mu_{12})\big) & \frac{\kappa_{13} + \kappa_{24}}{2u^2}\\
      - & t \leftrightarrow u & \frac{\kappa_{14} + \kappa_{23}}{2 t^2}\\
    \end{array}\\
    &
    \begin{array}{l @{{}\big[{}} c @{{}\big]{}} l}
      - & \ell_2^u\eps(1,2,3,\bar{\ell}_1) + u\, \eps(1,2,\bar{\ell}_1,\bar{\ell}_2) & \frac{4i(\kappa_{13} - \kappa_{24})}{u^2}\\
      - & t \leftrightarrow u & \frac{4i(\kappa_{14} - \kappa_{23})}{t^2},
    \end{array}
  \end{aligned}\\
  & n_{13} \equiv n\bigg(\usegraph{9}{nTriBarTriBar}\bigg) = n\bigg(\usegraph{9}{nBoxBoxBar}\bigg) \equiv n_2,\\
  & n_{14} \equiv n\bigg(\usegraph{9}{nHexBub}\bigg) = 4 \ell_2{\cdot}\ell_3(\kappa_{12} + \kappa_{34} + \kappa_{13} + \kappa_{24} + \kappa_{14} + \kappa_{23}),\\
  & n_{16} \equiv  n\bigg(\usegraph{9}{nHexBubRBar}\bigg) = - 2 \ell_2 {\cdot} \ell_3(\kappa_{12} + \kappa_{34} + \kappa_{13} + \kappa_{24} + \kappa_{14} + \kappa_{23}),\\
  & n_{17} \equiv n\bigg(\!\usegraph{9}{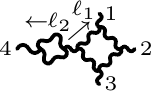}\!\bigg) = 4 \ell_2{\cdot}(\ell_2 - p_4)(\kappa_{12} + \kappa_{34} + \kappa_{13} + \kappa_{24} + \kappa_{14} + \kappa_{23}),\\
  & n_{18} \equiv n\bigg(\!\usegraph{9}{nBoxBubbleBar}\!\bigg) = 2 \ell_2{\cdot}(p_4 - \ell_2)(\kappa_{12} + \kappa_{34} + \kappa_{13} + \kappa_{24} + \kappa_{14} + \kappa_{23}),\\
  & n_{20} \equiv n\bigg(\!\!\usegraph{12}{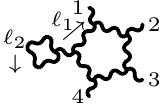}\bigg) = 8 \ell_1{\cdot}\ell_2(\kappa_{12} + \kappa_{34} + \kappa_{13} + \kappa_{24} + \kappa_{14} + \kappa_{23}),\\
  & n_{21} \equiv n\bigg(\!\!\usegraph{12}{nPentaTadBar}\bigg) = - 4 \ell_1{\cdot}\ell_2(\kappa_{12} + \kappa_{34} + \kappa_{13} + \kappa_{24} + \kappa_{14} + \kappa_{23}),\\
  & n_{23} \equiv n\bigg(\!\!\usegraph{12}{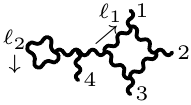}\!\bigg) = -8 p_4{\cdot}\ell_2(\kappa_{12} + \kappa_{34} + \kappa_{13} + \kappa_{24} + \kappa_{14} + \kappa_{23}),\\
  & n_{24} \equiv n\bigg(\!\!\usegraph{12}{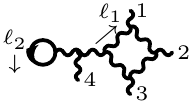}\!\bigg) = 4 \ell_2{\cdot} p_4(\kappa_{12} + \kappa_{34} + \kappa_{13} + \kappa_{24} + \kappa_{14} + \kappa_{23}).
\end{align*}
\endgroup

\makeatletter
\newcommand{\Biggg}{\bBigg@{3.5}}
\newcommand{\Bigggg}{\bBigg@{4.5}}
\makeatother

\subsection{Second solution}
\label{sec:MHV_sol2}
Another possible presentation of the two-loop color-dual $\cN=2$ SQCD amplitude in the MHV sector is found by demanding that singular diagrams corresponding to graphs with tadpoles or bubbles on external legs vanish in the on-shell limit. We keep two non-singular bubble-on-external-leg graphs (17 and 18): these numerators contain factors of $p_4^2$ that cancel the singular $1/p_4^2$ propagators coming from the external lines. This solution consists of 18 non-vanishing color-dual numerators:
\begingroup
\allowdisplaybreaks 
\begin{align*}
  &
  \begin{aligned}
    \MoveEqLeft[1] n_1 \equiv \, n\bigg(\usegraph{9}{nBoxBox}\bigg) = \\
    & (\ell_1^s-\ell_2^s-2 \ell_1^2-2\ell_2^2+s-6\mu_{13}-6\mu_{22})\tfrac{\kappa_{12}+\kappa_{34}}{3}+2 i \eps(\mu_1,\mu_2)(\kappa_{12}-\kappa_{34})\\
    &
    \begin{array}{l c l}
      +\!\Biggg[ & 
        \begin{multlined}[9cm]
          3 u\big((\ell_1^s)^2-(\ell_1^t)^2+(\ell_1^u)^2+\ell_1^s\ell_2^s+(\ell_2^s)^2\\[-12pt]
          +(\ell_1^u-\ell_1^t-\ell_2^t+\ell_2^u)(\ell_2^t+\ell_2^u)\big)
          +2u^2(\ell_1^s-\ell_2^s-2\ell_1^2-2\ell_2^2)\\
          +2s\big(u^2+ 3(\ell_1^u\ell_3^u+(\ell_2^u)^2)-6 u(\ell_1{\cdot}\ell_3+\ell_2^2+\mu_{13}+\mu_{22})\big)
        \end{multlined} 
        & \!\Biggg]\frac{\kappa_{13}+\kappa_{24}}{6u^2}\\
      + \big[& t \leftrightarrow u & \big] \frac{\kappa_{14}+\kappa_{23}}{6t^2}
    \end{array}\\
    &+\!\big[(\ell_1^u + 2 p_3{\cdot}\ell_2)\eps(1,2,3,\bar{\ell}_1) + (\ell_2^u + 2 p_1{\cdot}\ell_1)\eps(1,2,3,\bar{\ell}_2) + s \, \eps(1,3,\bar{\ell}_1,\bar{\ell}_2) \big]\tfrac{4i(\kappa_{13} - \kappa_{24})}{u^2}\\
    &+\!\big[\ell_3^t\eps(1,2,3,\bar{\ell}_1) + \ell_2^t \eps(1,2,3,\bar{\ell}_2) + t\,\eps(1,2,\bar{\ell}_1,\bar{\ell}_2)\big]\tfrac{4i(\kappa_{14} - \kappa_{23})}{t^2},
  \end{aligned}\\
  &
  \begin{aligned}
    \MoveEqLeft[1] n_2 \equiv n\bigg(\usegraph{9}{nBoxBoxBar}\bigg) =\\
    &\big(\ell_2^s-\ell_1^s+2\ell_1^2+2\ell_2^2-2s-15\mu_{12}\big)\tfrac{\kappa_{12}+\kappa_{34}}{15}+i \eps(\mu_1,\mu_2)(\kappa_{12}-\kappa_{34})\\
    &
    \begin{array}{l c l}
      +\!\Bigg[ & 
        \begin{multlined}[9cm]
          15u\big(\ell_1^s\ell_2^s+(\ell_1^u-\ell_1^t)(\ell_2^t+\ell_2^u)\big)+4u^2\big(\ell_2^s-\ell_1^s+2\ell_1^2+2\ell_2^2\big)\\[-12pt]
          +s\big(8u^2 + 30 \ell_1^u \ell_2^u - 60 u(\ell_1{\cdot}\ell_2+\mu_{12})\big)
        \end{multlined} 
        & \!\Bigg]\frac{\kappa_{13}+\kappa_{24}}{60u^2}\\
      + \big[& t \leftrightarrow u & \big] \frac{\kappa_{14}+\kappa_{23}}{60t^2}
    \end{array}\\
    &
    \begin{array}{l @{{}\big[{}} c @{{}\big]{}} l}
      + &\ell_2^u \eps(1,2,3,\bar{\ell}_1) + u \, \eps(1,2,\bar{\ell}_1,\bar{\ell}_2) & \frac{2i(\kappa_{13} - \kappa_{24})}{u^2} \\
      + & t \leftrightarrow u & \frac{2i(\kappa_{14} - \kappa_{23})}{t^2}
    \end{array},
  \end{aligned}\\
  &
  \begin{aligned}
    \MoveEqLeft[1] n_3 \equiv n\bigg(\usegraph{9}{nBoxBarBox}\bigg) =\\
    &\big(4s-5\ell_1^s-\ell_2^s+10\ell_1^2-2\ell_2^2+30\mu_{13}\big)\tfrac{\kappa_{12}+\kappa_{34}}{30} - i \eps(\mu_1,\mu_2)(\kappa_{12}-\kappa_{34})\\
    &
    \begin{array}{l c l}
      +\!\Bigg[ & 
        \begin{multlined}[9cm]
          15u\big(\ell_1^s\ell_3^s+(\ell_1^u-\ell_1^t)(\ell_3^t+\ell_3^u)\big)+2u^2(5\ell_1^s+\ell_2^s-10\ell_1^2+2\ell_2^2)\\[-12pt]
          -s\big(8u^2-30\ell_1^u\ell_3^u+60u(\ell_1{\cdot}\ell_3+\mu_{13})\big)
        \end{multlined} 
        & \!\Bigg]\frac{\kappa_{13}+\kappa_{24}}{60u^2}\\
      + \big[& t \leftrightarrow u & \big] \frac{\kappa_{14}+\kappa_{23}}{60t^2}
    \end{array}\\
    &
    \begin{array}{l @{{}\big[{}} c @{{}\big]{}} l}
      - & \ell_3^u\eps(1,2,3,\bar{\ell}_1) + u \, \eps(1,2,\bar{\ell}_1,\bar{\ell}_2) & \frac{2i(\kappa_{13}-\kappa_{24})}{u^2}\\
      - & t \leftrightarrow u & \frac{2i(\kappa_{14}-\kappa_{23})}{t^2},
    \end{array}
  \end{aligned}\\
  &
  \begin{aligned}
    \MoveEqLeft[1] n_4 \equiv n\bigg(\!\usegraph{9}{nPentaTri}\bigg) =\\
    &
    \begin{array}{l @{{}\big[{}} c @{{}\big]{}} l}
      -&3\big((\ell_1^t+\ell_1^u)\ell_2^s-\ell_1^s(\ell_2^t+\ell_2^u)\big)-4s(p_4{\cdot}\ell_1+2p_4{\cdot}\ell_2+\ell_1{\cdot}\ell_2)& \frac{\kappa_{12}+\kappa_{34}}{6s}\\
      & s \leftrightarrow u & \frac{\kappa_{13}+\kappa_{24}}{6u}\\
      & s \leftrightarrow t & \frac{\kappa_{14}+\kappa_{23}}{6t}
    \end{array}\\
    &+4i\big(\eps(3,4,\bar{\ell}_1,\bar{\ell}_2)\tfrac{\kappa_{12}-\kappa_{34}}{s} + \eps(2,4,\bar{\ell}_1,\bar{\ell}_2)\tfrac{\kappa_{13}-\kappa_{24}}{u} - \eps(1,4,\bar{\ell}_1,\bar{\ell}_2)\tfrac{\kappa_{14}-\kappa_{23}}{t}\big)\\
    &+2i\eps(\mu_1,\mu_2)(\kappa_{12} - \kappa_{34} + \kappa_{13} - \kappa_{24} - \kappa_{14} + \kappa_{23}),
  \end{aligned}\\
  & n_5 \equiv  n\bigg(\!\usegraph{9}{nPentaTriBar}\bigg) =
  \begin{multlined}[t][9cm]
    \frac{1}{15}(4\ell_1{\cdot}\ell_2+3\ell_2^2-2p_4{\cdot}\ell_1-p_4{\cdot}\ell_2)\\
    \times(\kappa_{12}+\kappa_{13}+\kappa_{14}+\kappa_{23}+\kappa_{24}+\kappa_{34}),
  \end{multlined}\\
  & n_6 \equiv n\bigg(\!\usegraph{9}{nPentaBarTri}\bigg) = \frac{1}{15}\big(p_4{\cdot}(\ell_1-\ell_2)-2\ell_1{\cdot}\ell_2\big)(\kappa_{12}+\kappa_{13}+\kappa_{14}+\kappa_{23}+\kappa_{24}+\kappa_{34}),\\
  &
  \begin{aligned}
    \MoveEqLeft[1] n_7 \equiv  n\bigg(\!\usegraph{9}{nPentaTriRBar}\bigg) =\\
    &
    \begin{array}{l @{{}\big[{}} c @{{}\big]{}} l}
      +&3\big((\ell_1^t+\ell_1^u)\ell_2^s-\ell_1^s(\ell_2^t+\ell_2^u)\big)-4s\big(p_4{\cdot}\ell_1+2p_4{\cdot}\ell_2)+\ell_1{\cdot}\ell_2\big)&\frac{\kappa_{12}+\kappa_{34}}{12s}\\
      +& s \leftrightarrow u & \frac{\kappa_{13}+\kappa_{24}}{12u}\\
      +& s \leftrightarrow t & \frac{\kappa_{14}+\kappa_{23}}{12t}
    \end{array}\\
    &-\!\!\big[\ell_1^s\eps(1,2,3,\bar{\ell}_2) - \ell_2^s\eps(1,2,3,\bar{\ell}_1) + s \big(\eps(1,2,\bar{\ell}_1,\bar{\ell}_2) + 2 \eps(1,3,\bar{\ell}_1,\bar{\ell}_2)\big)\big]\tfrac{2i(\kappa_{12}-\kappa_{34})}{s^2}\\
    &-\!\!\big[\ell_1^t\eps(1,2,3,\bar{\ell}_2) - \ell_2^t\eps(1,2,3,\bar{\ell}_1) + t \big(\eps(2,3,\bar{\ell}_1,\bar{\ell}_2) + 2 \eps(1,3,\bar{\ell}_1,\bar{\ell}_2)\big)\big]\tfrac{2i(\kappa_{14}-\kappa_{23})}{t^2}\\
    &- \eps(2,4,\bar{\ell}_1,\bar{\ell}_2)\tfrac{2i(\kappa_{13} - \kappa_{24})}{u}\\
    &-i \eps(\mu_1,\mu_2)(\kappa_{12} - \kappa_{34} + \kappa_{13} - \kappa_{24} - \kappa_{14} + \kappa_{23}),
  \end{aligned}\\
  &
  \begin{aligned}
    \MoveEqLeft[1] n_8 \equiv n\bigg(\!\!\usegraph{9}{nBoxBoxNP}\bigg) = \\
    &\left[
      \begin{multlined}[9cm]
        2s^2+3\big((\ell_1^t+\ell_1^u)\ell_2^s-\ell_1^s(\ell_2^t+\ell_2^u)\big)\\
        +2 s\big(\ell_1^s+\ell_2^t+\ell_2^u-2(p_4{\cdot}\ell_1+\ell_1{\cdot}\ell_3+\ell_2^2)-6(\mu_{13}+\mu_{22})\big)
      \end{multlined}\right]\tfrac{\kappa_{12}+\kappa{34}}{6s}\\
    &
    \begin{array}{l c l}
      +\!\Bigggg[ & 
        \begin{multlined}[9cm]
          3u\big((\ell_1^s)^2-(\ell_1^t)^2+(\ell_1^u)^2+(\ell_2^s)^2-(\ell_2^t)^2+(\ell_2^u)^2\\[-12pt]
          +\ell_1^s\ell_2^s-\ell_1^t\ell_2^t+\ell_1^u\ell_2^u+\ell_1^s\ell_2^u-\ell_1^u\ell_2^s\big)\\
          + 2u^2\big(\ell_1^s+\ell_2^t+\ell_2^u-2(p_4{\cdot}\ell_1+\ell_1{\cdot}\ell_3+\ell_2^2)\big)\\
          + 2s\big(u^2+3(\ell_1^u\ell_3^u+(\ell_2^u)^2)-6u(\ell_1{\cdot}\ell_3+\ell_2^2+\mu_{13}+\mu{22})\big)
        \end{multlined} 
        & \!\Bigggg]\frac{\kappa_{13}+\kappa_{24}}{6u^2}\\
      + \big[& t \leftrightarrow u & \big] \frac{\kappa_{14}+\kappa_{23}}{6t^2}
    \end{array}\\
    & - \eps(3,4,\bar{\ell}_1,\bar{\ell}_2) \tfrac{4i(\kappa_{12} - \kappa_{34})}{s}\\
    &+\!\! \bigg[
      \begin{multlined}[9cm]
        (\ell_1^u + 2 p_3{\cdot}\ell_2)\eps(1,2,3,\bar{\ell}_1) +(\ell_2^u + 2 p_1{\cdot}\ell_1)\eps(1,2,3,\bar{\ell}_2) \\
        + s\, \eps(1,3,\bar{\ell}_1,\bar{\ell}_2) - u\, \eps(2,4,\bar{\ell}_1,\bar{\ell}_2) 
      \end{multlined}
      \bigg] \tfrac{4i(\kappa_{13} - \kappa_{24})}{u^2}\\
    &+\!\!\big[\ell_1^t \eps(1,2,3,\bar{\ell}_1) + \ell_3^t \eps(1,2,3,\bar{\ell}_2) - t\, \eps(2,4,\bar{\ell}_1,\bar{\ell}_2) \big] \tfrac{4i(\kappa_{14} - \kappa_{23})}{t^2}\\
    &-2i\eps(\mu_1,\mu_2)(\kappa_{13} - \kappa_{24} - \kappa_{14} + \kappa_{23}),
  \end{aligned}\\
  &
  \begin{aligned}
    \MoveEqLeft[1] n_9 \equiv  n\bigg(\!\!\usegraph{9}{nBoxBoxBarNP}\bigg) = \\
    &\big(8s-6\ell_1^s-2\ell_1^t-2\ell_1^u+3\ell_2^s-\ell_2^t-\ell_2^u+8\ell_1^2-16\ell_1{\cdot}\ell_2-4\ell_2^2-60\mu_{12}\big)\tfrac{\kappa_{12}+\kappa_{34}}{60}\\
    &+i \eps(\mu_1,\mu_2)(\kappa_{12}-\kappa_{34})\\
    &
    \begin{array}{l c l}
      +\!\Biggg[ & 
        \begin{multlined}[9cm]
          15u\big(\ell_1^s\ell_2^s+(\ell_1^u-\ell_1^t)(\ell_2^t+\ell_2^u)\big)\\[-12pt]
          -u^2\big(6\ell_1^s+2\ell_1^u+2\ell_1^t-3\ell_2^s+\ell_2^t+\ell_2^u-8\ell_1^2+16\ell_1{\cdot}\ell_2+4\ell_2^2\big)\\
          +s\big(8u^2+30\ell_1^u\ell_2^u-60u(\ell_1{\cdot}\ell_2+\mu_{12})\big)
        \end{multlined}      
        & \!\Biggg]\frac{\kappa_{13}+\kappa_{24}}{60u^2}\\
      + \big[& t \leftrightarrow u & \big] \frac{\kappa_{14}+\kappa_{23}}{60t^2}
    \end{array}\\
    &
    \begin{array}{l @{{}\big[{}} c @{{}\big]{}} l}
      + &\ell_2^u \eps(1,2,3,\bar{\ell}_1) + u \, \eps(1,2,\bar{\ell}_1,\bar{\ell}_2) & \frac{2i(\kappa_{13} - \kappa_{24})}{u^2} \\
      + & t \leftrightarrow u & \frac{2i(\kappa_{14} - \kappa_{23})}{t^2},
    \end{array}
  \end{aligned}\\
  &
  \begin{aligned}
    \MoveEqLeft[1] n_{10} \equiv  n\bigg(\!\!\usegraph{9}{nBoxBarBoxNP2}\bigg) =\\
    &
    \left[
      \begin{multlined}
        15\big(\ell_1^s(\ell_2^t+\ell_2^u)-(\ell_1^t+\ell_1^u)\ell_2^s\big)\\
        +s\big(8s-3\ell_1^s-5(\ell_1^t+\ell_1^u+2(\ell_2^t+\ell_2^u))-4\ell_1^2+20\ell_2{\cdot}\ell_3+60\mu_{23}\big)
      \end{multlined}
      \right]\tfrac{\kappa_{12}+\kappa_{34}}{60s}\\
    &
    \begin{array}{l c l}
      +\!\Biggg[ & 
        \begin{multlined}[9cm]
          -15u\big(\ell_2^s(\ell_3^s-\ell_1^u)-\ell_2^t\ell_3^t+\ell_1^s\ell_2^u+\ell_2^u\ell_3^u\big)\\[-12pt]
          -u^2\big(3\ell_1^s+5(\ell_2^t+\ell_2^u+\ell_3^t+\ell_3^u)+4\ell_1^2-20\ell_2{\cdot}\ell_3\big)\\
          +s\big(8u^2-30\ell_2^u\ell_3^u+60u(\ell_2{\cdot}\ell_3+\mu_{23})\big)
        \end{multlined}      
        & \!\Biggg]\frac{\kappa_{13}+\kappa_{24}}{60u^2}\\
      + \big[& t \leftrightarrow u & \big] \frac{\kappa_{14}+\kappa_{23}}{60t^2}
    \end{array}\\
    &-\!\!\big[
      \ell_1^s\eps(1,2,3,\bar{\ell}_2) - \ell_2^s\eps(1,2,3,\bar{\ell}_1) + s\big(\eps(1,2,\bar{\ell}_1,\bar{\ell}_2) - 2\eps(2,3,\bar{\ell}_1,\bar{\ell}_2)\big)\big] \tfrac{2i(\kappa_{12} - \kappa_{34})}{s^2}\\
    &-\!\!\left[\tfrac{2i}{u^2}\big(\ell_2^u\eps(1,2,3,\bar{\ell}_3) + u \,\eps(2,3,\bar{\ell}_1,\bar{\ell}_2)\big) -i \eps(\mu_1,\mu_2)\right](\kappa_{13} - \kappa_{24})\\
    &-\!\!\left[\tfrac{2i}{t^2}\big(\ell_3^t\eps(1,2,3,\bar{\ell}_2) - t \,\eps(2,4,\bar{\ell}_1,\bar{\ell}_2)\big) +i \eps(\mu_1,\mu_2)\right](\kappa_{14} - \kappa_{23}),
  \end{aligned}\\
  &
  \begin{aligned}
    \MoveEqLeft[1] n_{11} \equiv n\bigg(\usegraph{9}{nTriTri}\bigg) = \\
    & - 4 \mu_{12}(\kappa_{12} + \kappa_{34}) + 4i\eps(\mu_1,\mu_2)(\kappa_{12} - \kappa_{34})\\
    &
    \begin{array}{l @{{}\big[{}} c @{{}\big]{}} l}
      + & u \big(\ell_1^s \ell_2^s + (\ell_1^u - \ell_1^t)(\ell_2^t + \ell_2^u)\big) + 2 s \big(\ell_1^u \ell_2^u - 2 u (\ell_1{\cdot}\ell_2 + \mu_{12})\big) & \frac{\kappa_{13} + \kappa_{24}}{u^2}\\
      + & t \leftrightarrow u & \frac{\kappa_{13} + \kappa_{24}}{t^2}
    \end{array}\\
    & +\!\!\bigg[
      \begin{multlined}[9cm]
        (\ell_2^u + 2 p_3{\cdot}\ell_2)\eps(1,2,3,\bar{\ell}_1) + 2 p_1{\cdot}\ell_1 \eps(1,2,3,\bar{\ell}_2) \\
        + u\, \eps(1,2,\bar{\ell}_1,\bar{\ell}_2) + s\, \eps(1,3,\bar{\ell}_1,\bar{\ell}_2)
      \end{multlined}
      \bigg] \tfrac{4i(\kappa_{13} - \kappa_{24})}{u^2}\\
    & +\!\!\bigg[
      \begin{multlined}[9cm]
        (\ell_2^t + 2 p_3{\cdot}\ell_2)\eps(1,2,3,\bar{\ell}_1) + 2 p_2{\cdot}\ell_1 \eps(1,2,3,\bar{\ell}_2) \\
        + t\, \eps(1,2,\bar{\ell}_1,\bar{\ell}_2) - s\, \eps(2,3,\bar{\ell}_1,\bar{\ell}_2)
      \end{multlined}\bigg] \tfrac{4i(\kappa_{14} - \kappa_{23})}{t^2},
  \end{aligned}\\
  &
  \begin{aligned}
    \MoveEqLeft[1] n_{12} \equiv  n\bigg(\usegraph{9}{nTriBarTri}\bigg) =\\
    &2 \mu_{12}(\kappa_{12} + \kappa_{34}) - 2i \eps(\mu_1,\mu_2)(\kappa_{12} - \kappa_{34})\\
    &
    \begin{array}{l @{{}\big[{}} c @{{}\big]{}} l}
      - & u\big(\ell_1^s \ell_2^s + (\ell_1^u - \ell_1^t)(\ell_2^u + \ell_2^t)\big) +2s\big(\ell_1^u\ell_2^u - 2 u(\ell_1{\cdot}\ell_2 + \mu_{12})\big) & \frac{\kappa_{13} + \kappa_{24}}{2u^2}\\
      - & t \leftrightarrow u & \frac{\kappa_{14} + \kappa_{23}}{2 t^2}\\
    \end{array}\\
    &
    \begin{array}{l @{{}\big[{}} c @{{}\big]{}} l}
      - & \ell_2^u\eps(1,2,3,\bar{\ell}_1) + u\, \eps(1,2,\bar{\ell}_1,\bar{\ell}_2) & \frac{4i(\kappa_{13} - \kappa_{24})}{u^2}\\
      - & t \leftrightarrow u & \frac{4i(\kappa_{14} - \kappa_{23})}{t^2},
    \end{array}
  \end{aligned}\\
  &
  \begin{aligned}
    \MoveEqLeft[1] n_{13} \equiv n\bigg(\usegraph{9}{nTriBarTriBar}\bigg) = \\
    &-\mu_{12}(\kappa_{12} + \kappa_{34}) +i \eps(\mu_1,\mu_2) (\kappa_{12} - \kappa_{34}) \\
    &
    \begin{array}{l @{{}\big[{}} c @{{}\big]{}} l}
      + & u \big(\ell_1^s\ell_2^s+(\ell_1^u-\ell_1^t)(\ell_2^t+\ell_2^u)\big) + 2 s \big(\ell_1^u\ell_2^u - 2 u (\ell_1 {\cdot} \ell_2 + \mu_{12})\big) & \frac{\kappa_{13} + \kappa_{24}}{4u^2}\\
      + & t \leftrightarrow u &\frac{\kappa_{14} + \kappa_{23}}{4 t^2}
    \end{array}\\
    &
    \begin{array}{l @{{}\big[{}} c @{{}\big]{}} l}
      + &\ell_2^u \eps(1,2,3,\bar{\ell}_1) + u \, \eps(1,2,\bar{\ell}_1,\bar{\ell}_2) & \frac{2i(\kappa_{13} - \kappa_{24})}{u^2} \\
      + & t \leftrightarrow u & \frac{2i(\kappa_{14} - \kappa_{23})}{t^2},
    \end{array}
  \end{aligned}\\
  & n_{14} \equiv n\bigg(\usegraph{9}{nHexBub}\bigg) = -\frac{2}{3}\ell_1^2 \, (\kappa_{12}+\kappa_{13}+\kappa_{14}+\kappa_{23}+\kappa_{24}+\kappa_{34}),\\
  & n_{15} \equiv n\bigg(\usegraph{9}{nHexBubBar}\bigg) = \frac{1}{15}(3 \ell_3^2 - \ell_1^2)(\kappa_{12}+\kappa_{13}+\kappa_{14}+\kappa_{23}+\kappa_{24}+\kappa_{34}),\\
  & n_{16} \equiv  n\bigg(\usegraph{9}{nHexBubRBar}\bigg) = \frac{1}{3}\ell_1^2(\kappa_{12}+\kappa_{13}+\kappa_{14}+\kappa_{23}+\kappa_{24}+\kappa_{34}),\\
  & n_{17} \equiv n\bigg(\!\usegraph{9}{nBoxBubble}\!\bigg) = \frac{4}{3} p_4^2 \, (\kappa_{12}+\kappa_{13}+\kappa_{14}+\kappa_{23}+\kappa_{24}+\kappa_{34}),\\
  & n_{18} \equiv n\bigg(\!\usegraph{9}{nBoxBubbleBar}\!\bigg) =-\frac{2}{3} p_4^2 \, (\kappa_{12}+\kappa_{13}+\kappa_{14}+\kappa_{23}+\kappa_{24}+\kappa_{34}).
\end{align*}
\endgroup

\subsection{Color}
The color factors corresponding to the $\cN=2$ SQCD diagrams in fig.~\ref{fig:masterdiags} are given here in terms of structure constants~$\tf^{abc}=\tr([T^a,T^b]T^c)$ and traces over generators~$T^a$, normalized as $\tr(T^aT^b)=\delta^{ab}$:
\begingroup
\allowdisplaybreaks 
\begin{equation}
  \begin{aligned}
    c_1 &= \tf^{a_1bc}\tf^{a_2bd}\tf^{a_3ef}\tf^{a_4eg}\tf^{dfh}\tf^{gch}, & c_{13} &= \tr(T^{a_1}T^{a_2}T^b)\tr(T^{a_3}T^{a_4}T^b),\\
    c_2 &= \tr(T^{a_1}T^{a_2}T^bT^{a_3}T^{a_4}T^b), & c_{14} &= \tf^{a_1bc}\tf^{a_2db}\tf^{a_3ed}\tf^{a_4fe}\tf^{fgh}\tf^{gch},\\
    c_3 &= \tf^{a_3bc}\tf^{a_4db}\tr(T^{a_1}T^{a_2}T^cT^d), & c_{15} &= \tr(T^{a_1}T^{a_2}T^{a_3}T^{a_4}T^aT^a),\\
    c_4 &= \tf^{a_1bc}\tf^{a_2bd}\tf^{a_3de}\tf^{a_4fg}\tf^{egh}\tf^{fch}, & c_{16} &= \tf^{a_1bc}\tf^{a_2db}\tf^{a_3ed}\tf^{a_4fe}\tr(T^cT^f),\\
    c_5 &= \tr(T^{a_1}T^{a_2}T^{a_3}T^bT^{a_4}T^b), & c_{17} &= \tf^{a_1bc}\tf^{a_2db}\tf^{a_3ed}\tf^{a_4fg}\tf^{ehc}\tf^{hgf},\\
    c_6 &= \tf^{a_4ab}\tr(T^{a_1}T^{a_2}T^{a_3}T^bT^a), & c_{18} &= \tf^{a_1bc}\tf^{a_2db}\tf^{a_3ed}\tf^{efc}\tr(T^{a_4}T^f),\\
    c_7 &= \tf^{a_1bc}\tf^{a_2db}\tf^{a_3ed}\tr(T^{a_4}T^cT^e), & c_{19} &= \tr(T^{a_1}T^{a_2}T^{a_3}T^b)\tr(T^{a_4}T^b),\\
    c_8 &= \tf^{a_1bc}\tf^{a_2db}\tf^{a_3ef}\tf^{a_4gh}\tf^{dhe}\tf^{gcf}, & c_{20} &= 0,\\
    c_9 &= \tf^{a_3bc}\tr(T^{a_1}T^{a_2}T^bT^{a_4}T^c), & c_{21} &= \tf^{a_1bc}\tf^{a_2db}\tf^{a_3ed}\tf^{a_4fe}\tf^{fgc}\tr(T^g),\\
    c_{10} &= \tf^{a_1bc}\tf^{a_2db}\tr(T^{a_3}T^dT^{a_4}T^c), & c_{22} &= \tr(T^{a_1}T^{a_2}T^{a_3}T^{a_4}T^b)\tr(T^b),\\
    c_{11} &= \tf^{a_1bc}\tf^{a_2db}\tf^{a_3ef}\tf^{a_4ge}\tf^{dhc}\tf^{hfg}, & c_{23} &= 0,\\
    c_{12} &= \tf^{a_3bc}\tf^{a_4db}\tf^{ecd}\tr(T^{a_1}T^{a_2}T^e), & c_{24} &= \tf^{a_1bc}\tf^{a_2db}\tf^{a_3ed}\tf^{a_4fg}\tf^{egc}\tr(T^f).
  \end{aligned}
\end{equation}
\endgroup

\bibliographystyle{JHEP}
\bibliography{references}

\end{document}